\documentclass[10pt,letterpaper]{article}
\usepackage[top=0.85in,left=2.75in,footskip=0.75in]{geometry} 

\usepackage{changepage}

\usepackage[utf8x]{inputenc}

\usepackage{textcomp,marvosym}

\usepackage{fixltx2e}

\usepackage{amsmath,amssymb}

\usepackage{cite}

\usepackage{nameref,hyperref}

\usepackage[right]{lineno}

\usepackage{color}

 \newcommand{\Kcomment}[1]{{\color{blue}{}}}
 \newcommand{\Ecomment}[1]{{\color{cyan}{}}}
 \newcommand{\Gcomment}[1]{{\color{magenta}{}}}
 \newcommand{\Mcomment}[1]{{\color{green}{}}}

\usepackage{tikz}
\usetikzlibrary{arrows}

\usetikzlibrary{decorations.markings}
\usetikzlibrary{decorations.pathmorphing}
\tikzstyle arrowstyle=[scale=1]
\tikzstyle directed=[{postaction={decorate,decoration={markings,
    mark=at position .65 with {\arrow[arrowstyle]{stealth}}}}, ultra thick}]
\tikzstyle directed_dotted=[{postaction={decorate,decoration={markings,
    mark=at position .65 with {\arrow[arrowstyle]{stealth}}}}, ultra thick,dotted}]
\tikzstyle synapse=[{decorate,decoration={snake}, ultra thick}]
\tikzstyle gray=[{circle,fill=gray,minimum size=10pt}]
\tikzstyle black=[{circle,fill=black,minimum size=10pt}]
\tikzstyle white=[{circle,draw,fill=white,minimum size=10pt}]
\tikzstyle math=[{circle,fill=none}]

\usepackage{microtype}
\DisableLigatures[f]{encoding = *, family = * }

\raggedright
\usepackage[aboveskip=1pt,labelfont=bf,labelsep=period,justification=raggedright,singlelinecheck=off]{caption}

\makeatletter
\renewcommand{\@biblabel}[1]{\quad#1.}
\makeatother

\date{}

\makeatletter
\newcommand*{\centerfloat}{%
  \parindent \z@
  \leftskip \z@ \@plus 1fil \@minus \textwidth
  \rightskip\leftskip
  \parfillskip \z@skip}
\makeatother

\begin{document}
\vspace*{0.2in}

\begin{flushleft}
{\Large
\textbf\newline{Linking structure and activity in nonlinear spiking networks} 
}
\newline
\\
Gabriel Koch Ocker \textsuperscript{1},
Kre\v{s}imir Josi\'c  \textsuperscript{2,3},
Eric Shea-Brown \textsuperscript{4,1,5},
Michael A. Buice* \textsuperscript{1,4},
\\
\bigskip
\textbf{1} Allen Institute for Brain Science, Seattle, WA, USA
\\
\textbf{2} Department of Mathematics and Department of Biology and Biochemistry, University of Houston, Houston, TX, USA
\\
\textbf{3} Department of BioSciences, Rice University, Houston, TX, USA
\\
\textbf{4} Department of Applied Mathematics, University of Washington, Seattle, WA, USA
\\
\textbf{5} Department of Physiology and Biophysics, and UW Institute of Neuroengineering, University of Washington, Seattle, WA, USA
\\
\bigskip

%
%





* michaelbu@alleninstitute.org

\end{flushleft}

\section*{Abstract}
Recent experimental advances are producing an avalanche of data on both neural connectivity and neural activity.  To take full advantage of these two emerging datasets we need a framework that links them, revealing how collective neural activity arises from the structure of neural connectivity and intrinsic neural dynamics.  This problem of {\it structure-driven activity} has drawn major interest in computational neuroscience.  Existing methods for relating activity and architecture in spiking networks rely on linearizing activity around a central operating point and thus fail to capture the nonlinear responses of individual neurons that are the hallmark of neural information processing.  Here, we overcome this limitation and present a new relationship between connectivity and activity in networks of nonlinear spiking neurons by developing a diagrammatic fluctuation expansion based on statistical field theory. We explicitly show how recurrent network structure produces pairwise and higher-order correlated activity, and how nonlinearities impact the networks' spiking activity. 
Our findings open new avenues to investigating how single-neuron nonlinearities---including those of different cell types---combine with connectivity to shape population activity and function.

\section*{Introduction}
\paragraph{}
A fundamental goal in computational neuroscience is to understand how network connectivity and intrinsic neuronal dynamics relate to collective neural activity, and  in turn drive neural computation.  Experimental advances are vastly expanding both the scale and the resolution with which we can  measure both neural connectivity and neural activity. Simultaneously, a wealth of new data suggests a possible partitioning of neurons into cell types with both distinct dynamical properties and distinct patterns of connectivity.  What is needed is a way to link these three types of data:  How is it that patterns of connectivity are translated into patterns of activity through neuronal dynamics?

\paragraph{}
Any model of neural activity should also capture the often-strong variability in spike trains across time or experimental trials. This variablity in spiking is often  coordinated, or {\it correlated}, across cells which has a variety of implications. First, correlations play an essential role in plasticity of network structure \cite{sejnowski_storing_1977, bienenstock_theory_1982, gerstner_neuronal_1996, pfister_triplets_2006}. Theories that describe spiking correlations allow for a self-consistent description of the coevolution of recurrent network structure and activity \cite{ocker_self-organization_2015, tannenbaum_shaping_2016}.
Second, correlations between synaptic inputs control their effect on postsynaptic neurons: inputs that arrive simultaneously can produce stronger responses than those arriving separately. This has been referred to as ``synergy'' or ``synchronous gain'' in early work \cite{abeles_role_1982}, and the magnitude of this synergy has been measured in the LGN by Usrey, Reppas \& Reid \cite{usrey_paired-spike_1998} and cortex by Bruno \& Sakmann \cite{bruno_cortex_2006} (but see \cite{histed_cortical_2014}).  Indeed, the level of correlation in an upstream population has been shown to act as a gain knob for firing rates downstream \cite{salinas_impact_2000}.  
Finally, correlated fluctuations in activity can impact the fidelity with which populations can encode information \cite{averbeck_neural_2006, panzeri_neural_2015}.  Importantly, the coding impact depends on a subtle interplay of how signals impact firing rates in a neural population and of how noise correlations occur across the population \cite{series_tuning_2004, josic_stimulus-dependent_2009, moreno-bote_information-limiting_2014, zylberberg_direction-selective_2016, franke_structures_2016}.  An accurate description of how network connectivity determines the individual and joint activity of neural populations is thus important for the understanding of neural activity, plasticity and coding.

\paragraph{}
Many studies of collective activity in spiking systems can be traced to the early work of Hawkes on self- or mutually-exciting point processes \cite{hawkes_spectra_1971, brillinger_estimation_1976}. 
The Hawkes model is also closely related to the linear response theory that can be used to describe correlations in integrate-and-fire networks \cite{doiron_oscillatory_2004, trousdale_impact_2012}. Here, each neuron and synapse is linearized around a central ``operating point'', and modes of collective activity are computed around that point \cite{ginzburg_theory_1994, brunel_dynamics_2000, mattia_population_2002}.
Including a nonlinear transfer of inputs to rates in the Hawkes model gives a generalized linear model, which has been applied with considerable success to multi-neuron spike train data \cite{kass_analysis_2014}. 

\paragraph{}
While analyses based on computing modes of collecting activity based on linearized dynamics have led to significant insights, they also impose a limitation.
While shifts of the operating point can modulate the linearized dynamics of biophysical models \cite{doiron_mechanics_2016}, this approach cannot capture the impact of nonlinear neural dynamics at the operating point.  

\paragraph{}
Here, we present a systematic method for computing correlations of any order for {\it nonlinear} networks of excitatory and inhibitory neurons.  
Nonlinear input-rate transfer couples higher-order spike train statistics to lower-order ones in a manner that depends on the order of the nonlinearity.  In its simplest form, this coupling shows how pairwise--correlated inputs modulate output firing rates.  This generalizes the effects of pairwise correlations on neural gain in single-neurons \cite{abeles_role_1982, salinas_impact_2000, usrey_paired-spike_1998} and feedforward circuits \cite{abeles_corticonics:_1991, diesmann_stable_1999, reyes_synchrony-dependent_2003, kumar_spiking_2010} to networks with high levels of recurrence and feedback.  

\paragraph{}
We begin with simple models and progress to nonlinearly interacting networks of spiking neurons.  Our method is {\it diagrammatic}, in the sense that the interplay of network connectivity and neural dynamics in determining network statistics is expressed and understood via a systematic series of graphical terms.  Such graphs are commonly referred to as ``Feynman diagrams" for Richard Feynman, who invented them.
We use this diagrammatic expansion to make and explain three main scientific points.  
First, we show how neural dynamics lead to spike correlations modulating firing rates in a recurrent, nonlinear network. Second, we illustrate an additional role of the prominent `heavy-tailed'' feature of neural connectivity where some neurons have many more connections than others, and some connections are much stronger than others. We show how this feature interacts with nonlinearities to control network activity.  And third, we show how different single-neuron nonlinearities affect the dependence of firing rates on correlations.

%

\section*{Results}


\section*{Diagrammatic expansion for spike train statistics}
\paragraph{}
We will show that any coupled point process model, even one with nonlinearities or negative interactions, has an associated expansion for all spike train cumulants organized by the strength of coupling of higher statistical moments with lower ones (e.g. the influence of the two-point correlation on the mean). The full model we aim to describe is one where each neuron generates a spike train which is conditionally renewal with intensity:
\begin{align}
r_i(t) = \phi_i\left(\sum_j \left(\mathbf{g}_{ij} \ast \frac{dN_j}{dt}\right)(t) + \lambda_i(t)\right)
\end{align}
Here $\mathbf{g}_{ij}(t)$ is a matrix of interaction filters, $\lambda_i(t)$ is the baseline point process drive to neuron $i$ and $\ast$ denotes convolution: $(g \ast f)(t) = \int_{t_0}^{\infty} dt' g(t - t') f(t')$ (with the integral starting at the initial time for the realization). $\phi_i$ is the transfer function of neuron $i$. Neuron $j$'s spike train is $\frac{dN_j}{dt} = \sum_k \delta(t-t_j^k)$, a sum over Dirac deltas at each of the $k$ spike times. We will take the spike trains to be conditionally Poisson given the input, so that in each time window $(t, t+dt)$, the probability of neuron $i$ generating $m$ spikes is $\left(r_i(t) dt\right)^m/m! \exp\left(-r_i(t) dt\right)$. This corresponds to a generalized linear point process model (GLM), or nonlinear multivariate Hawkes process \cite{bremaud_stability_1996}. In contrast to biophysical or integrate-and-fire models in which spike trains are generated deterministically given the membrane potential (which might, however, depend on noisy input), this model with ``escape noise" generates spike trains stochastically with a rate that depends on the ``free membrane potential" (i.e. with action potentials removed) \cite{gerstner_neuronal_2014}.  \Ecomment{Couple things ... can we refer to both GLM and Hawkes type models (first), and then the Gerstner approach?  I know this is equiv to escape noise but that is a bit specialized and I think could be confusing to a few readers if we ref to it first.  Also I got hung up on the text after the free potential, as it looks like this contains, e.g., spike generating currents, if I allow $g_{ii}$ terms, for example.}

\paragraph{}
Current methods for the analysis of single- and joint spiking statistics rely on linear response techniques: using a self-consistent mean field theory to compute mean firing rates, and then linearizing the spiking around those rates to determine the stability and correlations. We begin with a simple example highlighting the need to account for nonlinear effects. We take an excitatory-inhibitory network of $N_E = 200$ excitatory ($E$) neurons and $N_I = 40$ inhibitory ($I$) neurons, all with threshold-quadratic transfer functions $\phi_i (x) \equiv \phi(x)= \alpha \lfloor x \rfloor_+^2$. In this example we took network connectivity to be totally random (Erd\"os-R\'enyi), with connection probabilities $p_{EE}=0.2$ and $p_{EI} = p_{IE} = p_{II} = 0.5$.  For simplicity, we took the magnitude of all connections of a given type  ($E-E$, etc.)  was taken to be the same.  Furthermore, the time course of all network interactions is governed by the same filter $g(t) = \frac{t}{\tau^2}\exp{(-t/\tau)}$  (with $\tau = 10$ ms), so that $\mathbf{g}_{ij}(t) = \mathbf{W}_{ij}g(t)$.  $\mathbf{W}$ is a matrix of synaptic weights with units of mV, so that the input to $\phi$ can be interpreted as the free membrane potential. We set the strength of interactions such that the net inhibitory input weight on to a neuron was, on average, twice that of the net excitatory input weight so that for sufficiently strong interactions, the network was in an inhibitory-stabilized regime \cite{tsodyks_paradoxical_1997}.

\begin{figure}[!h]
\includegraphics[width=5in]{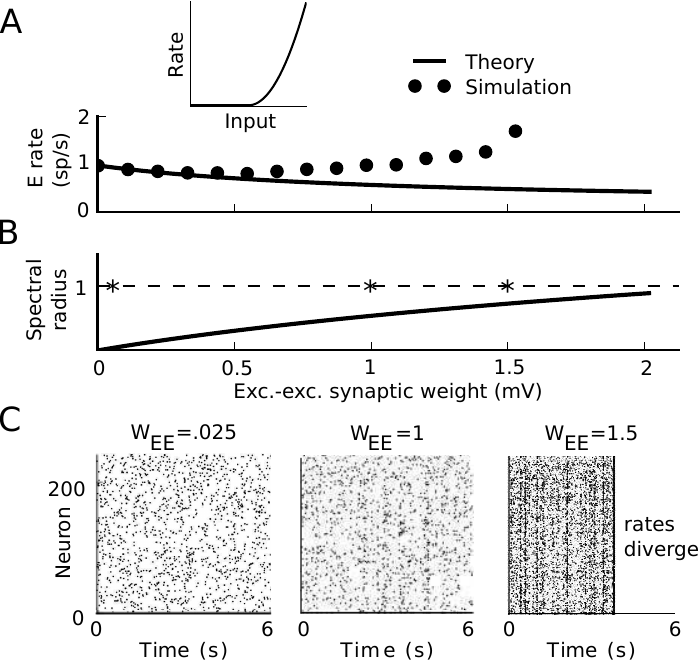} \\
\caption{{\bf Dynamics approaching the firing-rate instability in threshold-quadratic networks.}  A) Average firing rate of the excitatory neurons as synaptic weights are scaled. While the ordinate axis shows the excitatory-excitatory synaptic weight, all other weights are scaled with it. Solid lines: prediction of mean field theory. Dots: result of simulation. Inset: threshold-quadratic transfer function. B) Spectral radius of the stability matrix of mean field theory as synaptic weights are scaled. Stars indicate the weight values for the simulations below. C) Example realizations of activity for three different interaction strengths. As synapses become stronger, correlated activity becomes apparent. When synapses are strong enough the activity becomes unstable, even though the mean field theory is stable. All plotted firing rates in A) are averaged over the time period before the rates diverged (if they did). Left: $(W_{EE}, W_{EI}, W_{IE}, W_{II})$ = (.025, -.1, .01, -.1)  mV. C). Middle: $(W_{EE}, W_{EI}, W_{IE}, W_{II}) = (1, -4, .4, -4) $ mV. Right: $(W_{EE}, W_{EI}, W_{IE}, W_{II})$ = $(1.5, -6, .6, -6) $ mV.}
\label{fig:intro}
\end{figure}

\paragraph{}
We examined the magnitude and stability of firing rates as we increase the strength of synaptic coupling. We used mean-field theory to predict the firing rates, and predicted their linear stability by the standard measure of the spectral radius of the stability matrix $\mathbf{\Psi}_{ij} = \phi^{(1)}_i \mathbf{g}_{ij}$. ($\phi^{(1)}_i$ denotes the first derivative of neuron $i$'s transfer function with respect to its input.) As the strength of interactions increases, the mean field prediction for the firing rates loses accuracy (Fig. \ref{fig:intro} A). This occurs well before the mean field theory crosses the stability boundary $\left | \mathbf{\Psi} \right| = 1$ (Fig. \ref{fig:intro}B). Examining simulations as the weights are increased reveals an even more fundamental failure of the theory: before the synaptic weights are strong enough for the mean field theory to be unstable, the simulations reveal divergent firing rates (Fig. \ref{fig:intro}C; the raster stops when the instantaneous rates diverge).

\paragraph{}
Rather than restricting theoretical analysis to regimes of extremely low coupling strength or linear models, we here develop a framework for activity cumulants that can apply to models with nonlinear input-spike transfer. This will allow us to properly predict both the spiking cumulants and the location of the rate instability in the nonlinear network above. Thus, we develop a framework for activity cumulants that can apply to models with nonlinear input-spike transfer for strongly coupled networks. The mean field and linear response predictions for spiking cumulants correspond to the lowest order terms of this expansion, and are good whenever that lowest-order truncation is valid.

\paragraph{}
We will build this framework systematically: We begin with statistics of the drive $\lambda_i(t)$, then consider a filtered point process, $g \ast \lambda(t)$. In these simple models we will introduce the method and terminology which we will use to understand the more complicated models. We continue by considering the linearly self-exciting Hawkes process, taking a single neuron so $\mathbf{g} = g$ and $\phi(x) = x$, before proceeding to arbitrary nonlinearities $\phi$. Finally, we introduce an arbitrary network structure $\mathbf{g}$. This model is intimately related to the popular generalized linear models (GLMs) for spiking activity, where the nonlinearity $\phi$ is commonly taken to be exponential, refractory dynamics can be embedded in the diagonal elements of $\mathbf{g}$, and $\lambda$ corresponds to the filtered stimulus \cite{kass_analysis_2014}. The common use of GLMs it to fit them to recorded spike trains, and then ask about the structure of the inferred functional connectivity $\mathbf{g}$. In contrast, we interpret $\mathbf{g}$ as reflecting the structural connectivity and synaptic and membrane dynamics of a specified model network and aim to compute statistics of its neurons' spike trains. The derivation given here will be heuristic.  A more rigorous derivation of the expansion is given in \hyperref[methods:pathIntegral]{Methods: Path integral representation}.

\subsection*{Introduction to the general framework: Poisson process}
\paragraph{}
An inhomogeneous Poisson process generates counts within a window $dt$ independently with a Poisson distribution at rate $\lambda(t)$.  A spike train produced by this process is 
\begin{eqnarray}
	\frac{dN}{dt}(t) = \sum_k \delta (t - t_k)
\end{eqnarray} 
where $t_k$ is the $k$th spike time, and $N(t)$ is the spike count. The mean and autocovariance for this process are given by the familiar formulas:
\begin{eqnarray}
	\left \langle \frac{dN}{dt}(t) \right \rangle &=& \lambda (t) \\
	\left \langle \frac{dN}{dt}(t) \frac{dN}{dt}(t') \right \rangle_c &=& \lambda (t)\delta(t-t'),
\end{eqnarray}
where angular brackets denote the expectation across realizations and the subscript $c$ denotes a cumulant, not the moment (i.e. we have subtracted all terms which factor into products of lower moments) \cite{risken_fokker-planck_1996}. The delta function arises because the process is independent at each time step, so that there is no correlation between events from one time $t$ and any other time $t'$. In fact, because the events are generated independently at each time point, all of the cumulants of this process can be written as 
\begin{eqnarray}
	\left \langle  \prod_i \frac{dN}{dt}(t_i) \right \rangle_c &=& \int dt \lambda (t)\prod_i \delta(t_i - t)
	\label{eq:poissonCumulant}
\end{eqnarray}
where integrating out one of the delta functions puts the second cumulant in the above form. We can interpret this equation as describing a source of events appearing at rate $\lambda(t)$ at time $t$ that propagate to times $t_i$.  In this case, because the events are generated independently at each $t$, the events only propagate to the same time $t$.  For a general point process, cumulants measured at the collection of times $\left \{ t_i \right\}$ could be affected by events occurring at any past time, so that we would have to account for how events propagate across time.

\paragraph{}
The expansion for cumulants we will develop has a natural construction in terms of graphs (``Feynman diagrams"), wherein components of the graph represent factors in each term.  A set of defined rules dictate how each term in the expansion is formed from those graphs.  While this graphical representation is not necessary to understand the inhomogeneous Poisson process, we describe it in detail in this simple case to develop intuition and introduce terminology. We use cumulants in this construction because they provide the fundamental building blocks of moments; any $n$-th order moment can be constructed from up to $n$-th order cumulants). This also simplifies the developed expansion.

\paragraph{}
To begin, the $n$th cumulant has $n$ arguments $\left\{t_i, i=1, \ldots , n\right\}$, one for each factor of $\frac{dN}{dt}$ in Eq. \eqref{eq:poissonCumulant}.  We represent each of these by an open white vertex in the graph  labeled with the time point, $t_i$, and we represent the source term $\lambda(t)$ with a gray vertex.  The white vertices are called ``external" vertices whereas the gray vertex is called ``internal". 

\paragraph{}
The internal gray vertex represents the intensity of the underlying stochastic process generating events with rate $\lambda(t)$.  The white external vertices represent the spike trains whose statistics we measure at  times $\left \{ t_i \right \}$. For each delta function, $\delta (t-t_i),$ in Eq. \ref{eq:poissonCumulant}, we place an edge drawn as a dotted line from the gray vertex to the white vertex, i.e. from the event-generating process to the spike train's measurement times (Fig. \ref{fig:Poisson}). More generally, events generated by the source propagate through the graph to affect the external vertices. 

\paragraph{}
In order to construct the term associated with each diagram, we multiply the factors corresponding to edges (delta functions linking $t$ and $t_i$) or the internal vertex ($\lambda(t)$), and then integrate over the time $t$ associated with the internal vertex. This links the generation of events by $\lambda (t)$ to their joint measurement at times $\left\{ t_i \right\}$ through the propagator (here $\delta(t-t_i)$). For the diagrams shown in Fig. \ref{fig:Poisson}, these rules reproduce the cumulant terms in Eq. \ref{eq:poissonCumulant}.  Note that these graphs are directed, since we only consider causal systems where measured cumulants are only influenced by past events.

\paragraph{}
In general, a given moment will be the sum of terms associated with many different graphs.  For example the second moment is given by $\left \langle \frac{dN}{dt}(t_1) \frac{dN}{dt}(t_2) \right \rangle =\left \langle \frac{dN}{dt}(t_1) \frac{dN}{dt}(t_2) \right \rangle_c + \left \langle \frac{dN}{dt}(t_1) \right \rangle \left \langle \frac{dN}{dt}(t_2) \right \rangle$.  Each term on the right hand side will have a corresponding graph.  Moreover, the graph for the second term will include two disconnected components, one for each factor of the mean rate, which appears as in Figure~\ref{fig:Poisson}.  The graphs for the cumulant will always be described by connected graphs. 
\begin{figure}
\begin{center}
\begin{tikzpicture}

	\node[math](1) at (-8,2) {$\left \langle \frac{dN}{dt}(t)  \right \rangle$};
	\node[math](2) at (-8,0) {$\left \langle \frac{dN}{dt}(t_1) \frac{dN}{dt}(t_2) \right \rangle_c$};
	\node[math](3) at (-8,-3) {$\left \langle \frac{dN}{dt}(t_1) \frac{dN}{dt}(t_2) \frac{dN}{dt}(t_3) \right \rangle_c$};
	
	\node[math](4) at (-6,2) {$=$};
	\node[math](5) at (-6,0) {$=$};
	\node[math](6) at (-6,-3) {$=$};
	
	\node[white](7) at (-5,2) {$t$};
	\node[gray](8)  at (-3,2) {$t'$};
	\node[math](p) at (-4, 2.5) {$\delta(t-t')$};
	\node[math](p) at (-2.25, 2) {$\lambda (t')$};
	
	\draw[directed_dotted] (8)--(7);
	
	\node[white](9) at (-5,.5) {$t_1$};
	\node[white](10) at (-5,-.5) {$t_2$};
	\node[gray](11) at (-3,0) {$t'$};
	
	\draw[directed_dotted] (11)--(9);
	\draw[directed_dotted] (11)--(10);
	
	\node[white](12) at (-5,-2) {$t_1$};
	\node[white](13) at (-5,-3) {$t_2$};
	\node[white](14) at (-5,-4) {$t_3$};
	\node[gray](15) at (-3,-3) {$t'$};
	
	\draw[directed_dotted] (15)--(12);
	\draw[directed_dotted] (15)--(13);
	\draw[directed_dotted] (15)--(14);

\end{tikzpicture}
\end{center}
\caption{Feynman diagrams for the first three cumulants of the inhomogeneous Poisson process. Each dotted edge corresponds to a delta-function connecting the time indices of its two nodes. White nodes denote the measurement times, while gray nodes denote the times at which spikes are generated. The cumulants are constructed by convolving the term corresponding to the gray node with the product of all outgoing edges' terms (Eq. \eqref{eq:poissonCumulant}).}
\label{fig:Poisson}
\end{figure}

\subsection*{Filtered Poisson process}
\paragraph{}
We proceed to a simple model of synaptic input: a presynaptic Poisson spike train, with count $N(t)$ and intensity $\lambda(t)$, drives postsynaptic potentials with shape $g(t)$:
\begin{align}
	\nu(t)= \epsilon \left(g \ast \frac{dN}{dt}\right)(t), 
\end{align}
where $\ast$ denotes convolution: $(g \ast f)(t) = \int_{t_0}^{\infty} dt' g(t - t') f(t')$ (with the integral starting at the initial time for the realization). We assume $g$ is normalized, $\int_{-\infty}^\infty g(t) dt = 1$, so that $\epsilon$ gives the magnitude of the filtering.

\paragraph{}
The cumulants of the postsynaptic potential $\nu(t)$ can be calculated directly. 
In general, they are given by:
\begin{align}
	\left \langle  \prod_i \nu(t_i) \right \rangle_c &= \int dt \lambda (t) \prod_i \epsilon \left(g \ast \delta \right)(t_i-t)
	\label{eq:filtPoissonCumulant}
\end{align}
where the input spikes are generated at times $t$, arrive at times given by the delta functions and influence the cumulant measured at $\{t_i\}$ through $g$. Eq. \ref{eq:filtPoissonCumulant} is the same as that for the inhomogeneous Poisson process but with factors of $g\ast$.
This provides a simple interpretation of Eq. \ref{eq:filtPoissonCumulant}: cumulants of the filtered Poisson process are given by taking the cumulants of the underlying Poisson process and examining how they can be filtered through the system at hand.



\paragraph{}
Similarly to the case for the Poisson process, we can represent the cumulants graphically. We again represent each measurement time on the left hand side of Eq. \eqref{eq:filtPoissonCumulant} by an external vertex (Fig. \ref{fig:filtPoisson}a). The convolution of $\delta$ and $g$ in Eq.~\ref{eq:filtPoissonCumulant} corresponds to an internal time point which we integrate over (denoted by primes). We also represent these internal time points with white vertices that carry a factor of $\epsilon$, the magnitude of the filter. We represent the source term $\lambda(t)$ with a gray vertex.  All vertices that are not ``external" are called internal. Every internal vertex also carries its own time index, $t'$. 

\paragraph{}
The internal gray vertex again represents the intensity of the underlying stochastic process, $\lambda(t)$. The white external vertices represent the processes whose statistics we measure at  times $\left \{ t_i \right \}$. For each delta function, $\delta (t'-t),$ in Eq. \ref{eq:filtPoissonCumulant}, we place an edge drawn as a dotted line from the gray vertex to the white vertex, \emph{i.e.} from the event-generating process to the arrival time of the event $t'$. In this example an event ``arrives'' as soon as it is generated.
A wavy edge corresponds to the filter, $g$, and represents the effect of a spike arriving at time $t'$ on the output process measured at time $t_i$ (Fig. \ref{fig:filtPoisson}b). Events generated by the source thus propagate through the graph to affect the observed, external vertices. 

\paragraph{}
In order to construct the expression associated with each diagram, we again multiply the factors corresponding to each edge in the diagram (e.g. $\delta(t'-t)$ or $g(t_i-t')$) or internal vertex ($\epsilon$ or $\lambda(t)$), and then integrate over the times associated with the internal vertices. Note that integration over the internal times $t'$, $t''$, etc, results in the convolutions $\epsilon (g \ast \delta)(t_i - t)$. Integration over the time $t$ associated with the source term corresponds to the outermost integral in Eq.~\eqref{eq:filtPoissonCumulant} 
This links the generation of events by $\lambda (t)$ to their joint measurement at times $\left\{ t_i \right\}$ through their arrival times (via $\delta(t-t')$), and temporal filtering ($g(t_i-t')$). For the diagrams shown in Fig. \ref{fig:filtPoisson}, these rules reproduce the cumulant terms in Eq. \ref{eq:filtPoissonCumulant}.  Note that the graphs are directed, as for the expansion we describe the ``propagator" term will be causal. 

\paragraph{}
We can simplify the cumulants of this process (and the corresponding diagrammatic representations) by considering the propagator of $\nu(t)$ (also known as the linear response or impulse response). The propagator measures the change in $\left \langle \nu(t) \right \rangle$ in response to the addition of one input spike in $N(t)$. We can compute it by taking a functional derivative with respect to the input intensity $\lambda(t)$:
\begin{align}
	\Delta(t, t') &= \frac{\delta}{\delta \lambda(t')} \left \langle \nu(t) \right \rangle \nonumber \\
	&= \frac{\delta}{\delta \lambda(t')} \left( \epsilon \int_{t_0}^\infty dt'' \, g(t-t'') \left \langle \frac{dN(t'')}{dt}\right \rangle \right) \nonumber \\
	&= \epsilon \int_{t_0}^\infty dt'' \, g(t-t'') \frac{\delta \lambda(t'')}{\delta \lambda(t')} \nonumber \\
	&= \epsilon (g \ast \delta)(t-t')
	\label{eq:filtPoissonProp}
\end{align}

Since the dynamics are linear, this is also equivalent to the change of the expected rate with the addition of one spike to the input, \emph{i.e.} taking $\lambda(t) \leftarrow \lambda(t) + \delta(t'-t)$ and $\left \langle \nu(t)\right \rangle_c \leftarrow \left \langle \nu(t)\right \rangle_c + \Delta \ast \delta(t)$ (or equivalently the Green's function of the expected rate). This allows us to rewrite the cumulants in terms of the input rate and the propagator:
\begin{align}
	\left \langle  \prod_i \nu(t_i) \right \rangle_c &= \int dt \lambda (t)\prod_i \Delta(t_i, t)
	\label{eq:filtPoissonCumulant_prop}
\end{align}
which can be represented graphically by introducing a solid, directed edge for $\Delta(t, t')$ (Fig. \ref{fig:filtPoisson}c). The propagator will be a central feature of the expansion for cumulants in more complicated models involving connections among neurons.  

\Kcomment{The need for a propagator here is somewhat mysterious.  A somewhat artificial example to motivate 
is to think of a source neuron firing at rate $\lambda(t)$, and making multiple synaptic contacts with a post-synaptic neuron with 
spikes arriving with different latencies. Maybe just worth adding as a discussion point?}

\begin{figure}
\begin{center}
\begin{tikzpicture}

  \tikzstyle{internalVertex}=[circle,fill=gray,minimum size=10pt,inner sep=0pt]
  \tikzstyle{externalVertex}=[circle,draw, fill=white,minimum size=10pt,inner sep=0pt]
 \tikzstyle{plus}=[circle,minimum size=0pt,inner sep=0pt]

	\node[math](p) at (-9, 3) {$\mathbf{A)}$};
	\node[math](1) at (-8,2) {$\left \langle \frac{dM}{dt} (t) \right \rangle$};
	\node[math](2) at (-8,0) {$\left \langle \frac{dM}{dt}(t_1) \frac{dM}{dt}(t_2) \right \rangle_c$};
	\node[math](3) at (-8,-3) {$\left \langle \frac{dM}{dt}(t_1) \frac{dM}{dt}(t_2) \frac{dM}{dt}(t_3) \right \rangle_c$};
	
	\node[math](4) at (-6,2) {$=$};
	\node[math](5) at (-6,0) {$=$};
	\node[math](6) at (-6,-3) {$=$};
	
	\node[math](p) at (-5.5, 3) {$\mathbf{B)}$};
	\node[white](7a) at (-5, 2) {$t$};
	\node[white](7) at (-3,2) {$t'$};
	\node[gray](8)  at (-1,2) {$t''$};
	\node[math](p) at (-2, 2.5) {$\delta(t'-t'')$};
	\node[math](p) at (-1, 1.45) {$\lambda (t'')$};
	\node[math](p) at (-4, 2.5) {$g(t-t')$};
	\node[math]([) at (-3, 1.45) {$\epsilon$};
	
	\draw[directed_dotted] (8)--(7);
	\draw[synapse] (7)--(7a);
	
	\node[math] (p) at (0.5, 3) {$\mathbf{C)}$};
	\node[math] (p) at (0, 2) {$=$};
	\node[white] (7aa) at (1, 2) {$t$};
	\node[gray] (8aa) at (3, 2) {$t'$};
	\draw[directed] (8aa)--(7aa);
	\node[math] (p) at (2, 2.5) {$\Delta(t - t')$};
	
	\node[white](9a) at (-5,.5) {$t_1$};
	\node[white](10a) at (-5, -.5) {$t_2$};
	\node[white](9) at (-3,.5) {$t'$};
	\node[white](10) at (-3,-.5) {$t''$};
	\node[gray](11) at (-1,0) {$t'''$};
	
	\draw[synapse] (9)--(9a);
	\draw[synapse] (10)--(10a);
	\draw[directed_dotted] (11)--(9);
	\draw[directed_dotted] (11)--(10);
	
	\node[math] (p) at (0, 0) {$=$};
	\node[white] (9aa) at (1, .5) {$t_1$};
	\node[white] (10aa) at (1, -.5) {$t_2$};
	\node[gray] (11aa) at (3, 0) {$t'$};
	\draw[directed] (11aa)--(9aa);
	\draw[directed] (11aa)--(10aa);
	
	\node[white](12a) at (-5, -2) {$t_1$};
	\node[white](13a) at (-5, -3) {$t_2$};
	\node[white](14a) at (-5, -4) {$t_3$};
	\node[white](12) at (-3,-2) {$t'$};
	\node[white](13) at (-3,-3) {$t''$};
	\node[white](14) at (-3,-4) {$t'''$};
	\node[gray](15) at (-1,-3) {$t''''$};
	
	\draw[synapse] (12)--(12a);
	\draw[synapse] (13)--(13a);
	\draw[synapse] (14)--(14a);
	\draw[directed_dotted] (15)--(12);
	\draw[directed_dotted] (15)--(13);
	\draw[directed_dotted] (15)--(14);

	\node[math] (p) at (0,-3) {$=$};
	\node[white] (12aa) at (1, -2) {$t_1$};
	\node[white] (13aa) at (1, -3) {$t_2$};
	\node[white] (14aa) at (1, -4) {$t_3$};
	\node[gray] (15aa) at (3, -3) {$t'$};
	\draw[directed] (15aa)--(12aa);
	\draw[directed] (15aa)--(13aa);
	\draw[directed] (15aa)--(14aa);

\end{tikzpicture}
\end{center}
\caption{Feynman diagrams for the first three cumulants of the filtered inhomogeneous Poisson process. A) Cumulant corresponding to the graph. B) Diagrammatic expressions using the filter and the underlying Poisson process. Each dotted edge corresponds to a delta-function connecting the time indices of its two nodes. Each wavy edge corresponds to the filter $g$ connecting the time indices of its two nodes. C) Diagrammatic expressions using the propagator. In all graphs, external white nodes (leaves of the graph) denote measurement times. Gray nodes denote the times at which spikes are generated in the input spike train. Internal white nodes (with time indices $t'$) denote the times at which input spikes arrive at the postsynaptic neuron. The cumulants are constructed by convolving the term corresponding to the gray node with the product of all outgoing edges' terms (Eq. \eqref{eq:filtPoissonCumulant}). }
\label{fig:filtPoisson}
\end{figure}

\subsection*{Impact of self-excitation on activity statistics of any order: linearly self-exciting process}
\paragraph{}
In order to generalize the graphical representation of Poisson cumulants, we begin with a linearly self-exciting process as considered by Hawkes \cite{hawkes_spectra_1971}. Let the rate be a linear function of the instantaneous event rate (that is to say the firing rate conditioned on a particular realization of the event history)
\begin{align}
	r(t) = \epsilon \left(g \ast \frac{dN}{dt}\right)(t) + \lambda(t)  \label{eq:selfExciteRate}.
\end{align}
We assume that $g(\tau)$ and $\lambda(t)$ are such that $r(t) > 0$, and $\int_{-\infty}^\infty d\tau g(\tau) = 1$. If $\epsilon < 1$, then an event will generate less than one event on average, and the rate will not diverge. The history dependence of the firing rate will now enter into our calculations.  
We can compute the expected rate using the self-consistency equation:
\begin{align}
	\bar{r}(t) \equiv \left \langle \frac{dN}{dt}(t) \right \rangle =  \epsilon \left(g \ast \left \langle \frac{dN}{dt} \right \rangle \right)(t)  + \lambda(t)  = \epsilon (g \ast \bar{r})(t) + \lambda(t) \label{eq:selfExciteRateExp}
\end{align}

\paragraph{}
We provide an alternate derivation of this result that will prove useful below: We construct a perturbative expansion of  the mean firing rate and show how this expansion can be re-summed to yield the full rate of the self-exciting process. This procedure can also be applied to obtain cumulants of arbitrary order for this process. 

\paragraph{}
We will begin with a recursive formulation of the self-exciting process. In contrast to the filtered Poisson process of the previous section, here the process with count $N$ generates events, which then influence its own rate, $dN/dt.$ Each event can thus generate others in turn. In the case of a linear filter, $g$, the following approach \Kcomment{I changed `this' to `the following approach'} is equivalent to the Poisson cluster expansion \cite{hawkes_cluster_1974, saichev_generating_2011, jovanovic_cumulants_2015} and similar to the construction of previous linear response theories for spike train covariances \cite{trousdale_impact_2012}. 
Define the $n$th order self-exciting process, $N_n(t)$, to be the inhomogeneous Poisson process given by:
\begin{align}
	 \frac{dN_n(t)}{dt}= \frac{dN_0(t)}{dt} + \frac{dM_{n-1}(t)}{dt},
\label{eq:selfExciteRateRec}
\end{align}
where 
$N_0(t)$ and $M_n(t)$ are inhomogeneous Poisson processes with rates $\lambda(t)$ and
 $\nu_n(t),$ respectively, where
\begin{align}
	\nu_n(t) =  \epsilon \left(g \ast \frac{dN_{n}}{dt}\right)(t).
\label{eq:selfExciteNu}
\end{align}
, so $\nu_0(t) = \epsilon \left(g \ast \frac{dN_0}{dt}\right)(t)$. $M_n(t)$ is a process with \emph{intensity} that depends on a stochastic realization of $N_n(t)$; $M_0(t)$ is a ``doubly-stochastic" process. We can generate these processes recursively: To generate  $N_n(t)$, we use a realization of $N_{n-1}(t)$ to compute the rate $\nu_{n-1}$ and generate a realization of  $M_{n-1}(t)$.  These are added to events generated from the original inhomogeneous Poisson process with rate $\lambda(t)$ to produce $N_n(t)$. We can use this recursive procedure to develop an expansion for the cumulants of the process at a given order in $\epsilon$ (thus a given order in the self-convolution of $g$). 

\paragraph{}
Let us compute the value of $\left \langle \frac{dN}{dt}(t) \right \rangle$ in powers of $\epsilon$ using our recursive approach.  
The zeroth order solution, $\left \langle \frac{dN_0}{dt}(t) \right \rangle$, is the rate of the inhomogeneous Poisson process $\lambda(t)$. At order $n$, we compute $\left \langle \frac{dN_n}{dt}(t) \right \rangle$ using the $(n-1)$st order solution in the right hand side of Eq.~(\ref{eq:selfExciteRateRec}).  At first order, using the Poisson solution for $ \langle \frac{dN_0(t)}{dt} \rangle$ we get 
\begin{align}
	\left \langle \frac{dN_1}{dt}(t) \right \rangle &= \left \langle \frac{dN_0}{dt}(t) \right \rangle + \left \langle \frac{dM_0}{dt}(t) \right \rangle  \\
	&= \lambda(t) + \epsilon \left(g \ast \left \langle \frac{dN_0}{dt} \right \rangle\right)(t)  \\
	&= \lambda(t) + \epsilon	(g \ast \lambda)(t) \\
 	& = \lambda(t) + \epsilon \int_{t_0}^\infty g(t-t') \lambda(t')dt'
\end{align}
At second order we similarly arrive at
\begin{align}
	\left \langle \frac{dN_2}{dt}(t) \right \rangle &= \left \langle \frac{dN_0}{dt}(t) \right \rangle + \left \langle \frac{dM_1}{dt}(t) \right \rangle  \\
	&= \lambda(t) + \epsilon \left(g \ast \left \langle \frac{dN_1}{dt} \right \rangle\right)(t)  \\
	& = \lambda(t) + \epsilon \int_{t_0}^\infty dt' g(t-t') \lambda(t') + \epsilon^2 \int_{- \infty}^\infty dt' g(t-t') \int_{t_0}^\infty dt'' g(t' - t'') \lambda(t'')
\end{align}
At higher orders we would obtain further terms with additional convolutions with $g$.  

\paragraph{}
It will be useful to write these expansions in another way, which will allow their form to generalize to non-linear processes: we will construct the cumulants from the baseline rate and the propagator. \Kcomment{Maybe want to justify this -- ie here the propagator is trivial, but a spike may arrive after some delay say -- see my comment above.} We can always replace
\begin{eqnarray}
	\lambda(t) = \int_{t_0}^\infty dt' \delta(t-t')\lambda(t')
\end{eqnarray}
resulting in
\begin{align}
	\left \langle \frac{dN_1}{dt}(t) \right \rangle 	& = \int_{t_0}^\infty dt' \delta(t-t')\lambda(t') + \epsilon \int_{t_0}^\infty dt' g(t-t') \int_{t_0}^\infty dt'' \delta(t'-t'')\lambda(t'') \label{eq:meanSelfExciteExpansion}
\end{align}

\paragraph{}
Fig. \ref{fig:selfExciteMeanFeynman}a shows the graphical representation of this expansion.  As before, the order of the moment is given by the number of external vertices and each external vertex carries a measurement time $t_i$. We have three types of internal vertices: two open white vertices that carry factors of $\epsilon$ (one type has one wavy incoming and one wavy outgoing line - the other has one incoming dotted line and one wavy outgoing line) and one gray vertex (that has one outgoing dotted line). As before, each gray internal vertex corresponds to the source term, and thus represents the factor $\lambda(t)$.   The white internal vertices, and their edges represent how the events generated by the source are propagated through the filter $g$. Each white vertex corresponds to a possible past event time, $t'$. To construct the cumulant corresponding to a diagram we integrate over all these possible internal times, weighting each by their influence on the current spiking rate. These weights are given by the filters, $g$, represented by the wavy edges. The graphical representation of $\left \langle \frac{dN_1}{dt}(t) \right \rangle$ (using the delta function as in Eq. \ref{eq:meanSelfExciteExpansion}) is shown in Figure \ref{fig:selfExciteMeanFeynman}a.


\paragraph{}
We can compute the firing rate of the self-exciting process $\bar{r}(t)$ as the limit of the $n$th order self-exciting processes, continuing the process outlined for Eq. \ref{eq:selfExciteRateRec}: 
\begin{align}
\bar{r}(t) &= \sum_{n=0}^\infty \epsilon^{n} (g^{(n)} * \delta * \lambda)(t),
\end{align}
where $g^{(n)}$ is the $n$-fold convolution of $g$ with itself and $g^{(0)}(t) = \delta(t)$.  
Indeed, we can see that this expression for $\bar{r}(t)$ yields the same recursive self-consistency condition as above:
\begin{align}
\bar{r}(t) &= \epsilon^0(g^{(0)}\ast \delta \ast \lambda)(t) + \sum_{n=1}^\infty \epsilon^{n} (g^{(n)} * \delta * \lambda)(t) \nonumber \\
	&= \lambda(t) + \epsilon \left(g * \sum_{n=1}^\infty \epsilon^{n-1} g^{(n-1)} * \delta * \lambda \right)(t) \nonumber \\
    &= \lambda(t) + \epsilon (g * \bar{r})(t).
\label{eq:meanSelfExciteFullExpansion}
\end{align}


\paragraph{}
We can also represent this recursive relation graphically as in Fig.~\ref{fig:selfExciteMeanFeynman}b, using a black vertex to denote the mean-field rate $\bar{r}(t)$. The infinite sum defined by Eq.~\eqref{eq:selfExciteRateRec} has a specific graphical representation: Notice that the leftmost vertex and wavy line in the right-hand side of Figure \ref{fig:selfExciteMeanFeynman}b (top) can be detached and factored, with the remaining series of diagrams corresponding exactly to those of the mean. This series of subgraphs on the right hand side sums to $\left \langle \frac{dN}{dt}(t) \right \rangle$, leading to the recursion relation in Eq. \ref{eq:meanSelfExciteFullExpansion} (Fig. \ref{fig:selfExciteMeanFeynman}b). This graphical representation is equivalent to the recursion relation.

\begin{figure}
\begin{adjustwidth}{-1 in}{0in}
\begin{center}
\begin{tikzpicture}
  
  \tikzstyle{internalVertex}=[circle,fill=gray,minimum size=10pt,inner sep=0pt]
  \tikzstyle{externalVertex}=[circle,draw, fill=white,minimum size=10pt,inner sep=0pt]
 \tikzstyle{plus}=[circle,minimum size=0pt,inner sep=0pt]
 
 \node[plus](0) at (-4,1) {$\left \langle \frac{dN_1}{dt}(t) \right \rangle$};
 \node[plus](0) at (-2.85, 1) {$=$};
 \node[plus](0) at (-4,-.75) {$\left \langle \frac{dN_1}{dt}(t) \right \rangle $};
 \node[plus](0) at (-2.85, -.75) {$=$};
  
\node[plus](0a) at (-5,2) {\textbf{a)}};

\node[plus](3b) at (0,1.5) {$\lambda(t')$};
\node[gray](3) at (0,1) {};
\node[plus](1b) at (-2,2) {};
\node[white] (1) at (-2,1) {$t$};
\node[plus](1bc) at (-1,.5) {$\delta(t-t')$};
 
\node[plus](1a) at (-1,-.75) {$\int_{t_0}^\infty dt' \delta(t-t') \lambda (t') $};
\node[plus](2a) at (4.5,-.75) {$\epsilon \int_{t_0}^\infty dt'\; g(t-t') \int_{t_0}^\infty dt''\; \delta(t'- t'') \lambda (t'')$};

\node[plus](1bc) at (4,1.5) {$\epsilon$};		
\node[white](4) at (2,1) {$t$};
\node[plus](3b) at (6,1.5) {$\lambda(t'')$};
\node[plus](3b) at (4,2) {};
\node[white](6) at (4,1) {};
\node[gray](7) at (6,1) {};
 
\node[plus](1bc) at (5,.5) {$\delta(t'-t'')$};
\node[plus](1bc) at (3,.5) {$g(t-t')$};

\draw[directed_dotted] (3) -- (1);
\draw[synapse] (6) -- (4);
\draw[directed_dotted] (7) -- (6);
 	
\node[plus](p) at (1,1) {+};
\node[plus](p) at (1,-.75) {+};

\node[plus](2.1a) at (-5,-2) {\textbf{b)}};
\node[plus](2.2a) at (-4,-2.9) {$\left \langle \frac{dN}{dt}(t) \right \rangle$};
 \node[plus](2.2a) at (-2.85, -2.9) {$=$};
  
\node[internalVertex](2_3) at (0,-2.9) {};
\node[white] (2_1) at (-2,-2.9) {$t$};

\node[white](2_4) at (2,-2.9) {$t$};
\node[externalVertex](2_6) at (4,-2.9) {};
\node[internalVertex](2_6b) at (6,-2.9) {};
 
\node[white](2_7) at (2,-3.5) {$t$};
\node[externalVertex](2_9) at (4,-3.5) {}; 
\node[white](2_10) at (6,-3.5) {}; 
\node[gray](2_11) at (8,-3.5) {};

\node[plus](p) at (1,-2.9) {+};
\node[plus](p) at (1,-3.5) {+};
\node[plus](p) at (9,-3.5) {+};
\node[plus](p) at (10,-3.5) {$\cdots$};

\node[plus](p) at (-2.85, -4.25) {$=$};
\node[white](2_ext_fact) at (2,-4.25) {$t$};
\node[plus](par) at (4, -4.25) {$\Big [$};
\draw[synapse] (2_ext_fact) -- (par);
\node[plus](p) at (1, -4.25) {$+$};

\node[white](w) at (-2,-4.25) {$t$};
\node[gray](g) at (0, -4.25) {};
\draw[directed_dotted] (g) -- (w);

\node[white](2_int_white_fact) at (4.25,-4.25) {};
\node[internalVertex](2_int_gray_fact) at (6.25,-4.25) {};
\draw[directed_dotted]  (2_int_gray_fact) -- (2_int_white_fact);
\node[plus](p) at (7.25, -4.25) {+};
\node[plus](p) at (8.25, -4.25) {$\cdots \Big]$};

\draw[directed_dotted] (2_3) -- (2_1);
\draw[synapse] (2_6) -- (2_4);
\draw[directed_dotted] (2_6b) -- (2_6);
\draw[directed_dotted] (2_11) -- (2_10);
 
\draw[synapse] (2_10) -- (2_9);
\draw[synapse] (2_9) -- (2_7); 
 
\node[plus](2.2a) at (-2.85,-5) {$=$}; 
\node[internalVertex](3_3) at (0,-5) {};
\node[white] (3_1) at (-2,-5) {$t$};
 
\node[white](3_4) at (2,-5) {$t$};
\node[externalVertex,fill=black](3_6) at (4,-5) {};
 
\node[internalVertex,fill=black](2.2a) at (-4,-5) {};
 
\node[plus](p) at (1,-5) {+};
\draw[directed_dotted] (3_3) -- (3_1);
\draw[synapse] (3_6) --(3_4);

\node[plus](2.1a) at (-5,-6) {\textbf{c)}};
\node[plus](2.2a) at (-4,-7) {$\Delta(t,t') $};
\node[plus](p) at (-2.5, -7) {$=$};
 
\node[internalVertex,fill=none](2_3) at (0,-7) {};
\node[internalVertex,fill=none] (2_1) at (-2,-7) {};

\node[internalVertex,fill=none](2_4) at (2,-7) {};
\node[externalVertex](2_6) at (4,-7) {};
\node[internalVertex,fill=none](2_6b) at (6,-7) {};
 
\node[internalVertex,fill=none](2_7) at (2,-7.5) {};
\node[externalVertex](2_9) at (4,-7.5) {}; 
\node[white](2_10) at (6,-7.5) {}; 
\node[gray,fill=none](2_11) at (8,-7.5) {};

\node[plus](p) at (1,-7) {+};
\node[plus](p) at (1,-7.5) {+};
\node[plus](p) at (9,-7.5) {+};
\node[plus](p) at (10,-7.5) {$\cdots$};

\node[plus](p) at (-2.5, -8.25) {$=$};
\node[internalVertex,fill=none](2_x) at (-2, -8.25) {};
\node[internalVertex,fill=none](2_y) at (0, -8.25) {};
\draw[directed_dotted] (2_y) -- (2_x);

\node[internalVertex,fill=none](fact_hang) at (2, -8.25) {};
\node[white](par) at (4, -8.25) {};
\node[plus](p) at (4.25, -8.25) {\Big [ };
\draw[synapse] (par) -- (fact_hang);
\node[internalVertex,fill=none](fact_int) at (4.5, -8.25) {};
\node[internalVertex,fill=none](fact_int2) at (6.75, -8.25) {};
\draw[directed_dotted] (fact_int2) -- (fact_int);
\node[plus](p) at (7.25, -8.25) {+};
\node[plus](p) at (8.25, -8.25) {$\cdots \Big]$};

\draw[directed_dotted] (2_3) -- (2_1);
\draw[synapse] (2_6) -- (2_4);
\draw[directed_dotted] (2_6b) -- (2_6);
\draw[directed_dotted] (2_11) -- (2_10);
 
\draw[synapse] (2_10) -- (2_9);
\draw[synapse] (2_9) -- (2_7); 
 
\node[plus](2.2a) at (-2.5,-9) {$=$}; 
\node[internalVertex,fill=none](3_3) at (0,-9) {};
\node[internalVertex,fill=none] (3_1) at (-2,-9) {};
 
\node[internalVertex,fill=none](3_4) at (2,-9) {};
\node[externalVertex,fill=white](3_6) at (4,-9) {};
\node[internalVertex,fill=none](3_7) at (6,-9) {};
 
\node[internalVertex,fill=none](2_2ab) at (-3,-9) {};
\node[internalVertex,fill=none](2_2ac) at (-5,-9) {};
\draw[directed] (2_2ab) --(2_2ac);
 
\node[plus](p) at (1, -9) {+};
\node[plus](p) at (1,-8.25) {+};
\draw[directed_dotted] (3_3) -- (3_1);
\draw[synapse] (3_6) --(3_4);
\draw[directed] (3_7) -- (3_6);

\end{tikzpicture}

\end{center}
\caption{Diagrammatic expansion for the mean firing rate and linear response of the self-exciting process. A) First-order approximation of the firing rate. B) Diagrams corresponding to the re-summing of the expansion of the mean field rate (Eq. \ref{eq:meanSelfExciteFullExpansion}), which is represented by the black dot. C) Diagrams corresponding to the re-summing calculation of the propagator (Eq. \ref{eq:selfExcitePropagatorExpansion}), which is represented by the solid edge.  In all diagrams, time indices associated with internal vertices have been suppressed.  }
\label{fig:selfExciteMeanFeynman}
\end{adjustwidth}
\end{figure}


\paragraph{}
The propagator, $\Delta(t, t'),$ measures the fluctuation in the expected rate (around the mean-field value) in response to the addition of one spike at time $t'$ to the drive $\lambda(t)$.  Setting $\lambda(t) \leftarrow \lambda(t) + \delta(t-t')$ and $\bar{r}(t) \leftarrow \bar{r}(t) + (\Delta \ast \delta)(t, t')$ in Eq. \ref{eq:selfExciteRateExp} gives:
%
\begin{align}
	\bar{r}(t) + (\Delta \ast \delta)(t, t') &= \epsilon \big ( (g \ast \bar{r})(t)+ (g \ast \Delta \ast \delta)(t, t')\big)  + \lambda(t)  + \delta(t-t') \nonumber \\
	\Delta(t, t' )  &= \epsilon (g \ast \Delta)(t, t') + \delta(t - t')
	\label{eq:selfExciteProp}
\end{align}
where for convolutions involving $\Delta(t, t')$, we use the notation $(f \ast \Delta)(t, t') = \int dt''\; f(t-t'')\Delta(t'', t')$ and $(\Delta \ast f)(t, t') = \int dt''\; \Delta(t, t'') f(t''-t')$

\paragraph{}
As with the expected rate $\bar{r}(t)$, we can examine the propagators of the $n$-th order self-exciting processes. 
For the first-order process $N_1(t)$,
\begin{align}
\Delta_1(t, t') &= \delta(t-t') + \epsilon (g * \delta)(t-t')
\end{align}
The first term is the propagator of the inhomogeneous Poisson process. The second term of $\Delta_1$ is the propagator of the filtered Poisson process, Eq. \ref{eq:filtPoissonProp}. This equation can be represented by the same type of graphs as for the expected rate (Fig. \ref{fig:selfExciteMeanFeynman}c top), but stand for functions between two time points:  the initial time $t'$ of the perturbation, and the later time $t$, at which we are computing the rate of the process.  We don't represent these initial and final points as vertices, because the propagator is a function that connects two vertices. However, we still integrate over the times corresponding to the internal vertices since the propagator accounts for the total effect of a perturbation of the source on the observed activity.

\paragraph{}
In general, the propagator for the $n$th-order self-exciting process can be computed by taking a functional derivate of the rate with respect to the input rate $\lambda$:
\begin{align}
\Delta_n(t, t') &= \frac{\delta}{\delta \lambda(t')} \left( \lambda(t) + \epsilon g * \bar{r}_{n-1}(t) \right) \nonumber \\
&= \delta(t-t') + \epsilon \frac{\delta}{\delta \lambda(t')}\left(g * \bar{r}_{n-1}\right)(t)  \nonumber \\
&= \delta(t-t') + \epsilon \left(g * \sum_{k=0}^{n-1} \epsilon^k g^{(k)} * \Delta_k\right)(t, t')
\label{eq:selfExcitePropagatorExpansion}
\end{align}
This recursion relation can be expressed graphically just as for the mean rate (Fig. \ref{fig:selfExciteMeanFeynman}c, top). Factoring out $\epsilon g \ast$ corresponds to popping off an edge and vertex from the series (Fig. \ref{fig:selfExciteMeanFeynman}c, middle). Taking the limit $n \rightarrow \infty$ in Eq. \ref{eq:selfExcitePropagatorExpansion} yields the self-consistency condition for the full propagator $\Delta(t, t')$ given by Eq. \ref{eq:selfExciteProp}, and indicated by the solid black line in Fig. \ref{fig:selfExciteMeanFeynman}c (bottom).

\paragraph{}
These diagrammatic expansions may seem cumbersome for so simple a model. Even for the self-exciting Hawkes process, however, they allow the fast calculation of any order of spike train cumulant. Let us begin with the second cumulant of the instantaneous firing rate.  Again we will construct an expansion in $\epsilon$, i.e. powers of $g$.  To zeroth order, this is the inhomogeneous Poisson solution.  
To first order in $\epsilon$ we have
\begin{align}
	\left \langle \frac{dN_1}{dt}(t) \frac{dN_1}{dt}(t') \right \rangle_c
	&= \left \langle \left ( \frac{dN_0}{dt}(t) + \frac{dM_0}{dt}(t)    \right ) \left ( \frac{dN_0}{dt}(t') + \frac{dM_0}{dt}(t')    \right ) \right \rangle_c \nonumber \\
	&= \left \langle \frac{dN_0}{dt}(t)  \frac{dN_0}{dt}(t')    \right \rangle_c  + \left \langle \frac{dN_0}{dt}(t)  \frac{dM_0}{dt}(t')    \right \rangle_c \nonumber \\
	&+ \left \langle \frac{dM_0}{dt}(t)  \frac{dN_0}{dt}(t')    \right \rangle_c +  \left \langle   \frac{dM_0}{dt}(t)   \frac{dM_0}{dt}(t')    \right \rangle_c \nonumber \\
	& =  \int_{t_0}^\infty ds \; \delta(t - s) \delta(t'-s) \lambda(s)   \nonumber +  \epsilon \int_{t_0}^\infty ds \; \delta(t - s) (g \ast \delta)(t' - s) \lambda(s)  \nonumber \\
	&+  \epsilon \int_{t_0}^\infty ds \; \delta(t' - s) (g \ast \delta)(t - s) \lambda(s)  \nonumber \\
	&+  \epsilon \int_{t_0}^\infty ds \; \delta(t - s) \delta (t' - s) \left(g \ast  \int_{t_0}^\infty ds' \delta(s - s') \lambda(s')\right)(s) \label{eq:2pointSelfExcite}
\end{align}
The first term on the second line is the second cumulant of the inhomogenous Poisson process.  The other terms arise from the dependency of the processes $M_0(t)$ and $N_0(t)$. The expectation over $M_0(t)$ must be performed first, followed by that over $N_0(t)$, because the process $M_0(t)$ is conditionally dependent on the realization of $N_0(t)$, having intensity $\epsilon \left(g \ast \frac{dN_0}{dt}\right)(t)$ (Eq. ~\eqref{eq:selfExciteRateRec}). 
This decomposition relies on the linearity of the expectation operator.
 

\paragraph{}
 We can construct diagrams for each of these terms using the same rules as before, with the addition of two new internal vertices (Fig. \ref{fig:2pointSelfExciteFeynman}a). These new vertices are distinguished by their edges. The first has two outgoing dotted lines representing the zeroth-order propagator $\delta(t-t')$, as in the second cumulant of the inhomogeneous Poisson process.  It represents events that are generated by the drive $\lambda(t)$ and propagate jointly to the two measurement time points, The second new vertex has the same two outgoing lines and one incoming wavy line for the filter $g(t,t')$ - it represents the fourth term on the right hand side of Eq.~\eqref{eq:2pointSelfExcite}. This vertex carries a factor of $\epsilon$ and represents the filtering of past events that then propagate to the two measurement time points.

 

\paragraph{}
Continuing the computation of the second cumulant to any order in $\epsilon$ will result in higher order terms of the linear response and expected rate being added to the corresponding legs of the graph. At a given order $n$, one leg of each diagram will be associated with a particular term in the expansion, to order $n$, of the expected rate or the linear response. The second cumulant of $dN_2/dt$ would thus add diagrams with two filter edges to the diagrams of Fig. \ref{fig:2pointSelfExciteFeynman}a, either both on the same leg of the graph or distributed amongst the graph's three legs.

\paragraph{}
As with the filtered Poisson process, we can simplify this sum of four diagrams for the second cumulant of the first-order self-exciting process. Examining subgraphs of each term on the right-hand side of Fig. \ref{fig:2pointSelfExciteFeynman}A reveals a connection to the linear response and mean rate of the first-order self-exciting processes. On each leg emanating from the internal branching vertex, the four terms sum to the product of two copies of the linear responses of the first-order self-exciting process $N_1(t)$ (compare subgraphs on the top branch of the diagrams in Fig. \ref{fig:2pointSelfExciteFeynman}a with Fig. \ref{fig:selfExciteMeanFeynman}a). Similarly, the sum of the legs coming into the branching vertex is the firing rate of $N_1(t)$ (compare to Fig. \ref{fig:selfExciteMeanFeynman}b). So, we will group the terms on the legs of the graph into contributions to the linear response and the mean (Fig. \ref{fig:2pointSelfExciteFeynman}b middle).  

\paragraph{}
When we add the diagrams of up to order $n$ together, we can separately re-sum each of these expansions because of the distributivity of the expectation. So, we can replace the entire series \emph{to all orders in $\epsilon$} with simpler diagrams using the full representations for the linear response and expected rate (Fig. \ref{fig:2pointSelfExciteFeynman}b). This can be proved inductively, or by rigorously deriving the Feynman rules from the cumulant generating functional (\hyperref[methods:pathIntegral]{Methods: Path integral representation}). This yields the following result for the second cumulant, which corresponds to the final graph at the bottom far right of Fig. ~\ref{fig:2pointSelfExciteFeynman}b:
\begin{align}
\left \langle \frac{dN}{dt}(t) \frac{dN}{dt}(t') \right \rangle_c = \int_{t_0}^{\infty} ds \; \Delta(t - s) \Delta(t'- s) \bar{r}(s)
\label{eq:2pointSelfExciteFull}
\end{align}
This is the complete analytic result for the second cumulant of the self-exciting process for fluctuations around the mean field solution $\bar{r}(t)$ \cite{hawkes_spectra_1971}.  It can be represented by the single term on the right-hand side of Eq. \ref{eq:2pointSelfExciteFull} and the corresponding single diagram (Fig. \ref{fig:2pointSelfExciteFeynman}b, right). Compare this with the filtered Poisson process, which has a diagram of the same topology but with different constituent factors (Fig. \ref{fig:filtPoisson}C, middle row). The Feynman diagrams capture the form of the re-summed perturbative expansions for the cumulants, while the definitions of the vertices and edges capture the model-specific rate, $\bar{r}(t)$, and propagator, $\Delta(t,t')$. 

\paragraph{}
One might think that the higher cumulants are generated as simply by replacing each leg of the filtered inhomogeneous Poisson process with the correct propagator, along with the rate $\bar{r}(t)$.  This would mean that the general cumulant term would be given by:
\begin{eqnarray}
	\left \langle  \prod_i \frac{dN}{dt}(t_i) \right \rangle_c &=& \int_{t_0}^\infty dt \; \bar{r} (t)\prod_i \Delta(t_i, t)
	\label{eq:selfExciteCumulantNaive}
\end{eqnarray}
This is incorrect, as many important terms arising from the self-interaction would be lost.



\begin{figure}
\begin{adjustwidth}{-1 in}{0in}
	\begin{center}
	\begin{tikzpicture}
	
	\node[math](0a) at (-9,2) {\textbf{a)}};
	\node[math](1) at (-8,0) {$\left \langle \frac{dN_1}{dt}(t) \frac{dN_1}{dt}(t') \right \rangle_c$};
	\node[math](2) at(-6.5,0) {$=$};
	
	\node[gray](3) at (-5,0) {};
	\node[white](4) at (-6,1) {};
	\node[white](5) at (-6,-1) {};
		
	\draw[directed_dotted] (3)--(4);
	\draw[directed_dotted] (3)--(5);
	
	\node[math](6) at (-4,0) {$+$};
	
	\node[white](7) at (-2,0) {};
	\node[white](8) at (-3,1) {};
	\node[white](9) at (-3,-1) {};
	\node[white](10) at (-1,0) {};
	\node[gray](11) at (0,0) {};
		
	\draw[directed_dotted] (7)--(8);
	\draw[directed_dotted] (7)--(9);
	\draw[synapse] (10)--(7);
	\draw[directed_dotted] (11)--(10);
	
	\node[math](12) at (1,0) {$+$};
	
	\node[gray](13) at (3,0) {};
	\node[white](14) at (2,1) {};
	\node[white](15) at (2,-1) {};
	\node[white](16) at (1,2) {};
		
	\draw[directed_dotted] (13)--(14);
	\draw[directed_dotted] (13)--(15);
	\draw[synapse] (14) -- (16);
	
	\node[math](18) at (4,0) {$+$};
	
	\node[gray](19) at (6,0) {};
	\node[white](20) at (5,-1) {};
	\node[white](21) at (5,1) {};
	\node[white](22) at (4,-2) {};
		
	\draw[directed_dotted] (19)--(20);
	\draw[directed_dotted] (19)--(21);
	\draw[synapse] (20) -- (22);
	
	\node[math](0a) at (-9,-4) {\textbf{b)}};
	\node[math](24) at (-8,-6) {$\left \langle \frac{dN(t)}{dt} \frac{dN(t')}{dt} \right \rangle_c$};
	\node[math](25) at(-6.5,-6) {$=$};
	
	\node[white] at (-6, -3.5) {}; 
	\node[math,rotate=-45] at (-5.75, -3.75) {\Bigg[ };
	\node[gray,fill=none](26) at (-5.25, -3.25) {};
	\node[gray, fill=none](27) at (-4.25, -4.25) {};
	\draw[directed_dotted] (27) -- (26);
	\node[math, rotate=-45](p) at (-4, -4.5) {$+$};
	
	\node[gray,fill=none](26a) at (-5.75, -3.75) {};
	\node[white](26b) at (-4.75, -4.75) {};
	\draw[synapse] (26b) -- (26a);
	\node[gray,fill=none](26c) at (-4, -5.5) {};
	\draw[directed_dotted] (26c) -- (26b);
	
	\node[math,rotate=-45](p) at (-5.5,-5) {+ $\cdots$} ;
	\node[math, rotate=-45](p) at (-4, -5.5) {$\Bigg]$};
	
	\node[white] at (-6, -8.5) {};
	\node[math,rotate=45] at (-5.75, -8.25) {\Bigg[};
	\node[gray,fill=none](28) at (-6.25, -7.75) {};
	\node[gray,fill=none](29) at (-5.25, -6.75) {};
	\node[math,rotate=45](p) at (-5, -6.5) {$+$};
	\draw[directed_dotted] (29) -- (28);
	
	\node[gray,fill=none] (28a) at (-5.75, -8.25){};
	\node[white] (28b) at (-4.75, -7.25) {};
	\draw[synapse] (28b) -- (28a);
	\node[gray,fill=none] (28c) at (-4, -6.5) {};
	\draw[directed_dotted] (28c) -- (28b);
	\node[math,rotate=45](p) at (-4.75, -8.25) {$+ \cdots$};
	\node[math,rotate=45](p) at (-4, -6.5) {\Bigg]};
	
	\node[math] at (-3.5, -6) {\Big[ };
	\node[gray] at (-3, -6) {};
	\node[math] at (-2.5, -6) {+};
	\node[white](30) at (-2, -6) {};
	\node[white](31) at (-1, -6) {};
	\node[gray](32) at (0, -6) {};
	\draw[synapse] (31) -- (30);
	\draw[directed_dotted] (32) -- (31);
	\node[math] at (1, -6) {+ $\cdots$ \Big]};
	
	
	
	\node[math](34) at (3,-6) {$=$};

	\node[black](35) at (5,-6) {};
	\node[white](36) at (4,-5) {};
	\node[white](37) at (4,-7) {};
		
	\draw[directed] (35)--(36);
	\draw[directed] (35)--(37);
	
	\end{tikzpicture}
	\end{center}
	\caption{Diagrammatic expansion for the second cumulant for the self-exciting process. A) First-order approximation of the second cumulant. B) Re-summing to obtain the full second cumulant. Compare the expansions within the square brackets adjacent to external vertices to the expansion of the propagator, Fig. \ref{fig:selfExciteMeanFeynman}c, and compare the the expansion of the source term to that of the mean field rate, Fig.\ref{fig:selfExciteMeanFeynman}b.}
	\label{fig:2pointSelfExciteFeynman}
\end{adjustwidth}
\end{figure}

\paragraph{}
The reason this naive generalization fails is that it neglects the higher-order responses to perturbations in the event rate. For example, the second cumulant responds to perturbations in the rate; this quadratic response impacts the third cumulant. We can see this in the third cumulant of the first-order self-exciting process:
\begin{adjustwidth}{-1.8 in}{0in}
\begin{align}
	\left \langle \frac{dN_1}{dt}(t) \frac{dN_1}{dt}(t') \frac{dN_1}{dt}(t''') \right \rangle_c
	&= \left \langle \left ( \frac{dN_0}{dt}(t) + \frac{dM_0}{dt}(t)    \right ) \left ( \frac{dN_0}{dt}(t') + \frac{dM_0}{dt}(t')    \right )   \left ( \frac{dN_0}{dt}(t''') + \frac{dM_0}{dt}(t''')    \right ) \right \rangle_c \nonumber \\
	& =  \int_{t_0}^\infty ds \; \delta(t - s) \delta(t'-s) \delta(t''-s) \lambda(s)   \nonumber \\
	&+  \epsilon \int_{t_0}^\infty ds \; \delta(t - s) \delta(t'-s) (g \ast \delta)(t'' - s) \lambda(s)  + (t \leftrightarrow t' \leftrightarrow t'') \nonumber \\
	&+  \epsilon \int_{t_0}^\infty ds \; \delta(t - s) \delta (t' - s) \delta (t''- s) \left(g \ast  \int_{t_0}^\infty ds' \; \delta(s - s') \lambda(s')\right)(s)  \nonumber \\
	&+  \epsilon \int_{t_0}^\infty ds' \;  \delta (t''- s') \int_{t_0}^\infty ds \; \delta(t - s) \delta (t' - s)  (g \ast \delta)(s - s') \lambda(s') + (t \leftrightarrow t' \leftrightarrow t'') + \mathcal{O}(\epsilon^2) \label{eq:3pointSelfExcite}
\end{align}
\end{adjustwidth}
The first term is the third cumulant of the inhomogeneous Poisson process.  The second and third are generalizations of the terms found in the second cumulant (we have used $(t \leftrightarrow t' \leftrightarrow t'')$ to denote ``all other permutations of $t,t',t''$").  These terms are part of the naive expression in (\ref{eq:selfExciteCumulantNaive}).  The last term is the novel one that arises due to the ``quadratic response".  It appears when we compute 
\begin{align}
	 \left \langle \frac{dN_0}{dt}(t) \frac{dM_0}{dt}(t')    \frac{dM_0}{dt}(t'')  \right \rangle_c &= \epsilon \int_{t_0}^\infty ds\; \delta (t' - s) \delta(t'' - s) \left \langle \frac{dN_0}{dt}(t) \left(g \ast \frac{dN_0}{dt}\right)(s) \right \rangle_c
\end{align}
We have to take into account that the process $\frac{dN_0}{dt}(t)$ is correlated with the rate of the process $\frac{dM_0}{dt}(t)$ (since one is a linear function of the other!).  This produces a ``cascade" effect that results in the quadratic response. For the first-order process, only one step in the cascade is allowed.  By introducing branching internal vertices similar to those in Fig. \ref{fig:2pointSelfExciteFeynman}, we can express these somewhat unwieldy terms with diagrams. These are shown in Fig. ~\ref{fig:3pointSelfExciteFeynman}. The cascade of one source spike producing three spikes in the first-order process is represented by the second diagram of Fig. \ref{fig:3pointSelfExciteFeynman}a and the cascade of one source spike producing two spikes, one of which then produces another two spikes in the first-order process, is represented by the last diagram of Fig. \ref{fig:3pointSelfExciteFeynman}a.

\paragraph{}
As before, continuing to higher orders in the recursively self-exciting process would add diagrams with additional filter edges along the legs of the graphs in Fig. \ref{fig:3pointSelfExciteFeynman}a, corresponding to additional steps in the potential cascades of induced spikes. For example, the third cumulant of the second-order process, $\left \langle \frac{dN_2}{dt}(t)\frac{dN_2}{dt}(t')\frac{dN_2}{dt}(t'') \right \rangle_c$, would add diagrams with two filter edges to those of Fig. \ref{fig:3pointSelfExciteFeynman}a, with those two filter edges appearing either sequentially on the same leg of the graph or distributed amongst the legs of the graph. We can then use the same ideas that allowed us to resum the graphs representing the second cumulant.  As before, we identify the expansions of the mean-field rate, $\bar{r}$, and the linear response, $\Delta$, along individual legs of the graph and use the multilinearity of cumulants to resum those expansions to give the diagrams at the bottom of Fig. \ref{fig:3pointSelfExciteFeynman}. 

\paragraph{}
Considering the resummed graphs, we have the following result for the third cumulant:
\begin{align}
	\left \langle \frac{dN_1}{dt}(t) \frac{dN_1}{dt}(t') \frac{dN_1}{dt}(t'') \right \rangle_c
	& =  \int_{t_0}^\infty ds \; \Delta(t, s) \Delta(t', s) \Delta(t'', s) \bar{r}(s)   \nonumber \\
	&+   \int_{t_0}^\infty ds' \Delta (t'', s') \int_{t_0}^\infty ds \; \Delta(t, s) \Delta (t', s)  (g \ast \Delta)(s, s') \bar{r}(s') \nonumber \\
	&+ (t \leftrightarrow t' \leftrightarrow t'') \label{eq:3pointSelfExciteResum}
\end{align}

\paragraph{}
The types of diagram developed for up to the third cumulant encompass all the features that occur in the diagrammatic computations of cumulants of linearly self-exciting processes. The general rules for diagrammatically computing cumulants of this process are given in Fig. \ref{fig:selfExciteFeynmanRules}. They are derived in general in \hyperref[methods:pathIntegral]{Methods: Path integral representation}. The graphs generated with this algorithm correspond to the re-summed diagrams we computed above. 

\paragraph{}
For the $n$th cumulant, $\left \langle  \prod_i \frac{dN}{dt} (t_i) \right \rangle_c$, begin with $n$ white external vertices labelled $t_i$ for each $i$.  Construct all fully connected, directed graphs with the vertex and edge elements shown in  Figure~\ref{fig:selfExciteFeynmanRules}.  For each such fully connected directed graph constructed with the component vertices and edges, the associated mathematical term is constructed by taking the product of each associated factor, then integrating over the time points of internal vertices.  The $n$th cumulant is the sum of these terms. This produces cumulants of up to third order, as recently shown by Jovanovi\'c, Hertz \& Rotter \cite{jovanovic_cumulants_2015}, as well as cumulants of any order.  As we show next, this procedure can also be generalized to calculate cumulants in the presence of a nonlinearity, including both thresholds enforcing positive activity (as commonly disregarded in studies of the Hawkes process) and any nonlinear input-rate transfer function.




\begin{figure}
\begin{adjustwidth}{-1 in}{0in}
\begin{center}
	\begin{tikzpicture}
	\node[math](p) at (-11, 1) {$\mathbf{A)}$};
	\node[math](1) at (-9,0) {$\left \langle \frac{dN_1}{dt}(t) \frac{dN_1}{dt}(t') \frac{dN_1}{dt}(t''') \right \rangle_c$};
	\node[math](2) at(-6.5,0) {$=$};
	
	\node[gray](3) at (-5,0) {};
	\node[white](4) at (-6,1) {};
	\node[white](4b) at (-6,0) {};
	\node[white](5) at (-6,-1) {};
		
	\draw[directed_dotted] (3)--(4);
	\draw[directed_dotted] (3)--(5);
	\draw[directed_dotted] (3)--(4b);
	
	\node[math](6) at (-4,0) {$+$};
	
	\node[white](7) at (-2,0) {};
	\node[white](8) at (-3,1) {};
	\node[white](8b) at (-3,0) {};
	\node[white](9) at (-3,-1) {};
	\node[white](10) at (-1,0) {};
	\node[gray](11) at (0,0) {};
		
	\draw[directed_dotted] (7)--(8);
	\draw[directed_dotted] (7)--(9);
	\draw[directed_dotted] (7)--(8b);
	\draw[synapse] (10)--(7);
	\draw[directed_dotted] (11)--(10);
	
	\node[math](12) at (-9,-3.5) {$+$};
	
	\node[gray](13) at (-6,-4) {};
	\node[white](14) at (-7,-3) {};
	\node[white](14b) at (-7,-4) {};
	\node[white](15) at (-7,-5) {};
	\node[white](16) at (-8,-2) {};
		
	\draw[directed_dotted] (13)--(14);
	\draw[directed_dotted] (13)--(15);
	\draw[directed_dotted] (13)--(14b);
	\draw[synapse] (14) -- (16);
	
	\node[math](18) at (-5,-3.5) {$+$};
	\node[math](18b) at (-3.5,-3.5) {$(t \leftrightarrow t' \leftrightarrow t'')$};
	
	\node[math] (18c) at (-2,-3.5) {$+$};
	
	\node[white] (19) at (-1,-2.5) {};
	\node[white] (20) at (-1,-3.5) {};
	\node[white] (21) at (-1,-4.5) {};
	\node[white] (22) at (0,-3) {};
	\node[white] (23) at (1,-3) {};
	\node[gray] (23b) at (2,-3.5) {};
	
	\draw[directed_dotted] (22)--(19);
	\draw[directed_dotted] (22)--(20);
	\draw[directed_dotted] (23b)--(21);
	\draw[synapse] (23)--(22);
	\draw[directed_dotted] (23b)--(23);
	
	\node[math] (23c) at (2.5,-3.5) {$+$};
	\node[math](23d) at (4,-3.5) {$(t \leftrightarrow t' \leftrightarrow t'')$};
	
		
		
	\node[math](p) at (-11, -6) {$\mathbf{B)}$};
	\node[math](24) at (-9,-7) {$\left \langle \frac{dN}{dt}(t) \frac{dN}{dt}(t')\frac{dN}{dt}(t'')  \right \rangle_c$};
	\node[math](25) at(-6.5,-7) {$=$};
	
	\node[black](26) at (-5,-7) {};
	\node[white](27) at (-6,-6) {};
	\node[white](27b) at (-6,-7) {};
	\node[white](28) at (-6,-8) {};
		
	\draw[directed] (26)--(27);
	\draw[directed] (26)--(28);
	\draw[directed] (26)--(27b);
	
	\node[math](29) at (-4,-7) {$+$};
	
	\node[white] (30) at (-3,-6) {};
	\node[white] (31) at (-3,-7) {};
	\node[white] (32) at (-3,-8) {};
	\node[white] (33) at (-2,-6.5) {};
	\node[white] (34) at (-1,-6.5) {};
	\node[black] (35) at (0,-7) {};
	
	\draw[directed] (33)--(30);
	\draw[directed] (33)--(31);
	\draw[directed] (35)--(34);
	\draw[synapse] (34)--(33);
	\draw[directed] (35)--(32);
	
	\node[math] (36) at (0.5,-7) {$+$};
	\node[math](37) at (2,-7) {$(t \leftrightarrow t' \leftrightarrow t'')$};

		
	

		
	
	\end{tikzpicture}
	\end{center}

	\caption{Diagrams corresponding to third order cumulants. A) Diagrams corresponding to the third cumulant of the first-order self-exciting process. B) Diagrams corresponding to the third cumulant of the self-exciting process, after resumming the perturbative expansion. Nodes and edges correspond to the same terms as in Fig. \ref{fig:2pointSelfExciteFeynman}.}
	\label{fig:3pointSelfExciteFeynman}
\end{adjustwidth}
\end{figure}

\begin{figure}
	\begin{center}
	\begin{tikzpicture}
		\draw[gray,fill=none] (-1,-.5) rectangle (11,11);
		\draw (-1,10.2) -- (11,10.2);
		\node at (5,10.6) {Feynman Rules for the Self-Exciting Process};
		\node at (5,9.8) {For the $n$th cumulant, $\left \langle  \prod_i \frac{dN}{dt}(t_i) \right \rangle_c$:};  
		\node at (5,9.2) {I.  For each $i$, introduce a vertex labelled $t_i$.  These are \emph{external} vertices.};
		\node at (5,8.6) {II.  Construct all directed, connected graphs such that each vertex from I.};
		\node at (5,8.2) {has a single incoming propagator ($\Delta$) edge, using the vertices and edges };
		\node at (5, 7.8) { below. The time variables for each propagator or filter edge should match};
		\node at (5, 7.4) {those of its vertices. Filter edges can only impact internal vertices. Each };
		\node at (3.5, 7) { internal vertex has a unique associated time variable, $t'$.};
		\node at (5, 6.4) {III.  To construct the cumulant: for each graph, multiply the vertex or edge};
		\node at (5, 6) { factors together and integrate over the times for all internal vertices.};
		\node at (5,5.6) { Add the terms so obtained for each graph together.};
		\draw (-1,5.3) -- (11,5.3);
		\node at (2.5,5) {Internal Vertex or Edge};
		\node at (7.5,5.05) {Factor};
		\draw (-1,4.7) -- (11,4.7);
		\draw (5,-.5) -- (5,5.3);

		\node[black] (A) at (2.5,4) {}; 
		\node (B) at (1.5,4.5) {};
		\node (C) at (1.5,4) {};
		\node (D) at (1.5,3) {};
		\node[rotate=90] (E) at (1.75,3.75) {...};
		\draw[directed] (A) -- (B);
		\draw[directed] (A) -- (C);
		\draw[directed] (A) -- (D);
		
		\node at (7.5,4) {$\bar{r}(t)$};
		
		\node[white] (A1) at (2.5,2.5) {$t'$}; 
		\node (B1) at (1.5,3) {};
		\node (C1) at (1.5,2.5) {};
		\node (D1) at (1.5,1.5) {};
		\node[rotate=90] (E1) at (1.75,2.25) {...};
		\draw[directed] (A1) -- (B1);
		\draw[directed] (A1) -- (C1);
		\draw[directed] (A1) -- (D1);
		\node (F1) at (3.5,2.5) {};
		\draw[synapse] (A1) -- (F1);
		
		\node at (7.5,2.5) {$1$};
		
		\node (A2) at (1.5,1.25) {};
		\node[white] (B2) at (2.5,1.25) {$t'$};
		\node (C2) at (3.5,1.25) {};
		
		\draw[synapse] (B2) -- (A2);
		\draw[directed] (C2) -- (B2);
		
		\node at (7.5,1.25) {$1$};
		
		\node (A3) at (1.5,.5) {};
		\node (B3) at (3.5,.5) {};
		
		\draw[directed] (B3) -- (A3);
		
		\node at (7.5,.5) {$\Delta(t,t')$};
		
		\node (A4) at (1.5,0) {};
		\node (B4) at (3.5,0) {};
		
		\draw[synapse] (B4) -- (A4);
		
		\node at (7.5,0) {$g(t-t')$};
		
	\end{tikzpicture}
	\end{center}
	\caption{Feynman rules for the self exciting process.  These rules provide an algorithm for computing the expansion of the cumulants around the mean field solution $\bar{r}(t)$.  The dots between the legs of the first two vertices indicate that there are such vertices with any number of outgoing legs greater than or equal to two. \Kcomment{I tried to construct the graph corresponding to the fourth cumulant using these rules, and I got stuck at rule II. Can you please extend explanation?  What do I do with the available factors?  How do I combine them?} \Gcomment{Expanded explanation a bit; let me know if this is better.}}
	\label{fig:selfExciteFeynmanRules}
\end{figure}

\subsection*{Nonlinearities impose bidirectional coupling between different orders of activity: nonlinearly self-exciting process \label{sec:nonlinearity}}
Now we include a nonlinearity in the firing rate, so that the process produces events $dN/dt$ with a rate given by
\begin{align}
	r(t) =  \phi \left (\left(g \ast \frac{dN}{dt}\right)(t) + \lambda(t)  \right )  \label{eq:nonlinearRate}
\end{align}
We begin by considering the mean-field solution $\bar{r}$ which, if it exists, is self-consistently given by $\bar{r}(t) = \phi \left( (g \ast \bar{r})(t) + \lambda(t) \right)$.  Thus, as always the mean-field solution is given by neglecting second and higher-order cumulants of the spiking process.  Next, we consider the propagator, which as above is the linear response of the rate around the mean field, given by expanding Eq. \ref{eq:nonlinearRate} around the mean-field solution $\bar{r}(t)$ and examining the gain with respect to a perturbation of the rate. This propagator obeys:
\begin{align}
	\Delta(t,t') 	&= \phi^{(1)}\cdot \left((g \ast \Delta)(t,t')\right) + \delta(t - t') 
\end{align}
where $\phi^{(1)}$ is the first derivative of $\phi$ with respect to the input, evaluated at $g \ast \bar{r} + \lambda$. We will first develop a recursive formulation of the mean-field rate and propagator, which will be required for calculating cumulants of the full process. First, we Taylor expand $\phi$ about $\lambda$. For simplicity, consider a quadratic $\phi$ so that:
\begin{align}
r(t) = \lambda(t) + \epsilon_1 \left(\left(g \ast \frac{dN}{dt}\right)(t) + \lambda(t)\right) + \epsilon_2 \left(\left(g \ast \frac{dN}{dt}\right)(t) + \lambda(t)\right)^2
\end{align}
where $\epsilon_k$ is the $k$th Taylor coefficient, evaluated at $\lambda(t)$. We now develop the point process $dN/dt$ recursively at each order of the nonlinearity:
\begin{align}
	\frac{dN_{m,n}}{dt}(t) = \frac{dN_{0,0}}{dt}(t) + \frac{dM_{m-1, n}}{dt}(t) + \frac{dP_{m, n-1}}{dt}(t)
    \label{e.Nmn_first}
\end{align}
where $M_{m,n}$ is an inhomogeneous Poisson process with rate $\epsilon_1 \left(g \ast \frac{dN_{m,n}}{dt}\right)(t)$ and $P_{m,n}$ is an inhomogeneous Poisson process with rate $\epsilon_2 \left(g \ast \frac{dN_{m,n}}{dt} \right)^2(t)$ and $N_{0,0}(t)$ is an inhomogeneous Poisson process with rate $\lambda(t)$. To generate a set of events in $N$ at order $m$ in the linear term of $\phi$ and order $n$ in the quadratic term of $\phi$, we take events at each previous order, $(m-1, n)$ and $(m, n-1)$ and use those to develop $M_{m-1,n}(t)$ and $P_{m,n-1}(t)$. These, together with $N_{0,0}(t)$, give $\frac{dN_{m,n}}{dt}(t)$. In contrast to the linear self-exciting process, the quadratic process here is recursively defined on a lattice.

\paragraph{}
Similar to the case of the linearly self-exciting process, we can use this recursive definition to develop an expansion for the mean-field rate and propagator in powers of $\epsilon_1$ and $\epsilon_2$. When we calculate higher-order cumulants, we will identify the expansions of the mean-field firing rate and propagator which will allow us to use them to simplify the resulting diagrams. The mean-field rate to finite order in $m,n$ is once again given by neglecting second and higher-order cumulants of $N_{n,m}$ which allows us to take an expectation inside the quadratic term of Eq.~\eqref{e.Nmn_first}.  Taking the expectation of both sides of this equation in the mean field approach then yields:
\begin{align}
\bar{r}_{m,n}(t) = \lambda(t) + \epsilon_1(g \ast \bar{r}_{m-1, n})(t) + \epsilon_2 \left( g\ast \bar{r}_{m,n-1} \right)^2(t)
\end{align}
For example, 
\begin{align}
\bar{r}_{1,1} = \lambda(t) + \epsilon_1 (g \ast \bar{r}_{0,1})(t) + \epsilon_2 (g\ast \bar{r}_{1, 0})^2(t)
\end{align}
where
\begin{align}
\bar{r}_{1,0}(t) &= \lambda(t) + \epsilon_1 (g \ast \lambda)(t) \\
\bar{r}_{0,1}(t) &= \lambda(t) + \epsilon_2 \left(g \ast \lambda \right)^2(t) .
\label{e.r_epsilon}
\end{align}

\paragraph{}
Similarly, the propagator (for the dynamics of the recursive process, linearized around zero) is, to finite order in $m,n$:
\begin{align}
\Delta_{m,n}(t, t')  = \delta(t-t') + \epsilon_1(g \ast \Delta_{m-1,n})(t, t') + 2 \epsilon_2 \left(g \ast \bar{r}_{m,n-1}\right)(t) \left(g \ast \Delta_{m, n-1}\right)(t, t') 
\end{align}

\paragraph{}
To zeroth order in $\epsilon_2$, this yields an expansion of the mean-field rate $\bar{r}(t)$ which takes the same form as the expansion of the rate of the linearly self-exciting process, Eq. \ref{eq:selfExciteRateRec} and admits the same graphical representation (Fig. \ref{fig:selfExciteMeanFeynman}b). Similarly, a perturbative expansion of the linear response about the mean-field rate to zeroth order in $\epsilon_2$ takes the same form as for the linearly self-exciting process (Eq. \ref{eq:selfExcitePropagatorExpansion}) and admits the same graphical representation (Fig. \ref{fig:selfExciteMeanFeynman}c).

\paragraph{}
To account for the nonlinear terms arising at first order and greater in $\epsilon_2$, we will need to add another type of internal vertex in diagrammatic descriptions of the cumulants. These vertices, carrying factors of $\epsilon_2$, will have two incoming edges and any number of outgoing edges. Each incoming edge carries the operator $g \ast$ and the number of incoming edges corresponds to the order in the Taylor expansion of the nonlinearity. (It also corresponds to the order of cumulant influencing that vertex's activity. The number of outgoing edges corresponds to the order of cumulant being influenced, locally in that subgraph.) The factor of $\bar{r}(t)$ that appears in other vertices is modified to be consistent with the mean firing rate under the quadratic nonlinearity, and will thus obey Eq.(\ref{eq:nonlinearRate}) above.

\paragraph{}
The mean-field rate and propagator, to first order and greater in $\epsilon_2$, can be represented diagrammatically using the new vertex (e.g. Fig. \ref{fig:mftSelfExciteQuadratic}a, b). Notice that these directed graphs are tree-like, but with their leaves in the past. Repeating these calculations to the next order in $\epsilon_2$ can be accomplished by taking the basic structure of e.g. Fig. \ref{fig:mftSelfExciteQuadratic} and, along each leg entering the new vertex for $\epsilon_2$, inserting the previous-order graphs (Figs. \ref{fig:selfExciteMeanFeynman}a, \ref{fig:mftSelfExciteQuadratic}a). Including higher-order terms in $\epsilon_1$ would require inserting those graphs along the $\epsilon$-carrying vertices of Fig. \ref{fig:selfExciteMeanFeynman}a. 

\begin{figure}
	\begin{adjustwidth}{-1 in}{0in}
	\begin{center}
	\begin{tikzpicture}
	
		\node[math](p) at (-9, 1) {$\mathbf{A)}$};
		\node[gray] (A0) at (-5,0) {};
		\node[white] (B0) at (-6,0) {};
		\draw[directed_dotted] (A0)--(B0);
	
		\node[math] (p) at (-4, 0) {$+$};
	
		\node[gray] (A) at (-1,1) {};
		\node[gray] (B) at (-1,-1) {};
		\node[white] (C) at (-2, .5) {};
		\node[white] (D) at (-2, -.5) {};
		\node[white] (E) at (-3, 0) {};
        \node[math] (p) at (-2, 1.) {};
        \node[math] (p) at (-2, -1) {};
		\node[math] (p) at (-3, .5) {$\epsilon_2$};

		\draw[directed_dotted] (A)--(C);
		\draw[directed_dotted] (B)--(D);
		\draw[synapse] (C)--(E);
		\draw[synapse] (D)--(E);

		\node[math] (p) at (-8,0) {$\bar{r}_{0,1}(t)$};
		\node[math] (p) at (-7,0) {$=$};
		
		\node[math](p) at (-9, -2) {$\mathbf{B)}$};
		\node[math] (p) at (-8.25, -3) {$\Delta_{0,1}(t, t')$};
		\node[math] (p) at (-7, -3) {$=$};
		
		\node[] (G) at (-4.75, -3) {};
		\node[] (H) at (-6.25, -3) {};
		\draw[directed_dotted] (G)--(H);
		
		\node[math](p) at (-4, -3) {$+$};
		
		\node[white] (J) at (-2,-3) {};
		\node[] (K) at (-1, -2.5) {};
		\node[white] (L) at (-1, -3.5) {};
		\node[gray] (M) at (0, -4) {};
		\node[] (K0) at (-3, -3) {};
        \node[math] (p) at (-2, -2.5) {$\epsilon_2$};
        \node[math] (p) at (-1, -4) {}; 
        
		\draw[synapse] (L)--(J);
		\draw[directed_dotted] (M)--(L);
		\draw[synapse] (K)--(J);
        \draw[directed_dotted] (J)--(K0);
		
		\node[math](p) at (0, -3) {$+$};
		
		\node[white] (O) at (2, -3) {};
		\node[white] (P) at (3, -2.5) {};
		\node[gray] (Q) at (4, -2) {};
		\node[] (R) at (3, -3.5) {};
		\node[] (R0) at (1, -3) {};
        \node[math] (p) at (2, -2.5) {$\epsilon_2$};
        \node[math] (p) at (3, -2) {};
        
		\draw[synapse] (P)--(O);
		\draw[directed_dotted] (Q)--(P);
		\draw[synapse] (R)--(O);
		\draw[directed_dotted] (O)--(R0);
        

	\end{tikzpicture}
	\end{center}
	\end{adjustwidth}
	\caption{Expansion of the mean firing rate and propagator for the nonlinearly self-exciting process. A) One of the first nonlinear terms of the expansion of the mean-field firing rate, to first order in the quadratic term of the nonlinearity and zeroth order in the linear term.  The two diagrams shown correspond to the two terms in ~\eqref{e.r_epsilon}.  B) First nonlinear terms of the expansion of the propagator around the mean-field firing rate.}
	\label{fig:mftSelfExciteQuadratic}
\end{figure}  

\paragraph{}
We next consider the fluctuations of the firing rate around the mean-field:
\begin{align}
\left \langle \frac{dN}{dt}(t) \right \rangle = r(t) + \bar{r}(t) .
\end{align}
Again taking a quadratic nonlinearity, we have:
\begin{align}
r(t) = \epsilon_1 (g \ast r) (t) + \epsilon_2 \left( g \ast r \right)^2(t) .
\end{align}
where $\epsilon_k = \phi^{(k)} / k!$, evaluated at $(g \ast \bar{r}) (t) + \lambda(t)$. Similarly to before, we recursively define $dN/dt$:
\begin{align}
\frac{dN_{m,n}}{dt}(t) = \frac{dN_{0,0}}{dt}(t) + \frac{dM_{m-1, n}}{dt}(t) + \frac{dP_{m, n-1}}{dt}(t)
\end{align}
Once again, $M_{m,n}$ is an inhomogeneous Poisson process with rate $\epsilon_1 \left(g \ast \frac{dN_{m,n}}{dt}\right)(t)$ and $P_{m,n}$ is an inhomogeneous Poisson process with rate $\epsilon_2 \left(g \ast \frac{dN_{m,n}}{dt}\right)^2 (t)$.
$N_{0,0}(t)$ now has rate $\delta \lambda(t) = \lambda(t) - \bar{r}(t)$. We will begin calculating cumulants of $dN/dt$ to finite order in $m, n$. The first nonlinear correction to the firing rate appears at first order in $\epsilon_2$:
\begin{align}
\left \langle \frac{dN_{0,1}}{dt}(t) \right \rangle_c &= \delta\lambda(t) + \epsilon_2 \left \langle \left(g \ast \frac{dN_{0,0}}{dt}\right)^2(t) \right \rangle_c
\end{align}
which can be represented diagrammatically using the new vertex (Fig. \ref{fig:loopSelfExciteQuadratic}a). Notice that in contrast to the corresponding graph for the mean-field expansion (Fig. \ref{fig:mftSelfExciteQuadratic}a), this diagram has a ``loop" (a cycle were it an undirected graph). This reflects the dependence of the rate on the second cumulant of the baseline process $N_{0,0}$. This dependence of the firing rate on higher-order spiking cumulants is a fundamental feature of nonlinearities.

\begin{figure}
	\begin{center}
	\begin{tikzpicture}
	
		\node[math](p) at (-7, 1) {$\mathbf{A)}$};
		\node[gray] (A0) at (-3,0) {};
		\node[white] (B0) at (-4,0) {};
		\draw[directed_dotted] (A0)--(B0);
	
		\node[math] (p) at (-2, 0) {$+$};
	
		\node[gray] (A) at (1,0) {};
		\node[white] (B) at (0,1) {};
		\node[white] (C) at (0,-1) {};
		\node[white] (D) at (-1,0) {};
		\node[math] (G) at (-6,0) {$\left \langle r_{0,1} \right \rangle$};
		\node[math] (p) at (-5,0) {$=$};
		
		\draw[directed_dotted] (A)--(B);
		\draw[directed_dotted] (A)--(C);
		\draw[synapse] (B)--(D);
		\draw[synapse] (C)--(D);
		
		\node[math](p) at (-7, -2) {$\mathbf{B)}$};
	
		\node[black] (A) at (-1,-3) {};
		\node[white] (B) at (-2, -2) {};
		\node[white] (C) at (-2, -4) {};
		\node[white] (D) at (-3, -3) {};
		\node[white] (E) at (-4, -3) {};

		\node[math] (p) at (-6,-3) {$\left \langle r \right \rangle$};
		\node[math] (p) at (-5,-3) {$=$};
	
		\draw[directed] (A)-- (B);
		\draw[directed] (A) -- (C);
		\draw[synapse] (B) -- (D);
		\draw[synapse] (C) -- (D);
		\draw[directed] (D) -- (E);

	\end{tikzpicture}
	\end{center}
	\caption{Corrections to the mean-field firing rate of the nonlinearly self-exciting process, to quadratic order in the nonlinearity $\phi$.  A) The correction to mean-field theory for the firing rate to first order in the quadratic term of $\phi$. B) The full one-loop correction to the mean-field rate.}
	\label{fig:loopSelfExciteQuadratic}
\end{figure}  

\paragraph{}
Proceeding beyond the first order in both $\epsilon_1$ and $\epsilon_2$, we see that the expansion of each term of the nonlinearity depends on the other:
\begin{align}
\left \langle \frac{dN_{1,1}(t)}{dt} \right \rangle_c &= \delta\lambda(t) + \epsilon_1 g \ast \left \langle \frac{dN_{0,1}}{dt} (t)\right \rangle _c+ \epsilon_2 \left \langle \left(g \ast \frac{dN_{1,0}}{dt} \right)^2(t) \right \rangle_c
\end{align}
so that at each order in $\epsilon_2$ we must insert the solution at the previous order in $\epsilon_2$ and the same order in $\epsilon_1$ (and vice versa). This recursive mixing of expansions between the linear and nonlinear terms of $\phi$ seems intractable. However, this joint expansion can be re-summed to obtain the full correction to the firing rates \cite{zinn-justin_quantum_2002}; \hyperref[methods:pathIntegral]{Methods: Path integral representation}. The Feynman rules for the re-summed diagrams of the nonlinearly self-exciting process are given in Fig.~\ref{fig:selfExciteFeynmanRulesNonlinear}. For a quadratic nonlinearity, this yields the one-loop correction to the firing rate:
\begin{align}
r_1 &=  \int_{t_0}^t dt_1 \int_{t_0}^t dt_2 \; \Delta(t, t_1) \frac{\phi^{(2)}}{2} \left( (g \ast \Delta)(t_1, t_2)\right)^2  \bar{r}(t_2)
\end{align}
where we relabel $r$ as $r_1$ to denote that this is the one-loop correction and $\phi^{(2)}$ is evaluated at $(g \ast \bar{r})(t) + \lambda(t)$.

\paragraph{}
Similarly to the tree-level propagator, the propagator with an arbitrary nonlinearity $\phi$ can be calculated to one loop by taking the rate dynamics to one loop and expanding to linear order around the mean-field solution:
\begin{align}
\left \langle \frac{dN}{dt}(t) \right \rangle &\approx \phi\left( g \ast \bar{r}(t) + \lambda(t)\right) + \left(\Delta \ast \frac{\phi^{(2)}}{2}\left( g \ast \Delta\right)^2 \bar{r} \right)(t) \\
 &= \bar{r} + \phi^{(1)} \cdot \left(g \ast r(t) \right) + \left(\Delta \ast \frac{\phi^{(2)}}{2}\left( g \ast \Delta \right)^2 \left( \bar{r} + \phi^{(1)} \cdot \left(g \ast r \right) \right)\right)(t) .
\end{align}
Providing a perturbation to the rate fluctuation, $r(t) \rightarrow r(t) + \epsilon(t)$ and differentiating with respect to $\epsilon(t')$ yields:
\begin{align}
\Delta (t, t') &\equiv \frac{\partial r(t)}{\partial \epsilon(t')} \nonumber \\
&\approx\delta(t-t') + \phi^{(1)} \left(g \ast \Delta \right)(t, t')  + \left( \bar{\Delta} \ast \frac{\phi^{(2)}}{2}\left( g \ast \bar{\Delta} \right)^2 \left(\phi^{(1)} \left(g \ast \bar{\Delta} \right) \right)\right)(t, t') \label{eq:1loopProp}
\end{align}
where only keeping the first two terms defines the tree-level propagator, $\bar{\Delta}(t, t')$ and the third term is the one-loop correction to the propagator.

\paragraph{}
The appearance of a loop in the Feynman diagram for the mean rate of the quadratically self-exciting process is a general feature of nonlinearities.  It represents the influence of higher-order spike train cumulants on lower order ones. In order to measure that dependency, we can count the number of loops in the graphs. To do this, we add a bookkeeping variable, $h$.  We count the number of loops by multiplying factors of $h$ and $1/h$. Each internal vertex adds a factor of $1/h$ and each outgoing edge a factor of $h$. In this way every vertex with more than one outgoing edge will contribute a factor of $h$ for every edge beyond the first.  $h$ thus effectively counts the order of fluctuations contributed by the vertex.  For example, the mean for the linear self-exciting process has a graph with a single internal vertex and a single internal edge, so it is zeroth order in $h$ (Fig. \ref{fig:selfExciteMeanFeynman}b).  The two point function, however, having two edges and one internal vertex (Fig. \ref{fig:2pointSelfExciteFeynman}b), is first order in $h$.  Similarly, the tree-level diagrams will always contribute a total of $h^{n-1}$, where $n$ is the order of the cumulant.  

\paragraph{}
In terms of powers of $h$, a graph for a $n$th order cumulant with one loop will be equivalent to a graph for a $n+1$st order cumulant with one less loop. Consider cutting one of the lines that form the loop in Fig. \ref{fig:loopSelfExciteQuadratic}b at the internal vertex and leaving it hanging.  Now the graph for the one-loop correction to the mean rate appears to be a graph for a second cumulant - it has two endpoints. The power counting in terms of $h$, however, has not changed. The one-loop correction to the mean is of the same order in $h$ as the tree-level second cumulant.  In general, we will have that the order $h^m$ will be given by
\begin{align}
	m = n +l -1
\end{align}
where $n$ is the number of external vertices and $l$ is the number of internal loops.  The number of loops thus tracks the successive contributions of the higher order fluctuations.  This expansion is called the ``loop" expansion and is equivalent to a small-fluctuation expansion.  If one can control the size of the fluctuations, one can truncate the loop expansion as an approximation for the statistics of the system.  One way of doing this with networks is to insure that the interactions are $\mathcal{O}(1/N)$ so that $h \propto 1/N$ and the expansion becomes a system size expansion.  

\paragraph{}
The Taylor expansion of an arbitrary nonlinearity $\phi$ could have infinitely many terms. This would lead, in the recursive formulation, to infinitely many processes $\{M, P, \ldots \}$. Even after re-summing the recursive formulation, this would leave an infinite number of diagrams corresponding to any given cumulant.
There are two ways to approach this. The first is to insist on a perturbative expansion in the non-linear terms, e.g. only consider terms up to a fixed order in the Taylor expansion of the nonlinearity $\phi$.

\paragraph{}
The second approach to controlling the order of the loop expansion is to consider a regime in which mean field theory is stable as this will also control the fluctuations, limiting the magnitude of the loop contributions \cite{zinn-justin_quantum_2002}.  The expansion then breaks down in the regime of a bifurcation or ``critical point".  In this case, the linear response diverges, causing all loop diagrams to similarly diverge.  This is a fluctuation-dominated regime in which mean field theory, along with the fluctuation expansion around it, fails. In that case, renormalization arguments can allow discussion of the scaling behavior of correlations \cite{buice_field-theoretic_2007}.

\begin{figure}
	\begin{center}
	\begin{tikzpicture}
		\draw[gray,fill=none] (-1,0) rectangle (11,11.5);
		\draw (-1,10.7) -- (11,10.7);
		\node at (5,11.1) {Feynman Rules for the Nonlinearly Self-Exciting Process};
		\node at (5,10.3) {For the $n$th cumulant, $\left \langle  \prod_i \frac{dN}{dt}(t_i) \right \rangle_c$:};  
		\node at (5,9.7) {I.  For each $i$, introduce a vertex labelled $t_i$.  These are \emph{external} vertices.};
		\node at (5,9.1) {II.  Construct all directed, connected graphs such that each vertex from I.};
		\node at (5,8.7) {has a single incoming propagator ($\Delta$) edge, using the vertices and edges };
		\node at (5, 8.3) { below. The time variables for each propagator or filter edge should match};
		\node at (5, 7.9) {those of its vertices. Filter edges can only impact internal vertices. Each };
		\node at (3.5, 7.5) { internal vertex has a unique associated time variable, $t'$.};
		\node at (5, 6.9) {III.  To construct the cumulant: for each graph, multiply the vertex or edge};
		\node at (5, 6.5) { factors together and integrate over the times for all internal vertices.};
		\node at (5,6.1) { Add the terms so obtained for each graph together.};
		\draw (-1,5.8) -- (11,5.8);
		\node at (2.5,5.5) {Internal Vertex or Edge};
		\node at (7.5,5.55) {Factor};
		\draw (-1,5.2) -- (11,5.2);
		\draw (5,0) -- (5,5.8);		
		
		\node[black] (A) at (2.5,4.5) {}; 
		\node (B) at (1.5,5.0) {};
		\node (C) at (1.5,4.5) {};
		\node (D) at (1.5,3.5) {};
		\node[rotate=90] (E) at (1.75,4.25) {...};
		\draw[directed] (A) -- (B);
		\draw[directed] (A) -- (C);
		\draw[directed] (A) -- (D);
		
		\node at (7.5,4.5) {$\bar{r}(t)$};
		
		\node[white] (A1) at (2.5,3) {$t'$}; 
		\node (B1) at (1.5,3.5) {};
		\node (C1) at (1.5,3.) {};
		\node (D1) at (1.5,2.) {};
		\node[rotate=90] (E1) at (1.75,2.75) {...};
        		\node[math] at (1.5, 2.75) {$a$};
		\draw[directed] (A1) -- (B1);
		\draw[directed] (A1) -- (C1);
		\draw[directed] (A1) -- (D1);
		\node (F1) at (3.5,3) {};
		\node (F1c) at (3.5,2) {};
		\node[rotate=90] (E2) at (3.25,2.75) {...};
		\node[math] (E2b) at (3.5,2.75) {$b$};
		\draw[synapse] (A1) -- (F1);
		\draw[synapse] (A1) -- (F1c);
		
		\node at (7.5,3) {$\phi^{(b)}\left(( (g \ast \bar{r})(t) + \lambda(t)\right) / b!$};
		
		\node (A2) at (1.5,1.75) {};
		\node[white] (B2) at (2.5,1.75) {$t''$};
		\node (C2) at (3.5,1.75) {};
		
		\draw[synapse] (B2) -- (A2);
		\draw[directed] (C2) -- (B2);
		
		\node at (7.5,1.75) {$1$};
		
		\node (A3) at (1.5,1) {};
		\node (B3) at (3.5,1) {};
		
		\draw[directed] (B3) -- (A3);
		
		\node at (7.5,1) {$\Delta(t,t')$};
		
		\node (A4) at (1.5,.5) {};
		\node (B4) at (3.5,.5) {};
		
		\draw[synapse] (B4) -- (A4);
		
		\node at (7.5,.5) {$g(t-t')$};
		
	\end{tikzpicture}
	\end{center}
	\caption{Feynman rules for the nonlinearly self exciting process.  These rules provide an algorithm for computing the expansion of the cumulants around the mean field solution $\bar{r}(t)$.  The dots between the outgoing legs of the first vertex indicate that there any number of outgoing legs greater than or equal to two.   The number $b$ of incoming edges of the second vertex correspond to it's factor containing the $b$th derivative of $\phi$, evaluated at the mean field input. The $a$ dots between the outgoing edges of the second vertex indicate that it can have any number of outgoing edges such that $a+b \geq 3$.}
	\label{fig:selfExciteFeynmanRulesNonlinear}
\end{figure}

\section*{Interaction between single-neuron nonlinearities and network structure}
\paragraph{}
No new concepts are required in moving from a nonlinear self-exciting process to a network of interacting units.  Each external and internal vertex must now be associated with a unique neuron index $i$ and the integrations over time for the internal vertices must now be accompanied by summations over the indices of the internal vertices.  In addition, the filter $g(\tau)$ must be expanded to include coupling across units.  In general, this is given by $\mathbf{g}_{ij}(\tau)$ for the coupling from neuron $j$ to neuron $i$.  We will consider the general model of a network of units that generate conditionally Poisson-distributed events, given an input variable.  The conditional rate for unit $i$ is given by
\begin{align}
	r_i(t) &= \phi_i \left (\sum_j  \left(\mathbf{g}_{ij} \ast \frac{dN_j}{dt}\right)(t) + \lambda_i(t) \right )
	\label{eq:PoissonNetwork}
\end{align}
Similarly, the propagator now obeys
\begin{align}
	\Delta_{ij}(t,t') 	&= \phi^{(1)}_i \cdot \left( \sum_k  (g_{ik}  \ast \Delta_{kj}) (t,t') \right) + \delta_{ij} \delta(t - t') 
\end{align}
These dynamics are qualitatively the same as those of the nonlinearly self-exciting process (Eq. \ref{eq:nonlinearRate} but replace the neuron's own rate with the sum over its presynaptic inputs). Introducing these sums over neuron indices yields the complete set of rules for generating Feynman diagrams for the cumulants of this model, shown in Figure~\ref{fig:networkPoissonFeynman}.

\begin{figure}

		\begin{center}
	\begin{tikzpicture}
		\draw[gray,fill=none] (-1,0) rectangle (11,11.5);
		\draw (-1,10.7) -- (11,10.7);
		\node at (5,11.1) {Feynman Rules for the Networks of Linear-Nonlinear-Poisson Neurons};
		\node at (5,10.3) {For the $n$th cumulant, $\left \langle  \prod_i \frac{dN}{dt}(t_i) \right \rangle_c$:};  
		\node at (5,9.7) {I.  For each $i$, introduce a vertex labelled $t_i$.  These are \emph{external} vertices.};
		\node at (5,9.1) {II.  Construct all directed, connected graphs such that each vertex from I.};
		\node at (5,8.7) {has a single incoming propagator ($\Delta$) edge, using the vertices and edges };
		\node at (5, 8.3) { below. The time variables and neuron indices for each propagator or filter };
		\node at (5, 7.9) {edge should match its vertices. Filter edges can only impact internal vertices. };
		\node at (5, 7.5) {Each internal vertex has a unique associated time variable and neuron index.};
		\node at (5, 6.9) {III.  To construct the cumulant: for each graph, multiply the vertex or edge};
		\node at (5, 6.5) { factors together, integrate over the times and sum over neuron indices for};
		\node at (5, 6.1) {all internal vertices. Add the terms so obtained for each graph together.};
		\draw (-1,5.8) -- (11,5.8);
		\node at (2,5.5) {Internal Vertex or Edge};
		\node at (8,5.55) {Factor};
		\draw (-1,5.2) -- (11,5.2);
		\draw (5,0) -- (5,5.8);		
		
		\node[black] (A) at (2.5,4.5) {}; 
		\node (B) at (1.5,5.0) {};
		\node (C) at (1.5,4.5) {};
		\node (D) at (1.5,3.5) {};
		\node[rotate=90] (E) at (1.75,4.25) {...};
		\draw[directed] (A) -- (B);
		\draw[directed] (A) -- (C);
		\draw[directed] (A) -- (D);
		
		\node at (8,4.5) {$\bar{r}_i(t)$};
		
		\node[white] (A1) at (2.5,3) {$t', j$}; 
		\node (B1) at (1.5,3.5) {};
		\node (C1) at (1.5,3.) {};
		\node (D1) at (1.5,2.) {};
		\node[rotate=90] (E1) at (1.75,2.75) {...};
        		\node[math] at (1.5, 2.75) {$a$};
		\draw[directed] (A1) -- (B1);
		\draw[directed] (A1) -- (C1);
		\draw[directed] (A1) -- (D1);
		\node (F1) at (3.5,3) {};
		\node (F1c) at (3.5,2) {};
		\node[rotate=90] (E2) at (3.25,2.75) {...};
		\node[math] (E2b) at (3.5,2.75) {$b$};
		\draw[synapse] (A1) -- (F1);
		\draw[synapse] (A1) -- (F1c);
		
		\node at (8,3) {$\phi^{(b)}_j \left(\sum_k (\mathbf{g}_{jk} \ast \bar{r}_k )(t)+ \lambda_j(t) \right ) / b!$};
		
		\node (A2) at (1.5,1.75) {};
		\node[white] (B2) at (2.5,1.75) {$t'', k$};
		\node (C2) at (3.5,1.75) {};
		
		\draw[synapse] (B2) -- (A2);
		\draw[directed] (C2) -- (B2);
		
		\node at (8,1.75) {$1$};
		
		\node (A3) at (1.5,1) {};
		\node (B3) at (3.5,1) {};
		
		\draw[directed] (B3) -- (A3);
		
		\node at (8,1) {$\Delta_{ij}(t,t')$};
		
		\node (A4) at (1.5,.5) {};
		\node (B4) at (3.5,.5) {};
		
		\draw[synapse] (B4) -- (A4);
		
		\node at (8,.5) {$\mathbf{g}_{ij}(t-t')$};
		
	\end{tikzpicture}
	\end{center}
	\caption{Feynman rules for networks of stochastically spiking neurons with nonlinear input-rate transfer $\phi$.  These rules provide an algorithm for computing the expansion of the cumulants around the mean field solution $\bar{r}(t)$.  The dots between the outgoing legs of the first vertex indicate that there are any number of outgoing legs greater than or equal to two. The number $b$ of incoming edges of the second vertex correspond to it's factor containing the $b$th derivative of $\phi$, evaluated at the mean field input. The $a$ dots between the outgoing edges of the second vertex indicate that it can have any number of outgoing edges such that $a+b \geq 3$.}	\label{fig:networkPoissonFeynman}
\end{figure}

\newpage

\subsubsection*{Mid-course summary:  diagrammatic expansion reveals interplay of network structure and neural nonlinearities in driving neural activity}
\begin{figure}[!h]
\includegraphics[width=5in]{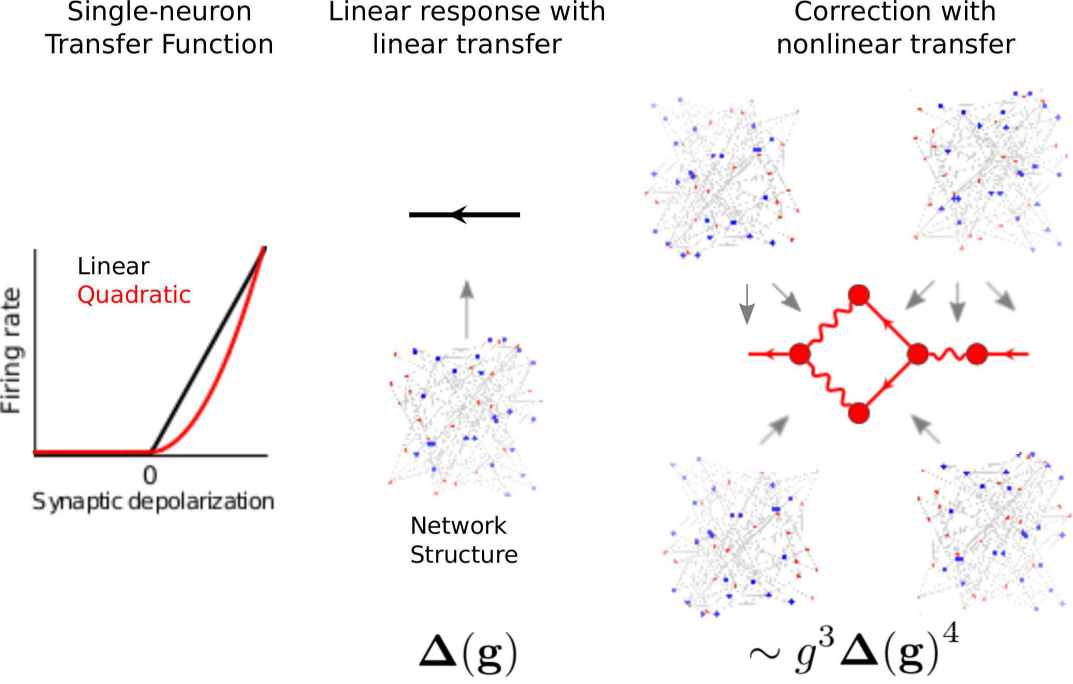} \\
\caption{{\bf Fluctuation expansion links single-neuron nonlinearities and network structure to determine network activity.} The linear response for linear neurons depends on the network structure both explicitly and implicitly, through the mean-field rates. The first nonlinear correction brings in additional explicit and implicit dependencies on the connectivity.}
\label{fig:schematic}
\end{figure}

\paragraph{}
To summarize, we now have in hand a set of tools---a fluctuation expansion---to compute spike train cumulants of arbitrary order in networks of linear-nonlinear-Poisson neurons. This expansion provides a systematic way to account for the synergistic dependence of lower-order activity on higher-order activity through the spiking nonlinearity, and naturally incorporates the full microstructure of the neuronal network. The order of the nonlinearity (for non-polynomials, the order of its Taylor expansion) determines the single-neuron transfer gain (Fig. \ref{fig:schematic} left). It \emph{also} determines how activity propagates through the network (Fig. \ref{fig:schematic}). 

\paragraph{}
We next provide an example of using this fluctuation expansion to compute a cumulant of spiking activity: the first nonlinear correction to the second cumulant. We will compute these using the Feynman rules (Fig. \ref{fig:networkPoissonFeynman}) to construct the corresponding diagrams, from which we will write the corresponding equation.

\paragraph{}
We begin by placing external vertices corresponding to the measurement times, each with one propagator edge coming into it. For the two-point cumulant there are two external vertices (Fig. \ref{fig:exampleFeynman}a). The propagators coming into those external vertices can, according to the Feynman rules, arise from either a source vertex or from an internal vertex with incoming filter edges (Fig. \ref{fig:networkPoissonFeynman}). If both propagators arise from a source vertex, we arrive at the tree-level diagram of Fig. \ref{fig:2pointSelfExciteFeynman}b, which provides the linear prediction for the two-point cumulant. To obtain the first nonlinear correction, we will begin by adding an internal vertex. There are two ways we can do this: with one internal vertex providing the propagators for both external vertices or with the internal vertex providing the propagator of just one external vertex (Fig. \ref{fig:exampleFeynman}b, top and bottom respectively). 

\paragraph{}
We next proceed to add another layer of vertices. The internal vertices added in each diagram of Fig. \ref{fig:exampleFeynman}b have incoming propagator edges. Those edges could emanate from other internal vertices or from source verticies. We will start by finishing the diagrams where they emanate from source vertices; placing these yields the top two, final diagrams of Fig. \ref{fig:exampleFeynman}c. We then continue constructing the diagram with an internal vertex providing the two propagators (Fig. \ref{fig:exampleFeynman}c, bottom). Note that if we added an internal vertex to the propagator hitting the $t_2$ external vertex, it would require at least two incoming filter edges (in order to obey $a+b \geq 3$ per the rules of Fig. \ref{fig:networkPoissonFeynman}) which would give rise to a second loop in the graph. 

\paragraph{}
This last diagram has a hanging filter edge, which must arise from an internal vertex with one incoming propagator edge. We finish the diagram with that internal vertex and the source vertex providing its propagator (Fig. \ref{fig:exampleFeynman}d). We could not add additional internal vertices along that path, since they would either violate $n+m \geq 3$ or give rise to more than one loop in the diagram (and thus belong at a higher order in the loop expansion).

\paragraph{}
Following the rules of Fig. \ref{fig:networkPoissonFeynman}, while restricting ourselves to graphs with a certain number of loops (here one) thus allowed us to construct the diagrams correspond to the first nonlinear correction to the two-point cumulant. We next write the equation corresponding to these diagrams. For each of the complete diagrams, we begin at the external vertices, and proceeding from left to right in the graph, multiply the factors corresponding to each edge and vertex together. We finish by summing over indices for all internal vertices and integrating over all internal time points. The contributions from each diagram are then added together. This yields:

\begin{adjustwidth}{-2 in}{0 in}
\begin{align}
\left \langle \frac{dN_i}{dt}(t_1) \frac{dN_j}{dt}(t_2) \right \rangle_{c, 1} &= \sum_{k, l, m, n} \int dt' \, \int dt'' \, \Delta_{ik}(t_1, t')\Delta_{jk}(t_2, t') \frac{\phi^{(2)}_k}{2}\left(g_{kl}\ast \Delta_{ln}\right)(t', t'') \left( g_{km}\ast \Delta_{mn} \right)(t', t'')\bar{r}_l(t'')  \nonumber \\
&+  \sum_{k, l, m, n} \int dt' \, \int dt'' \, \Delta_{ik}(t_1,t') \frac{\phi^{(2)}_k}{2}\left(g_{kl} \ast \Delta_{ln}\right)(t', t'')  \left(g_{km} \ast \Delta_{mn} \right)(t', t'')  \Delta_{jn}(t_2, t'') \bar{r}_n(t'') \nonumber \\
&+ \sum_{k, l, m, n} \int dt' \, \int dt'' \, \Delta_{ik}(t_2, t') \frac{\phi^{(2)}_k}{2}\left(g_{kl} \ast \Delta_{ln} \right)(t', t'') \left(g_{km} \ast \Delta_{mn}\right)(t', t'')   \Delta_{jn}(t_1, t'') \bar{r}_n(t'') \nonumber \\
&+ \sum_{k,l,m,n,o,p} \int dt' \int dt'' \int dt''' \,\Big[ \Delta_{ik}(t_1, t') \frac{\phi^{(2)}_k}{2}\left(g_{kl} \ast \Delta_{ln}\right)(t', t'')  \left(g_{km} \ast \Delta_{mn}\right)(t', t'')  \nonumber \\
&\;\;\;\;\;\;\;\; \cdot \phi^{(1)}_n \left(g_{no} \ast \Delta_{op}\right)(t'', t''') \Delta_{jp}(t_2, t''')\bar{r}_p(t''') \Big] \nonumber \\
&+ \sum_{k,l,m,n,o,p} \int dt' \int dt'' \int dt''' \,\Big[ \Delta_{ik}(t_2, t') \frac{\phi^{(2)}_k}{2}\left(g_{kl} \ast \Delta_{ln}\right)(t', t'')  \left(g_{km} \ast \Delta_{mn} \right)(t', t'') \nonumber \\
&\;\;\;\;\;\;\;\;  \phi^{(1)}_n \left(g_{no} \ast \Delta_{op}\right)(t'', t''') \Delta_{jp}(t_1, t''')\bar{r}_p(t''') \Big] \label{eq:2point1loop}
\end{align}
\end{adjustwidth}

\begin{figure}
\begin{adjustwidth}{-2 in}{0 in}
	\begin{center}
	\begin{tikzpicture}

		\node[math] at (-11, 11) {$\mathbf{a})$};
		\node[white] (A0) at (-10, 10) {$t_1$};
		\node[white] (B0) at (-10, 9) {$t_2$};
	
		\node at (-8.5, 10) (A1) {};
		\node at (-8.5, 9) (B1) {};
		
		\draw[directed] (A1)--(A0);
		\draw[directed] (B1)--(B0);
		
		\node[math] at (-8, 11) {$\mathbf{b})$};

		\node[white] (A0) at (-7.5, 10) {$t_1$};
		\node[white] (B0) at (-7.5, 9) {$t_2$};
	
		\node[white] (A1) at (-6.5, 9.5) {};
		\node (B1) at (-5.5, 9) {};
		\node (C1) at (-5.5, 10) {};
		
		\draw[directed] (A1)--(A0);
		\draw[directed] (A1)--(B0);
		\draw[synapse] (B1)--(A1);
		\draw[synapse] (C1)--(A1);

		\node[white] (A0) at (-7.5, 7.5) {$t_1$};
		\node[white] (B0) at (-7.5, 6) {$t_2$};
		\node[white] (A1) at (-6.5, 7.5) {};
		\node (B1) at (-5.5, 8) {};
		\node (C1) at (-5.5, 7) {};
		\node (D1) at (-6, 6) {};

		\draw[directed] (A1)--(A0);		
		\draw[directed] (D1)--(B0);
		\draw[synapse] (B1)--(A1);
		\draw[synapse] (C1)--(A1);
		
		\node[math] at (-7, 4) {$+t_1 \leftrightarrow t_2$};
		
		\node[math] at (-5, 11) {$\mathbf{c})$};
		\node[white] (A0) at (-4.5, 10) {$t_1$};
		\node[white] (B0) at (-4.5, 9) {$t_2$};
		
		\node[white] (A1) at (-3.5, 9.5) {};
		\node[white] (B1) at (-2.5, 9) {};
		\node[white] (C1) at (-2.5, 10) {};
		\node[black] (D1) at (-1.5, 9.5) {};
		
		\draw[directed] (A1)--(A0);
		\draw[directed] (A1)--(B0);
		\draw[synapse] (B1)--(A1);
		\draw[synapse] (C1)--(A1);
		\draw[directed] (D1)--(C1);
		\draw[directed] (D1)--(B1);
		
		\node[white] (A0) at (-4.5, 7.5) {$t_1$};
		\node[white] (B0) at (-4.5, 5.5) {$t_2$};
		\node[white] (A1) at (-3.5, 7) {};
		\node[white] (B1) at (-2.5, 7.25) {};
		\node[white] (C1) at (-3, 6.25) {};
		\node[black] (D1) at (-1.75, 6.25) {};
		
		\draw[directed] (A1)--(A0);
		\draw[synapse] (B1)--(A1);
		\draw[synapse] (C1)--(A1);
		\draw[directed] (D1)--(C1);
		\draw[directed] (D1)--(B1);
		\draw[directed] (D1)--(B0);
		
		\node[math] at (-3, 4) {$+t_1 \leftrightarrow t_2$};	
	
		\node[white] (A0) at (-4.5, 3) {$t_1$};
		\node[white] (B0) at (-4.5, 1) {$t_2$};
		\node[white] (A1) at (-3.5, 2.5) {};
		\node[white] (B1) at (-2.5, 2.75) {};
		\node[white] (C1) at (-3, 1.75) {};
		\node[white] (D1) at (-1.75, 1.75) {};
		\node (E1) at (-0.75, 1.25) {};
		\node (F1) at (-3, 1) {};
		
		\draw[directed] (A1)--(A0);
		\draw[synapse] (B1)--(A1);
		\draw[synapse] (C1)--(A1);
		\draw[directed] (D1)--(C1);
		\draw[directed] (D1)--(B1);
		\draw[synapse] (E1)--(D1);
		\draw[directed] (F1)--(B0);
		
		\node[math] at (-3, 0) {$+t_1 \leftrightarrow t_2$};
		
		\node[math] at (0, 11) {$\mathbf{d})$};
		\node[white] (A0) at (0.5, 4) {$t_1$};
		\node[white] (B0) at (0.5, 1) {$t_2$};
		\node[white] (A1) at (1.5, 3.5) {};
		\node[white] (B1) at (2.5, 3.75) {};
		\node[white] (C1) at (2, 2.75) {};
		\node[white] (D1) at (3.25, 2.75) {};
		\node[white] (E1) at (4.25, 2.25) {};
		\node[black] (F1) at (5.25, 1.75) {};
		
		\draw[directed] (A1)--(A0);
		\draw[synapse] (B1)--(A1);
		\draw[synapse] (C1)--(A1);
		\draw[directed] (D1)--(C1);
		\draw[directed] (D1)--(B1);
		\draw[synapse] (E1)--(D1);
		\draw[directed] (F1)--(E1);
		\draw[directed] (F1)--(B0);
		
		\node[math] at (2.5, 0) {$+t_1 \leftrightarrow t_2$};

	\end{tikzpicture}
	\end{center}
	\caption{Construction of Feynman diagrams for the first nonlinear correction to the two-point cumulant (graphs containing one loop). In each panel, we add a new layer of vertices to the diagrams, until we arrive at a source vertex. When there are multiple potential ways to add vertices, we add diagrams to account for each of those constructions. A) External vertices corresponding to the two measurement times, with incoming propagator ($\Delta$) edges. B) Diagrams with one internal vertex added. $t_1 \leftrightarrow t_2$ corresponds to switching the two external vertices in the bottom diagram; the top diagram is symmetric with respect to that switch. C) Diagrams with two layers of vertices. The top diagram finishes that of B, top. The second two arise from the second diagram of B, and each also have copies with $t_1 \leftrightarrow t_2$. D) Last diagrams containing one loop. The final diagrams corresponding to the one-loop correction to the second cumulant are the top two of C) and that of D).} \label{fig:exampleFeynman}
	\end{adjustwidth}
\end{figure}

Note: this construction neglects several diagrams (see \nameref{correction}). In the networks we study here, those neglected terms are small (of third order in the coupling strength).

\section*{Demonstrating  the interplay between single-neuron transfer, connectivity structure, and network dynamics}

Our methods predict how spike time statistics of all orders emerge from the interplay of single-neuron input-output properties and the structure of network connectivity.  Here we demonstrate how these methods can be used to predict key phenomena in recurrent spiking networks: the fluctuations and stability of population activity, and stimulus coding.  
First, we isolate the contributions of nonlinearities in single-neuron dynamics to network activity and coding as a whole.  We do so by computing ``one-loop'' correction terms; these correspond to the first structures in our diagrammatic expansion that arise from nonlinear neural transfer. The one-loop corrections provide for the dependence of $n$th order spiking cumulants on $n+1$st order cumulants.  Predictions that would be made by linearizing neural dynamics, as in classic approaches for predicting pairwise correlations \cite{hawkes_spectra_1971, pernice_how_2011, trousdale_impact_2012} and recent ones for higher-order correlations \cite{jovanovic_cumulants_2015}, are described as ``tree-level."  We show how these one-loop corrections, which give new, explicit links between network structure and dynamics (Fig. \ref{fig:schematic}), predict spiking statistics, stability, and the accuracy of coding in recurrent networks. 
\subsection*{1.  Recurrent spike-train correlations drive firing statistics and stability in nonlinear networks}

\paragraph{}
In our analysis of the impact of nonlinear neural transfer on network dynamics, a principal finding was that spike correlations could affect firing rates, as described by the one-loop correction to the mean-field firing rates.  In this section we illustrate the importance of this effect in a class of networks under intensive study in neuroscience:  randomly connected  networks of excitatory and inhibitory cells. We began with a network for which we expect classical theoretical tools to work well, taking the neurons to have threshold-linear transfer functions $\phi(x) = \alpha \lfloor x \rfloor$.  Here, as long as the neurons do not receive input fluctuations that push their rates below this threshold, the ``tree-level'' theory that takes transfer to be entirely linear should work well. We then move on to consider nonlinear effects.

\paragraph{}
As in our original motivational example, we took network connectivity to be totally random (Erd\"os-R\'enyi), with $p_{EE}=0.2$ and $p_{EI} = p_{IE} = p_{II} = 0.5$.   The magnitude of all connections of a given type  ($E-E$, etc.)  was taken to be the same and the time course of all network interactions was governed by the same filter $g(t) = \frac{t}{\tau^2}\exp{(-t/\tau)}$ (with $\tau = 10$ ms), so that $\mathbf{g}_{ij}(t) = \mathbf{W}_{ij}g(t)$. (The matrix $\mathbf{W}$ contains synaptic) The net inhibitory input weight on to a neuron was, on average, twice that of the net excitatory input weight.

\begin{figure}[!h]
\includegraphics[width=5in]{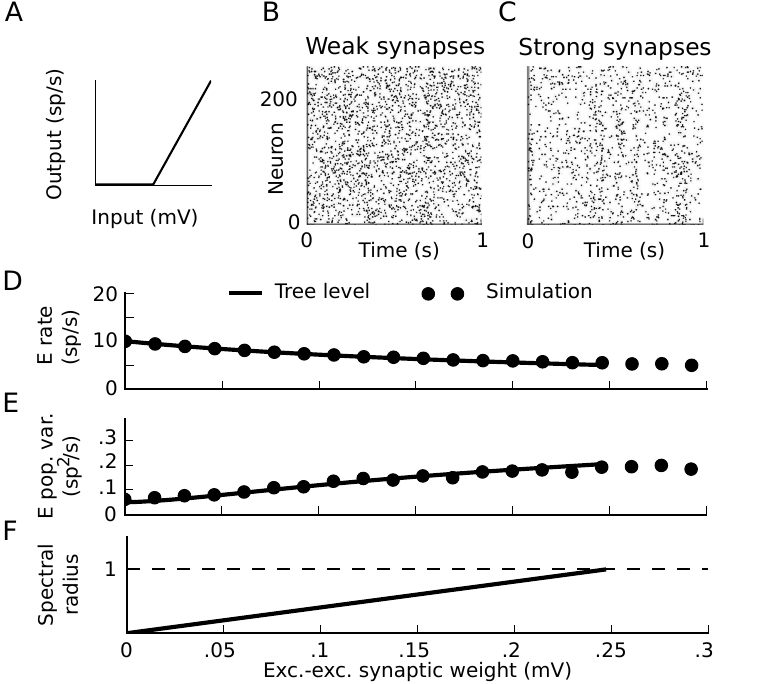} \\
\caption{{\bf Dynamics approaching the firing-rate instability in threshold-linear networks.}  A) Threshold-linear input-rate transfer function. B,C) Raster plots of 1 second realizations of activity for weak and strong synaptic weights. Neurons 0-199 are excitatory and 200-240 are inhibitory. B) $(W_{EE}, W_{EI}, W_{IE}, W_{II}) = (.025, -.1, .01, -.1) $ mV. C) $(W_{EE}, W_{EI}, W_{IE}, W_{II})$ = $(.2, -.8, .08, -0.8) $ mV. D-F)  Average firing rate of the excitatory neurons (D), integral of the auto-covariance function of the summed population spike train (E), and spectral radius of the stability matrix of mean-field theory. (F) vs excitatory-excitatory synaptic weight. While excitatory-excitatory weight is plotted on the horizontal axis, all other synaptic weights increase proportionally with it. Black lines: tree-level theory: mean-field firing rates and covariance computed by linearizing dynamics around it, for each value of synaptic weights. Dots: simulation. }
\label{fig:linear}
\end{figure}

\paragraph{}
We examined the spiking dynamics as the strength of synaptic weights proportionally increased (Fig. \ref{fig:linear}A), and studied network activity with using both theory and direct simulation. Due to the high relative strength of inhibitory synapses in the network, firing rates decreased with synaptic weight (Fig. \ref{fig:linear}D). The magnitude of spike train covariances (reflected by the integrated autocovariance of the summed excitatory population spike train) increased (Fig. \ref{fig:linear}E). These changes were also visible in raster plots of the network's activity (Fig. \ref{fig:linear}B,C).


\paragraph{}
At a critical value of the synaptic weights, the mean-field theory for the firing rates loses stability (Fig. \ref{fig:linear}F). The location of this critical point is predicted by the linear stability of the dynamics around the mean-field rate; the spectrum of the propagator $\mathbf{\Delta}(\omega) = \left(\mathbf{I}-\phi^{(1)} \mathbf{g}(\omega) \right)^{-1}$ diverges when the spectral radius of $\phi^{(1)} \mathbf{g}$ is $\geq 1$.  (This is also the point where the spectral radius of the inverse propagator crosses zero.) Until that critical point, however, the ``tree-level" predictions for both firing rates and spike train covariances (i.e. mean-field theory and linear response theory) provided accurate predictions (Fig. \ref{fig:linear}D,E). This combination of mean-field theory for firing rates and a linearization around it to predict spike train covariances has a long history in theoretical neuroscience (e.g. \cite{ginzburg_theory_1994, doiron_oscillatory_2004, ostojic_how_2009, pernice_how_2011, trousdale_impact_2012}).

\begin{figure}[!h]
\includegraphics[width=5in]{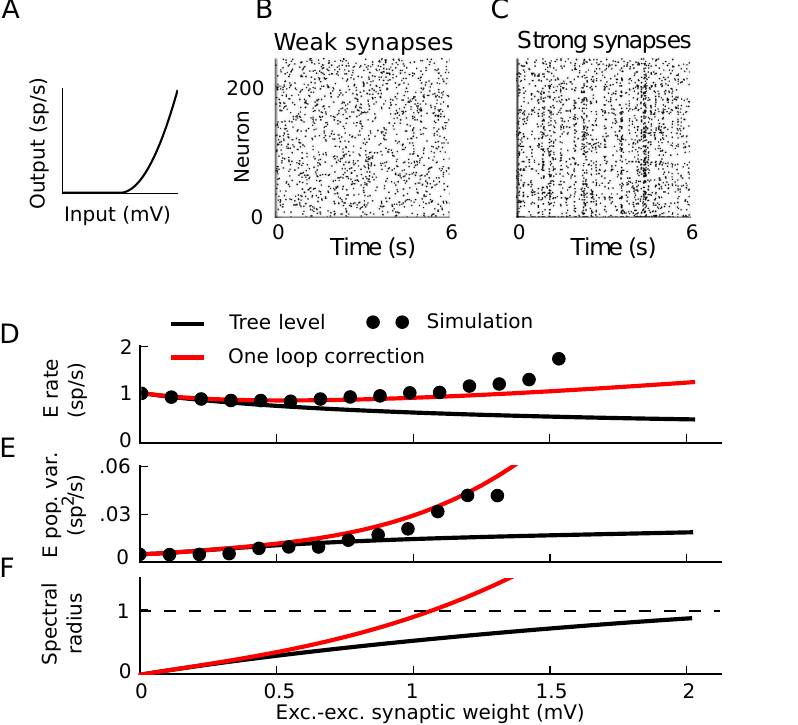} \\
\caption{{\bf Correlation-driven instability in nonlinear networks.} A) Threshold-quadratic input-rate transfer function. B,C) Raster plots of 6 second realizations of activity for weak and strong synaptic weights. Neurons 0-199 are excitatory and 200-240 are inhibitory. B) $(W_{EE}, W_{EI}, W_{IE}, W_{II}) =$  $(.025, -.1, .01, -.1)  $ mV. C) $(W_{EE}, W_{EI}, W_{IE}, W_{II}) =$ $(1.5, -6, .6, -6) $ mV. D-F)  Average firing rate of the excitatory neurons (D), integral of the auto-covariance function of the summed population spike train (E), and spectral radius of the stability matrix of mean-field theory (F) vs excitatory-excitatory synaptic weight. While excitatory-excitatory weight is plotted on the horizontal axis, all other synaptic weights increase proportionally with it. Black line: tree-level theory. Red line: one-loop correction accounting for impact of the next order (pairwise correlations' influence on mean and triplet correlations' influence on pairwise). Dots: simulation. All dots after the one-loop spectral radius crosses 1 represent results averaged over the time period before the activity diverges.}
\label{fig:nonlinear}
\end{figure}

\paragraph{}
We next give a simple example of how nonlinearities in neural transfer cause this standard  tree-level theory (mean-field formulas for rates and linear response theory for covariances) to fail -- and how tractable ``one-loop" corrections from our theory give a much-improved description of network dynamics.  We take the same network as above, but replace neurons' threshold-linear transfer functions with a rectified power law $\phi(x) = \alpha \lfloor x \rfloor^p$ (Fig. \ref{fig:nonlinear}A). This has been suggested as a good description of neural transfer near threshold \cite{heeger_normalization_1992, miller_kenneth_d._neural_2002, priebe_contribution_2004, priebe_direction_2005, priebe_mechanisms_2006}. For simplicity, we take the power law to be quadratic ($p=2$). As we increased synaptic weights, the tree-level theory qualitatively failed to predict the magnitude of spike train covariances and firing rates (Fig. \ref{fig:nonlinear}D,E black curve vs dots). This occurred well before the mean-field firing rates lost stability (Fig. \ref{fig:nonlinear}F, black).

\paragraph{}
Higher-order terms of the loop expansion described above (\hyperref[sec:nonlinear]{Nonlinearities impose bidirectional coupling between different orders of activity}) provide corrections to mean-field theory for both firing rates and spike train correlations. These corrections represent coupling of higher-order spike train cumulants to lower order cumulants. In the presence of an input-rate nonlinearity, for example, synchronous (correlated) presynaptic spike trains will more effectively drive postsynaptic activity \cite{abeles_role_1982, reid_r._clay_divergence_2001}. This effect is described by the one-loop correction to the firing rates (Fig. \ref{fig:loopSelfExciteQuadratic}). 

\paragraph{}
The one-loop correction for the mean field rate of neuron $i$ in a network is given by the same diagram as the one-loop correction for the nonlinearly self-exciting process, Fig. \ref{fig:loopSelfExciteQuadratic}, but interpreted using the network Feynman rules (Fig. \ref{fig:networkPoissonFeynman}). This yields:
\begin{align}
r_{i,1} = \int_{t_0}^t dt_1 \int_{t_0}^t dt_2 \sum_{j, k}\Delta_{ij}(t,t_1) \frac{1}{2}\phi^{(2)}_j \left(\sum_l g_{jl} \ast \Delta_{lk}(t_1, t_2)\right)^2 \bar{r}_k(t_2),
\end{align}
where $t_0$ is the initial time. This correction was, on average, positive (Fig. \ref{fig:nonlinear}D for excitatory neurons; also true for inhibitory neurons). Similarly to firing rates, the loop expansion provides corrections to higher-order spike train cumulants. The one-loop correction to the spike train covariances (Eq. \eqref{eq:2point1loop}, derived in Fig. \ref{fig:exampleFeynman}) accounts for the impact of triplet correlations (third joint spike train cumulants) on pairwise correlations and provided an improved prediction of the variance of the population spike train as synaptic weights increased (Fig. \ref{fig:nonlinear}E).

\paragraph{}
Since the one-loop correction to the firing rates could be large, we also asked whether it could impact the stability of the firing rates - that is, whether pairwise correlations could, through their influence on firing rates through the nonlinear transfer function, induce an instability. This is a question of when the eigenvalues of the propagator diverge---or equivalently, when the eigenvalues of the inverse propagator cross zero. The inverse of the one-loop propagator is given by the ``proper vertex" obtained by amputating the outside propagator edges of the one-loop correction to the propagator; or equivalently, calculated from the Legendre transform of the cumulant-generating functional \cite{zinn-justin_quantum_2002}. We can heuristically derive the one-loop stability correction as follows. 

\paragraph{}
The full propagator, $\Delta$, obeys the expansion 
\begin{align}
\mathbf{\Delta} &= \bar{\mathbf{\Delta}} + \mathbf{\Delta}_1 + \mathbf{\Delta}_2 + \ldots
 \label{eq:propExpansion}
\end{align}
where $\bar{\mathbf{\Delta}}$ is the tree-level propagator, $\Delta_1$ is the one-loop correction, two-loop corrections are collected in $\mathbf{\Delta}_2$, and so on (Fig. \ref{fig:loopStability}A). The one-loop correction is of the form $\bar{\mathbf{\Delta}} \mathbf{\Gamma}_1 \bar{\mathbf{\Delta}} $; the diagram begins and ends with the tree-level propagator, and we label the loop $\mathbf{\Gamma}_1$. The first two-loop correction is a chain of loops (Fig. \ref{fig:loopStability}A), and so can also be factored as $ \bar{\mathbf{\Delta}} \mathbf{\Gamma}_1 \bar{\mathbf{\Delta}} \mathbf{\Gamma}_1 \bar{\mathbf{\Delta}}$. We can represent this factorization diagrammatically by pulling out the tree-level propagator and the loop $\mathbf{\Gamma}_1$ (Fig. \ref{fig:loopStability}B). Just as at two-loop order we were able to factor out a factor $ \bar{\mathbf{\Delta}} \mathbf{\Gamma}_1$ and obtain the expansion of the propagator to one loop, continuing to higher-orders in the loop expansion of the full propagator would all the rest of the full propagator with factors of  $\bar{\mathbf{\Delta}} \mathbf{\Gamma}_1$ in front. The remaining terms would have factors starting with the two-loop correction, and so forth. 

\begin{figure}
\begin{adjustwidth}{-2.5 in}{0 in}
	\begin{center}
	\begin{tikzpicture}

		\node[math] at (-11, 12) {$\mathbf{A}$};
		\node[math] at (-11, 11) {$\mathbf{\Delta} = $};
		
		\node (a) at (-10.5, 11) {};
		\node (b) at (-9.5, 11) {};
		\draw[directed] (b)--(a);
		
		\node[math] at (-9, 11) {$+$};

		\node (A0) at (-3.5, 11) {};
		\node[white] (A1) at (-4.5, 11) {};
		\node[white] (A) at (-5.5,11) {};
		\node[white] (B) at (-6.5, 10.5) {};
		\node[white] (C) at (-6.5, 11.5) {};
		\node[white] (D) at (-7.5, 11) {};
		\node (E) at (-8.5, 11) {};
	
		\draw[directed] (A0)--(A1);
		\draw[synapse] (A1)--(A);
		\draw[directed] (A)-- (B);
		\draw[directed] (A) -- (C);
		\draw[synapse] (B) -- (D);
		\draw[synapse] (C) -- (D);
		\draw[directed] (D) -- (E);

		\node[math] at (-3, 11) {$+$};
		
		\node[white] (AA0) at (2.5, 11) {};
		\node[white] (AA1) at (1.5, 11) {};
		\node[white] (AA) at (.5,11) {};
		\node[white] (BB) at (-.5, 10.5) {};
		\node[white] (CC) at (-.5, 11.5) {};
		\node[white] (DD) at (-1.5, 11) {};
		\node (EE) at (-2.5, 11) {};
	
		\draw[directed] (AA0)--(AA1);
		\draw[synapse] (AA1)--(AA);
		\draw[directed] (AA)-- (BB);
		\draw[directed] (AA) -- (CC);
		\draw[synapse] (BB) -- (DD);
		\draw[synapse] (CC) -- (DD);
		\draw[directed] (DD) -- (EE);
	
		\node (EEE0) at (6.5, 11) {};
		\node[white] (AAA1) at (5.5, 11) {};
		\node[white] (AAA) at (4.5,11) {};
		\node[white] (BBB) at (3.5, 10.5) {};
		\node[white] (CCC) at (3.5, 11.5) {};
	
		\draw[directed] (EEE0)--(AAA1);
		\draw[synapse] (AAA1)--(AAA);
		\draw[directed] (AAA)-- (BBB);
		\draw[directed] (AAA) -- (CCC);
		\draw[synapse] (BBB) -- (AA0);
		\draw[synapse] (CCC) -- (AA0);

		\node[math] at (-9, 8.5) {$+$};
		\node (A) at (-8.5, 8.5) {};
		\node[white] (AA) at (-7.5, 8.5) {};
		\node[white] (BB0) at (-6.5, 8.5) {};
		\node[white] (BB1) at (-6.5, 9.5) {};
		\node[white] (BB2) at (-6.5, 7.5) {};
		\node[white] (C) at (-5.5, 8.5) {};
		\node[white] (D) at (-4.5, 8.5) {};
		\node (E) at (-3.5, 8.5) {};
		
		\draw[directed] (AA)--(A);
		\draw[synapse] (BB0)--(AA);
		\draw[synapse] (BB1)--(AA);
		\draw[synapse] (BB2)--(AA);
		\draw[directed] (C)--(BB0);
		\draw[directed] (C)--(BB1);
		\draw[directed] (C)--(BB2);
		\draw[synapse] (D)--(C);
		\draw[directed] (E)--(D);

		\node[math] at (-3, 8.5) {$+$};
		\node (A) at (-2.5, 8.5) {};
		\node[white] (AA) at (-1.5, 8.5) {};
		
		\node[white] (AA0) at (-.5, 9) {};
		\node[white] (AA1) at (.5, 9.5) {};
		\node[white] (AA2) at (1.5, 10) {};
		\node[white] (AA3) at (1.5, 9) {};
		\node[white] (AA4) at (2.5, 9.5) {};
		\node[white] (AA5) at (3.5, 9) {};
		\node[white] (AA6) at (4.5, 8.5) {};
		\node[white] (BB1) at (1.5, 8) {};
		\node[white] (CC1) at (5.5, 8.5) {};
		\node (DD1) at (6.5, 8.5) {};
		
		\draw[directed] (AA)--(A);
		\draw[synapse] (AA0)--(AA);
		\draw[directed] (AA1)--(AA0);
		\draw[synapse] (AA2)--(AA1);
		\draw[synapse] (AA3)--(AA1);
		\draw[directed] (AA4)--(AA2);
		\draw[directed] (AA4)--(AA3);
		\draw[synapse] (AA5)--(AA4);
		\draw[directed] (AA6)--(AA5);
		\draw[synapse] (BB1)--(AA);
		\draw[directed] (AA6)--(BB1);
		\draw[synapse] (CC1)--(AA6);
		\draw[directed] (DD1)--(CC1);
		
		
		\node[math] at (-9, 6) {$+$};
		
		\node (A) at (-8.5, 6) {};
		\node[white] (AA) at (-7.5, 6) {};
		
		\node[white] (AA0) at (-6.5, 5.5) {};
		\node[white] (AA1) at (-5.5, 5) {};
		\node[white] (AA2) at (-4.5, 5.5) {};
		\node[white] (AA3) at (-4.5, 4.5) {};
		\node[white] (AA4) at (-3.5, 5) {};
		\node[white] (AA5) at (-2.5, 5.5) {};
		\node[white] (AA6) at (-1.5, 6) {};
		\node[white] (BB1) at (-4.5, 6.5) {};
		\node[white] (CC1) at (-.5, 6) {};
		\node (DD1) at (.5, 6) {};
		
		\draw[directed] (AA)--(A);
		\draw[synapse] (AA0)--(AA);
		\draw[directed] (AA1)--(AA0);
		\draw[synapse] (AA2)--(AA1);
		\draw[synapse] (AA3)--(AA1);
		\draw[directed] (AA4)--(AA2);
		\draw[directed] (AA4)--(AA3);
		\draw[synapse] (AA5)--(AA4);
		\draw[directed] (AA6)--(AA5);
		\draw[synapse] (BB1)--(AA);
		\draw[directed] (AA6)--(BB1);
		\draw[synapse] (CC1)--(AA6);
		\draw[directed] (DD1)--(CC1);
		
		\node[math] at (1, 6) {$+$};
		\node[math] at (2.5, 6) {$\mathcal{O}(\mathrm{3 \; loops})$};
	
		\node[math] at (-11, 4) {$\mathbf{B}$};
		\node[math] at (-11, 3) {$\mathbf{\Delta}=$};
		\node (a) at (-10.5, 3) {};
		\node (b) at (-9.5, 3) {};
		
		\draw[directed] (b)--(a);
	
		\node[math] at (-9, 3) {$+$};
		
		\node (a) at (-8.5, 3) {};
		\node[white] (b) at (-7.5, 3) {};
		\node[white] (c1) at (-6.5, 3.5) {};
		\node[white] (c2) at (-6.5, 2.5) {};
		\node[white] (d) at (-5.5, 3) {};
		\node[white] (e) at (-4.5, 3) {};
		
		\draw[directed] (b)--(a);
		\draw[synapse] (c1)--(b);
		\draw[synapse] (c2)--(b);
		\draw[directed] (d)--(c1);
		\draw[directed] (d)--(c2);
		\draw[synapse] (e)--(d);
	
		\node[math] at (-4.25, 3) {$\Bigg[$};
	
		\node (a) at (-4.25, 3) {};
		\node (b) at (-3.25, 3) {};
		\draw[directed] (b)--(a);
		
		\node[math] at (-3, 3) {$+$};
	
		\node (A0) at (2.5, 3) {};
		\node[white] (A1) at (1.5, 3) {};
		\node[white] (A) at (0.5,3) {};
		\node[white] (B) at (-.5, 2.5) {};
		\node[white] (C) at (-.5, 3.5) {};
		\node[white] (D) at (-1.5, 3) {};
		\node (E) at (-2.5, 3) {};
	
		\draw[directed] (A0)--(A1);
		\draw[synapse] (A1)--(A);
		\draw[directed] (A)-- (B);
		\draw[directed] (A) -- (C);
		\draw[synapse] (B) -- (D);
		\draw[synapse] (C) -- (D);
		\draw[directed] (D) -- (E);
		
		\node[math] at (3, 3) {$+$};
		\node[math] at (4, 3) {$\ldots \Bigg]$};
		\node[math] at (5, 3) {$+$};
		\node[math] at (6.5, 3) {$\mathcal{O}(2 \; \mathrm{loops})$};

		\node[math] at (-10.75, 1) {$=$};
		\node (a) at (-10.5, 1) {};
		\node (b) at (-9.5, 1) {};
		
		\draw[directed] (b)--(a);
	
		\node[math] at (-9, 1) {$+$};
		
		\node (a) at (-8.5, 1) {};
		\node[white] (b) at (-7.5, 1) {};
		\node[white] (c1) at (-6.5, 1.5) {};
		\node[white] (c2) at (-6.5, 0.5) {};
		\node[white] (d) at (-5.5, 1) {};
		\node[white] (e) at (-4.5, 1) {};
		
		\draw[directed] (b)--(a);
		\draw[synapse] (c1)--(b);
		\draw[synapse] (c2)--(b);
		\draw[directed] (d)--(c1);
		\draw[directed] (d)--(c2);
		\draw[synapse] (e)--(d);
		
		\node[math] at (-4, 1) {$ \Big[ \mathbf{\Delta} \Big] $};
		\node[math] at (-3, 1) {$+$};
		\node[math] at (-1.5, 1) {$\mathcal{O}(2 \; \mathrm{loops})$};

	\end{tikzpicture}
	\end{center}
	\caption{Calculation of the one-loop stability correction. A) Loop expansion of the full propagator. B) Factorization of the loop and resumming of the full propagator after that factorization.} \label{fig:loopStability}
	\end{adjustwidth}
\end{figure}

\paragraph{}
Pulling out all terms of Eq. ~\eqref{eq:propExpansion} that begin with  $\bar{\mathbf{\Delta}} \mathbf{\Gamma}_1$ and summing them allows us to write (Fig. \ref{fig:loopStability}B):
\begin{align}
\mathbf{\Delta} = \bar{\mathbf{\Delta}} +  \bar{\mathbf{\Delta}} \mathbf{\Gamma}_1 \mathbf{\Delta} + \mathcal{O}(\mathrm{2 \; loops})
\end{align}
We now truncate at one loop, and operate on both sides with the inverse of the tree-level propagator:
\begin{align}
\mathbf{\Gamma}_0 \mathbf{\Delta} &\approx \mathbf{\Gamma}_0 \bar{\mathbf{\Delta}} +  \mathbf{\Gamma}_0 \bar{\mathbf{\Delta}} \mathbf{\Gamma}_1 \mathbf{\Delta} \\
&= \mathbf{I} \delta + \mathbf{\Gamma}_1\mathbf{\Delta}
\end{align}
, revealing that $-\mathbf{\Gamma}_1$ is the one-loop correction to the inverse propagator. From the Feynman rules (Fig. \ref{fig:networkPoissonFeynman}), that factor is:

\Kcomment{Where does this come from?  Also, can we see directly from this formula that the stability is lost earlier?}
\begin{align}
\mathbf{\Gamma}_{jm, 1} = \int_{t_0}^t dt_1 \int_{t_0}^t dt_2 \sum_{k} \frac{1}{2}\phi^{(2)}_j \left(\sum_l g_{jl} \ast \Delta_{lk}(t_1, t_2)\right)^2\phi^{(1)}_k g_{km}
 \end{align}
where $\phi^{(2)}_j$ denotes the second derivative of the transfer function of neuron $j$, evaluated at its mean-field input (and similar for $\phi^{(1)}_k$). The eigenvalues of this provide a correction to the stability analysis based on the tree-level propagator. This predicted that the firing rates should lose stability significantly before the bifurcation of the mean-field theory (Fig. \ref{fig:nonlinear}F, red vs black). Indeed, we saw in extended simulations that the spiking network could exhibit divergent activity even with synaptic weights which the mean-field theory predicted should be stable (Fig. \ref{fig:intro}C). In summary, 
mean-field theory can mis-predict the bifurcation of the rate of spiking models since it fails to capture the impact of correlations on firing rates through nonlinear transfer functions.
\subsection*{2.  Impact of connectivity structure on correlation-driven instabilities in nonlinear networks}
\paragraph{}
Recent work has shown that cortical networks are more structured than simple Erd\H{o}s-R\'enyi networks (e.g. \cite{song_highly_2005, perin_synaptic_2011, lee_anatomy_2016, yoshimura_excitatory_2005, yoshimura_fine-scale_2005, morgenstern_multilaminar_2016}).  One feature of cortical networks is a broad spread of neurons'  in- and out-degree distributions (i.e., the distributions of the number of synaptic inputs that each neuron receives or sends); another is broadly spread synaptic weights.  These network properties, in turn, can have a strong impact on population activity \cite{iyer_influence_2013, hu_motif_2013, hu_local_2014, roxin_role_2011, zhao_synchronization_2011}.  Here, we illustrate the link between network structure and activity in the presence of nonlinear neural transfer.  To generate structured networks, we began with the type of excitatory-inhibitory networks discussed in the previous section, but took the excitatory-excitatory coupling to have both heavy-tailed degree and weight distributions. Specifically, we took it to have truncated, correlated power law in- and out-degree distributions (\hyperref[methods:nonERnetwork]{Methods: non-Erd\H{o}s-R\'enyi network model}). We then took to the synaptic weights to be log-normally distributed  \cite{song_highly_2005, cossell_functional_2015}. For simplicity, we took the location and scale parameters of the weight distribution to be the same.

\begin{figure}[ht!]
\includegraphics[]{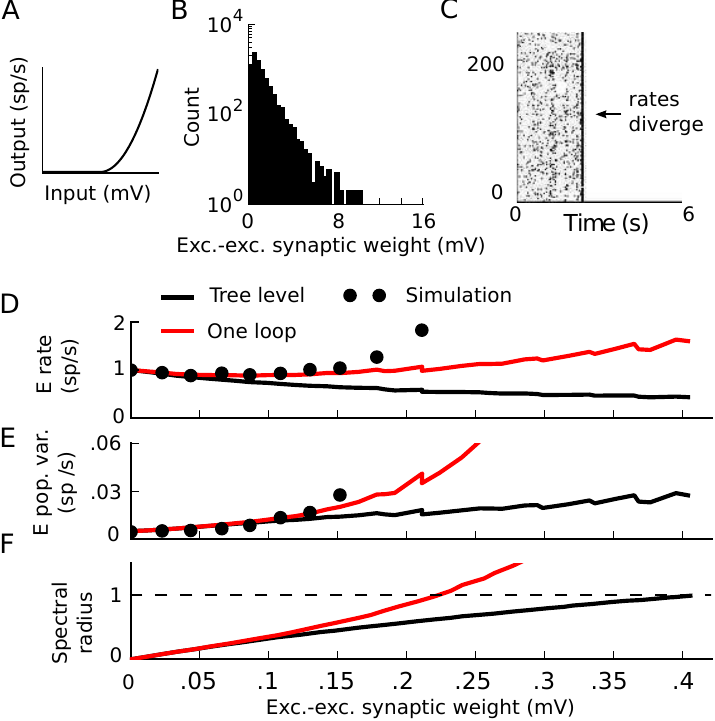} \\
\caption{{\bf Correlation-driven instability in a non-Erd\H{o}s-R\'enyi network with broadly distributed excitatory-excitatory weights.} A) Threshold-quadratic input-rate transfer function. B) Histogram of excitatory-excitatory synaptic weights with location parameter of $1.42$ (mean of $.29$ mV), corresponding to the simulation in panel C. C)  Raster plots of 6 second realizations of activity. Neurons 0-199 are excitatory and 200-240 are inhibitory.  $(W_{EE}, W_{EI}, W_{IE}, W_{II}) =$ $(1.125, -4.5, .45, -4.5) $ mV. D-F)  Average firing rate of the excitatory neurons (D), integral of the auto-covariance function of the summed population spike train (E), and spectral radius of the stability matrix of mean-field theory (F) vs excitatory-excitatory synaptic weight. While the mean excitatory-excitatory weight is plotted on the horizontal axis, all other synaptic weights increase proportionally with it. Black line: tree-level theory. Red line: one-loop correction. Dots: simulation. If a simulation exhibits divergent activity, the spike train statistics are averaged over the transient time before that divergence for visualization.}
\label{fig:nonlinear_nonER}
\end{figure}

\paragraph{}
We then examined the network dynamics as the location and scale of the excitatory-excitatory synaptic weights increased. For each mean weight, we sampled the excitatory-excitatory weights from a lognormal distribution with that mean and variance. The excitatory-inhibitory, inhibitory-excitatory and inhibitory-inhibitory weights remained delta-distributed. Each such network specified a weight matrix $\mathbf{W}$, which allowed the methods described previously for computing tree-level and one-loop rates, covariances and stability to be straightforwardly applied. For strong and broadly distributed synaptic weights (Fig. \ref{fig:nonlinear_nonER}B), the network exhibited a similar correlation-induced instability as observed in the Erd\H{o}s-R\'enyi network (Fig. \ref{fig:nonlinear_nonER}C) even though mean-field theory predicted that the firing rates should be stable (Fig. \ref{fig:nonlinear_nonER}F, black vs. red curves). As synaptic weights increased from zero, the mean-field theory for firing rates provided a misprediction (Fig. \ref{fig:nonlinear_nonER}D, black line vs dots) and the linear response prediction for the variance of the population spike train also broke down (Fig. \ref{fig:nonlinear_nonER}E, black line vs dots). The one-loop corrections, accounting for the impact of pairwise correlations on mean rates and of triplet correlations on pairwise correlations, yielded improved predictions (Fig. \ref{fig:nonlinear_nonER}D, E red lines) and a much more accurate prediction for when firing rates would lose stability (Fig. \ref{fig:nonlinear_nonER}C, F). These effects were similar to those seen in Er\H{o}s-R\'enyi networks (Fig. \ref{fig:nonlinear}), but the transition of the firing rates occurred sooner, both for the mean field (because of the effect of the weight and degree distributions on the eigenvalues of the weight matrix) and one loop theories (because of the impact of the correlations on the firing rates).

\subsection*{3. Exponential single-neuron nonlinearities}
\paragraph{}
In the previous section, we investigated how a non-Erd\H{o}s-R\`enyi network structure could amplify the one-loop corrections by increasing spike train correlations. We now examine a different single-neuron nonlinearity: $\phi(x) = \alpha e^x$, which is the canonical link function commonly used to fit GLM point process models to spiking data \cite{kass_analysis_2014}. The exponential has arbitrary-order derivatives, so there is no reason to expect the one-loop description to be a sufficient correction to mean-field theory.

\begin{figure}[ht!]
\includegraphics[]{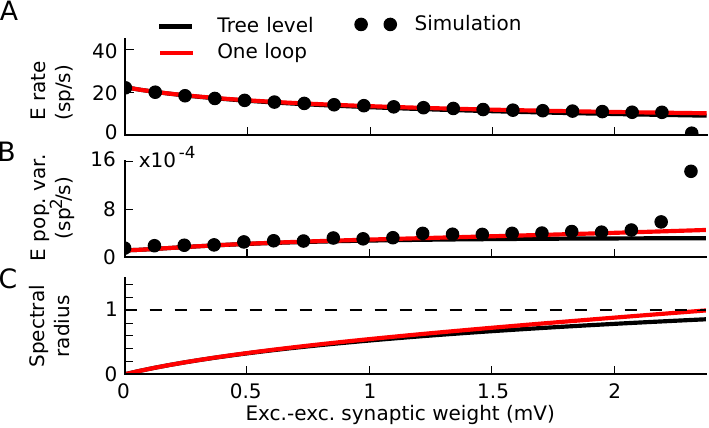} \\
\caption{{\bf Stability of a network with exponential transfer functions.}  A) Mean firing rate of the excitatory neurons. B) Integral of the auto-covariance function of the summed population spike train. C) Spectral radius of the stability matrix of mean-field theory, all (A-C) vs excitatory-excitatory synaptic weight. While the mean excitatory-excitatory weight is plotted on the horizontal axis, all other synaptic weights increase proportionally with it. Black line: tree-level theory. Red line: one-loop correction. Dots: simulation. If a simulation exhibits divergent activity, the spike train statistics are averaged over the transient time before that divergence for visualization.}
\label{fig:exponential_balanced}
\end{figure}

\paragraph{}
As before, we take the mean synaptic weight onto each neuron in the network to be 0. First, we take excitatory and inhibitory neurons to have the same baseline drive, $\lambda_E = \lambda_I = -1.5$. As we scale synaptic weights, we see that the one-loop correction is small compared to the tree-level theory for the firing rates, population variances and stability analysis (Fig. \ref{fig:exponential_balanced}A-C, red vs. black lines). It nevertheless provides an improved correction for the variance of the excitatory population spike train (Fig. \ref{fig:exponential_balanced}B, between 1.5 and 2 mV synaptic weights). The bifurcation of the one-loop theory is close to the bifurcation of the mean-field theory, and before that point the mean-field theory and one-loop corrections both lose accuracy (Fig. \ref{fig:exponential_balanced}A,B). This makes sense: when the mean-field theory fails, the only reason that the one-loop correction to the rates would be accurate is if all third- and higher-order spike train cumulants are small. Those higher-order correlations are not small near the instability.

\begin{figure}[ht!]
\includegraphics[]{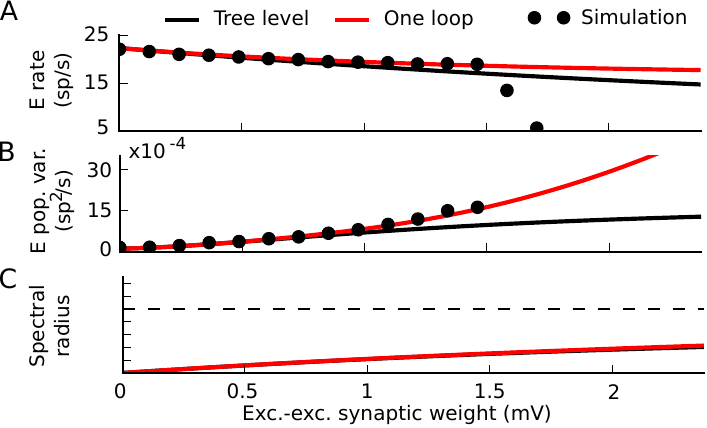} \\
\caption{{\bf Failure of one-loop corrections with exponential transfer functions.}  A) Mean firing rate of the excitatory neurons. B) Integral of the auto-covariance function of the summed population spike train. C) Spectral radius of the stability matrix of mean-field theory, all (A-C) vs excitatory-excitatory synaptic weight. While the mean excitatory-excitatory weight is plotted on the horizontal axis, all other synaptic weights increase proportionally with it. Black line: tree-level theory. Red line: one-loop correction. Dots: simulation. If a simulation exhibits divergent activity, the spike train statistics are averaged over the transient time before that divergence for visualization.}
\label{fig:exponential_unbalanced}
\end{figure}

\paragraph{}
Next, we broke the symmetry between excitatory and inhibitory neurons in the network by giving inhibitory neurons a lower baseline drive ($\lambda_I = -2.$). This shifted the bifurcation of the mean-field theory and the one-loop correction to much higher synaptic weights (Fig. \ref{fig:exponential_unbalanced}C). For intermediate synaptic weights, we saw that the one-loop correction provided a better match to simulations than the tree-level theory (Fig. \ref{fig:exponential_unbalanced}A, B, between 1 and 1.5 mV synaptic weights). For stronger synapses, however, the simulations diverged strongly from the tree-level and one-loop predictions (Fig. \ref{fig:exponential_unbalanced}A,B, around 1.5 mV synaptic weights). In principle, we could continue to calculate additional loop corrections in an attempt to control this phenomenon. The exponential has arbitrary-order derivatives, however, so we might need infinitely many orders of the loop expansion--suggesting a renormalization approach \cite{zinn-justin_quantum_2002}, which is beyond the scope of this article. In sum, with an exponential transfer function we saw that for intermediate synaptic weights, the one-loop correction improved on the tree-level theory. For strong enough synaptic weights, however, both failed to predict the simulations. How soon before the mean-field bifurcation this failure occurred depended on the specific model. 

\section*{Discussion}
\paragraph{}
Joint spiking activity between groups of neurons can control population coding and controls the evolution of network structure through plasticity. Theories for predicting the joint statistics of activity in model networks have been locally linear so far. We present a systematic and diagrammatic fluctuation expansion (or, in reference to those diagrams, loop expansion) for spike-train cumulants, which relies on a stochastic field theory for networks of stochastically spiking neurons. It allows the computation of arbitrary order joint cumulant functions of spiking activity and dynamical response functions which provide a window into the linear stability of the activity, as well as nonlinear corrections to all of those quantities.

\paragraph{}
Using this expansion, we 
investigated how nonlinear transfer can affect firing rates and fluctuations in population activity, imposing a dependence of rates on higher-order spiking cumulants. This coupling could significantly limit how strong synaptic interactions could become before firing rates lost stability. 

\subsection*{Convergence and truncation of the loop expansion}
\paragraph{}
The loop expansion is organized by the dependence of lower-order activity cumulants to higher-order ones. The first-order (''tree level") description of the $n$th activity cumulant does not depend on higher-order cumulants. One loop corrections correspond to dependence of the order $n$ cumulants on the tree-level $n+1$-order cumulants, two-loop corrections correspond to dependence of the order $n$ cumulant on the tree-level $n+1$ and $n+2$ order cumulants, and so on. This coupling arises from the nonlinearity of the single-neuron transfer function $\phi$ (\hyperref[sec:nonlinearity]{Results: Nonlinearities impose bidirectional coupling between different orders of activity: nonlinearly self-exciting process}; \hyperref[methods:pathIntegral]{Methods: Path integral representation}). When the transfer function is linear at the mean-field rates, the tree-level theory provides an accurate description of activity so long as the network is stable (Fig. \ref{fig:linear}). This corresponds to the 2nd and higher-order derivatives of the transfer function $\phi$ with respect to the total input, evaluated at the mean-field rates, being zero. When $\phi$ has non-zero $2nd$ or higher derivatives at the mean-field rates, orders of the loop expansion corresponding to those order derivatives can be important (with one loop corresponding to the second derivative, two loops corresponding to the third derivative, etc.) The magnitude of the $n$-loop correction depends on two things: the magnitude of the $n+1$-order tree-level activity cumulant and the magnitude of the $n+1$st derivative of $\phi$ at the mean-field rates (i.e. the strength of the coupling to that cumulant).

\paragraph{}
Recent work has shown that the the magnitude of order-$n$ activity cumulants depend on the motif structure of the network (Fig. \ref{fig:nonlinear_nonER}; \cite{pernice_how_2011, hu_motif_2013, hu_local_2014, jovanovic_interplay_2016}), as well as on the correlation structure of the inputs it receives. We also used a particular form of the interaction kernel, $g(t)$ and assumed that it had unit integral over $t$; any continuous $g(t)$ with finite integral over $t$ should work as well. This criterion on the integral of $g$ is necessary for the stability of mean-field theory with only excitatory interactions. If $g(t)$ did not have a well-defined integral, an appropriate balance between excitation and inhibition could perhaps still ensure a stable mean-field solution (similar to \cite{van_vreeswijk_chaotic_1998}). If the system lies close to a bifurcation of the mean-field theory, so that the eigenvalues of the propagator diverge, then the mean field theory and this expansion around it can also fail. In that case, renormalization arguments can allow the discussion of the scaling behavior of correlations \cite{buice_field-theoretic_2007}.

\subsection*{Relationship to other theoretical methods}
\paragraph{}
A classic and highly influential tool for analyzing the dynamics of neural rate models with Gaussian-distributed synaptic weights is dynamical mean field theory, which reveals a transition to chaotic rate fluctuations in networks of rate units with Gaussian connectivity \cite{sompolinsky_chaos_1988}. Dynamical mean field theory proceeds, briefly, by taking the limit of large networks and replacing interactions through the quenched heterogeneity of the synaptic weights by an effective Gaussian process mimicking their statistics. Recent extensions of dynamical mean field theory have incorporated a number of simple biological constraints, including positive-valued firing rates \cite{kadmon_transition_2015, harish_asynchronous_2015, mastrogiuseppe_intrinsically-generated_2016} and certain forms of cell type-specific connectivity \cite{aljadeff_transition_2015, kadmon_transition_2015, aljadeff_low-dimensional_2016}. In this framework, spiking is usually only described in the limit of slow synapses as additive noise in the rates which can shift the transition to chaotic rate fluctuations to higher coupling strengths and smooth the dynamics near the transition \cite{kadmon_transition_2015, goedeke_noise_2016}.

\paragraph{}
An alternative approach to dynamical mean-field theory is to start from the bottom up: to posit an inherently stochastic dynamics of single neurons and specify a finite-size network model, and from these derive a set of equations for statistics of the activity \cite{buice_beyond_2013, bressloff_path-integral_2015}. This approach provides a rigorous derivation of a finite-size rate model as the mean field of the underlying stochastic activity, as well as the opportunity to calculate higher-order activity statistics for the activity of a particular network \cite{ohira_master-equation_1993, buice_field-theoretic_2007, bressloff_stochastic_2009, buice_systematic_2010}. This is the approach taken here with the popular and biologically motivated class of linear-nonlinear-Poisson models. A similar approach underlies linear response theory for computing spike train covariances \cite{doiron_oscillatory_2004, de_la_rocha_correlation_2007}, which corresponds to the tree level of the loop expansion presented here. For integrate-and-fire neuron models receiving Poisson inputs, Fokker-Planck theory for the membrane potential distributions can be used to calculate the linear response function of an isolated neuron \cite{fourcaud_dynamics_2002}, which together with the synaptic filter and weight matrix determines the propagator.

\subsection*{Dynamics and stability in spiking networks}
\paragraph{Fluctuations in large spiking networks  }
Networks of excitatory and inhibitory neurons with instantaneous synapses have been shown, depending on their connectivity strengths and external drive, to exhibit a variety of dynamics, including the ``{classical}'' 
asynchronous state, oscillatory population activity, and strong, uncorrelated rate fluctuations \cite{brunel_dynamics_2000, ledoux_dynamics_2011, ostojic_two_2014}. The classical asynchronous state and oscillatory regimes exist in the presence of Poisson-like single-neuron activity, either due to external white noise or to internally generated high-dimensional chaotic fluctuations \cite{van_vreeswijk_chaos_1996, renart_asynchronous_2010}.  Transitions between these modes correspond to bifurcations in which a given state loses stability.  The present results allow one to compute these transition points with greater accuracy, by explicitly computing correlations of arbitrary order and, crucially, how these correlations ``feed back'' to impact firing rates and the stability of states with different rates.

\paragraph{Inhibitory-stabilized and supralinear-stabilized networks}
Beyond the overall stability of network states, an important general question is how firing rates depend on inputs in recurrent neural networks.  ``Inhibitory-stabilized'' networks can have surprising dependencies on inputs, with rates trending in opposite directions from what one would at first expect \cite{tsodyks_paradoxical_1997}.  Supra-linear input-rate transfer in inhibitory-stabilized networks can explain a variety of physiological observations \cite{ahmadian_analysis_2013, ozeki_inhibitory_2009, rubin_stabilized_2015}. Our results are therefore useful in predicting how correlations emerge and couple to firing rates in these networks.  The impact of cell type-specific dynamics on dynamics and coding remains to be fully elucidated \cite{litwin-kumar_inhibitory_2016}.

%

\begin{figure}[h!]
\begin{adjustwidth}{-1 in}{0 in}
\includegraphics[]{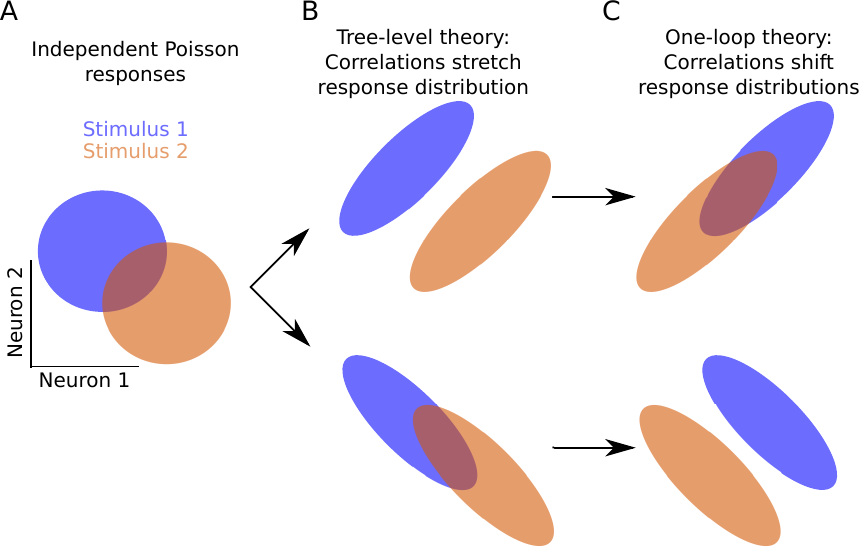} \\
\caption{{\bf Potential impacts of correlations on coding in a presence of a nonlinearity.} A) The independent Poisson assumption for neurons gives rise to uncorrelated distributions of population activity. B) Correlations could increase or decrease the overlap of those distributions by stretching them, decreasing or increasing the decoding performance (top and bottom, respectively). C) The impact of correlations on the mean responses can shift those distributions, potentially counteracting the impact of stretching the distributions (as shown), or exaggerating it.}
\label{fig:codingSchematic}
\end{adjustwidth}
\end{figure}

\paragraph{A new potential impact of correlations on population coding}
Many studies have examined the impact of ``noise correlations" on population coding, examining the vector of neural responses. If all responses are conditionally independent given the stimulus, the distribution of responses to a particular stimulus is spherical.  The discriminability of the responses to two stimuli corresponds to the area of overlap of those multi-neuron response distributions. To tree level in the loop expansion of the population responses, correlations stretch that response distribution. These correlations can either improve or lower coding performance, depending on how they relate to the stimulus-evoked responses \cite{averbeck_neural_2006, hu_sign_2014, moreno-bote_information-limiting_2014, zylberberg_direction-selective_2016, franke_structures_2016}.  In the presence of a nonlinear transfer function, a further potential impact of correlations is to change neurons' mean activities (Fig. \ref{fig:nonlinear}. This corresponds to a translation of the multi-neuron response distributions (Fig. \ref{fig:codingSchematic}, bottom) which could, in principle, either increase or decrease their discriminability (Fig. \ref{fig:codingSchematic}, bottom).

\section*{Materials and Methods}
\subsection*{non-Erd\H{o}s-R\'enyi network model} \label{methods:nonERnetwork}
\paragraph{}
For Fig. \ref{fig:nonlinear_nonER}, we generated the excitatory-excitatory connectivity with a truncated power-law degree distribution. The marginal distributions of the number of excitatory-excitatory synaptic inputs (in-degree) or outputs (out-degree) obeyed:
\begin{equation}
p(d) = 
\begin{cases}C_1d^{\gamma_1}, & 0\leq d \leq L_1 \\
			C_2d^{\gamma_2}, &L_1 \leq L_2 \\
			0, &\text{ else}
\end{cases}
\end{equation}
where $d$ is the in- or out-degree. Parameter values are contained in Table \ref{table:params}; $C_1$ and $C_2$ are normalization constants to make the degree distribution continuous at the cutoff $L_1$. The in- and out-degree distributions were then coupled by a Gaussian copula with correlation coefficient $\rho$ to generate in- and out-degree lists. These lists generated likelihoods for each possible connection proportional to the in-degree of the postsynaptic neuron and the out-degree of the presynaptic neuron. We then sampled the excitatory-excitatory connectivity according to those likelihoods.

\subsection*{Path integral representation \label{methods:pathIntegral}}
\paragraph{}
Here we outline the derivation of a path integral formalism for a network of processes with nonlinear input-rate transfer, following methods developed in nonequilibrium statistical mechanics \cite{martin_statistical_1973, doi_second_1976, doi_stochastic_1976, peliti_path_1985, tauber_field-theory_2007, cardy_non-equilibrium_2008}. We will begin by developing the formalism for a simple model, where a spike train is generated stochastically with intensity given by some input $\nu(t)$. We will specify the cumulant generating functional of the spike train given $\nu(t)$ below. 

\paragraph{}
The general strategy is to introduce an auxiliary variable, called the ``response variable'', whose dynamics will determine how information from a given configuration (e.g. the spike counts $n(t)$) at one time will effect future configurations of the dynamics. Introducing the response variable allows us to write the probability density functional for the process in an exponential form. The integrand of that exponential is called the ``action" for the model, which can then be split into a ``free" action (the piece bilinear in the configuration and response variables) and an ``interacting" one (the remainder). Cumulants of the process can then be computed, in brief, by taking expectations of the configuration and response variables and the interacting action against the probability distribution given by the free action. 

\paragraph{}
Let $n(t)$ be the number of spike events recorded since some fiducial time $t_0$. In a time bin $dt$, $\Delta n$ events are generated with some distribution $p(\Delta n)$ and added to $n(t)$. Let the events generated in any two time bins be conditionally independent given some inhomogeneous rate $\nu(t)$, so that $p(\Delta n) = p(\Delta n | \nu)$. So, assuming that initially $n(t_0)=0$, the probability density functional of the vector of events over $M$ time bins is:
\begin{align}
p[\Delta n(s): s \leq t ] = \prod_{i=1}^M p(\Delta n_i | \nu_i) = \prod_{i=1}^j \int \frac{ d\tilde{n}_i}{2\pi i} e^{-\tilde{n}_i \Delta n_i } P(\tilde{n}_i | \nu_i) = \prod_{i=1}^j \int  \frac{ d\tilde{n}_i}{2\pi i} e^{-\tilde{n}_i \Delta n_i + W[\tilde{n}_i | \nu_i]} 
\end{align}	
where $P(\tilde n_i | \nu_i)$ is the Laplace transform of $p(\Delta n_i | \nu_i)$ and $W[\tilde{n}_i | \nu_i]$ is the cumulant generating functional for the auxiliary variable. In the third step we have written the distribution of $p(\Delta n_i)$ as the inverse Laplace transform of the Laplace transform.  The Laplace transform variable $\tilde{n}_i$ is our auxiliary response variable. In the fourth step we identified the Laplace transform of the probability density functional as the moment generating functional, so that $W[\tilde{n}_i | \nu_i]$ is the cumulant generating functional of the spike count. Note that these are complex integrals. The contour for the integration over $\tilde{n}_i$ is parallel to the imaginary axis.  

\paragraph{}
Taking the continuum limit $M \rightarrow \infty, dt \rightarrow 0$ then yields the probability density functional of the spike train process $\dot{n}$:
\begin{align}
p[\dot{n}] = \int \mathcal{D}\tilde{n}(t) e^{-\int dt (\tilde{n}(t)\dot{n}(t) - W[\tilde{n}(t)] )}
\label{eq:pathIntegralLinear}
\end{align}
where $\mathcal{D}\tilde{n}(t) = \lim_{M \rightarrow \infty} \prod_{i=1}^M  \frac{d\tilde{n}_i}{2\pi i}$ and $\dot{n} = \frac{dn}{dt}$ and we suppress the conditional dependence of $\tilde{n}(t)$ on $\nu(t)$. In the continuum limit the integral is a functional or path integral over realizations of $\tilde{n}(t)$. We will call the negative exponent of the integrand in Eq. \ref{eq:pathIntegralLinear} the action:
\begin{align}
S[\tilde{n}, \dot{n}] = \int dt \big(\tilde{n}\dot{n} - W[\tilde{n}]  \big)
\end{align}
We have slightly abused notation here in that a factor of $1/dt$ has been absorbed into $W[\tilde{n}]$.  We will justify this below.

\paragraph{}
We have not yet specified the conditional distribution of the events given the input $\nu(t)$, leaving $W[\tilde{n}(t)]$ unspecified. Here, we will take the events to be conditionally Poisson \cite{lefevre_dynamics_2007}, so that 
\begin{align}
W[\tilde{n}] = \left(e^{\tilde{n}}-1\right)\nu(t)
\end{align}
(In the continuum limit, the rate $\nu(t)$ allowed us to absorb the factor of $1/dt$ into $W$. A finite size time bin would produce $\nu(t)dt$ events in bin $dt$.) 

\paragraph{}
This representation of the probability density functional yields the joint moment generating functional (MGF) of $n$ and $\tilde{n}$:
\begin{align}
Z[J, \tilde{J}] = \int \mathcal{D}\dot{n}(t)\, \int \mathcal{D}\tilde{n}(t)\, e^{-S[\tilde{n}, \dot{n}] + J \tilde{n} + \tilde{J}n}
\end{align}
and the moment generating functional of $\dot{n}$:
\begin{align}
Z[\tilde{J}] = \int \mathcal{D}\dot{n}(t)\, \int \mathcal{D}\tilde{n}(t)\, e^{-S[\tilde{n}, n] + \tilde{J}\dot{n}}
\end{align}

\paragraph{}
The above strictly applies only to the inhomogeneous Poisson process.  This formalism is adapted to the self-exciting process by introducing conditional dependence of the rate $\nu(t)$ on the \emph{previous} spiking history.  In the discrete case, before taking the limit $M\rightarrow \infty$, we say that the rate $\nu_i = \phi[n_{i-}]$, where $\phi$ is some positive function and $n_{i-}$ indicates all spiking activity up to but not including bin $i$.  This requirement is equivalent to an Ito interpretation for the measure on the stochastic process $\dot{n}(t)$.  Because of this assumption, the previous derivation holds and we can write
\begin{align}
W[\tilde{n}] = \left(e^{\tilde{n}(t)}-1\right)\phi(\dot{n}(< t))
\end{align}
where $\dot{n}(<t) = \dot{n}(s): s<t$. In the continuum limit, there is an ambiguity introduced by the appearance of the time variable $t$ in both $\tilde{n}(t)$ and $\dot{n}(t)$.  This is resolved in the definition of the measure for the functional integral, and affects the definition of the linear response (below).  Again, this is a manifestation of the Ito assumption for our process.

\paragraph{}
The specific model used in this paper assumes a particular form for the argument of $\phi$.  We assume that the input is given by
\begin{align}
     \nu(t) = \phi( (g \ast \dot{n})(t) + \lambda(t))
\end{align}
where $g(t)$ is a filter that defines the dynamics of the process in question and $\lambda(t)$ is an inhomogeneous rate function.  The result is that the action for non-linearly self-exciting process is given by
\begin{align}
S[\tilde{n}, \dot{n}] = \int dt  \; \Big(\tilde{n}\dot{n} - \left(e^{\tilde{n}(t)}-1\right)\phi \big( (g \ast \dot{n})(t) + \lambda(t) \big)  \Big)
\end{align}

\paragraph{}
The only extension required to move from the above action to the network model is to introduce indices labelling the neurons and couplings specific for each neuron pair.  Nothing of substance is altered in the above derivation and we are left with \Gcomment{Removed subscript indices on $\tilde{n}, n$}
\begin{align}
S[\tilde{n}, n] =\sum_i \int dt \; \Bigg(\tilde{n}_i\dot{n}_i - \left(e^{\tilde{n}_i (t)}-1\right)\phi \Big( \sum_j (\mathbf{g}_{ij} \ast \dot{n}_j)(t) + \lambda_i(t)\Big)  \Bigg)
\end{align}

\subsubsection*{Mean field expansion and derivation of Feynman rules. \label{methods:meanField}}
\paragraph{}
We could use the above action in order to derive Feynman rules for these processes.  The expansions so described would be equivalent to our initial expansions before resumming (the sets of diagrams that use dashed lines).  These would describe an expansion about $\dot{n}(t)=0$.  We can arrive at  this expansion by separating the action into two pieces, called the ``free'' action and the ``interacting'' action: $S[\tilde{n},n] = S_0[\tilde{n},n] + S_{V}[\tilde{n},n]$.  The free action, $S_0[\tilde{n},n]$ is defined by the bilinear component of $S[\tilde{n},n]$ in an expansion around $0$, i.e. \Gcomment{added indices on $n, \tilde{n}$ on either side of $K$}
\begin{align}
	S_0[\tilde{n},n] = - \sum_{i,j} \int dt dt' \tilde{n}_i(t) K_{ij}(t,t') n_j(t')
\end{align}
for some operator $K_{ij}(t,t')$. Define
\begin{align}
	\langle n_i(t) \tilde{n}_j(t') \rangle = \Delta_{ij}(t,t')
\end{align}
Taking the expectation with respect to the probability density given by the free action yields
\begin{align}
	\int ds \; K_{ik}(t,s)\Delta_{kj}(s,t') = \delta(t-t')\delta_{ij}
\end{align}
so that $K$ is the operator inverse of $\Delta$ under the free action. That expectation can be computed via the moment generating functional for the free action (which we denote $Z_0[\tilde{J}, J]$), and then completing the square in order to compute the integral. This leaves
\begin{align}
	Z_0[J, \tilde{J}] = e^{\sum_{i,j} \int dt dt' \; \tilde{J}_i(t) \Delta_{ij} (t,t')J_j(t')}  \label{eq:freeGenFunc}
\end{align}
which implies that $\left \langle \dot{n}_i(t)\tilde{n}_j(t') \right \rangle = \Delta_{ij}(t, t')$.  We have used the fact that $Z_0[\tilde{J}, J]=1$.

\paragraph{}
Computing moments requires functional integrals of the form
\begin{align}
	\langle \prod_i \dot{n}_i(t_i) \prod_j \tilde{n}_j(t_j) \rangle = \int \mathcal{D}\dot{n}(t) \mathcal{D}\tilde{n}(t) \; \prod_i \dot{n}_i(t_i) \prod_j \tilde{n}_j(t_j) e^{-S[\tilde{n}, \dot{n}]}  \label{eq:generalMomentPathIntegral}
\end{align}
We Taylor expand each neuron's nonlinearity $\phi$ (around its $\lambda_i(t)$) and expand the exponential arising from the cumulant generating functional of the spike counts (that in $\left(e^{\tilde{n}}-1\right)$) around zero. We then collect the terms with one power of $\tilde{n}_i$ and of $\dot{n}_i$ in the free action. This leaves the interacting action $S_V[\tilde{n}, \dot{n}]$ as:
\begin{align}
S_V = - \sum_i \int dt \sum^\infty_{\substack{{p, q = 0} \\ {\backslash(p=q=1)}}} \frac{1}{p!} \frac{\phi_i^{(q)}}{q!}\tilde{n}_i^p \left(\sum_j \mathbf{g}_{ij}\ast \dot{n}_j \right)^q 
\end{align}
Note that at each term in this expansion, each of the $p$ factors of $\tilde{n}_i$ and the $q$ factors of $\sum_j \mathbf{g}_{ij}\ast \dot{n}_j$ carries its own time variable, all of which are integrated over; we have suppressed these time variables and their integrals. Now the action can be written as:
\begin{align}
	S[\tilde{n}, \dot{n}] = - \tilde{n}_i K_{ij} \dot{n}_j - \sum^\infty_{\substack{{p, q = 0} \\ {\backslash(p=q=1)}}}\frac{1}{p!}V^i_{p, q} \tilde{n}_i^{p} \Big(\sum_j \mathbf{g}_{ij}\ast \dot{n}_j\Big)^{q} \label{eq:actionVertex}
\end{align}
where we have suppressed the sums over neuron indices and all time integrals.  We have defined the ``vertex factor" $V^i_{pq} = \phi_i^{(q)} / q!$ (the index $p$ recalls which power of $\tilde{n}$ it arrived with). Note that we have defined vertex factors with a minus sign relative to $S_V[\tilde{n}, \dot{n}]$. 
Introducing the shorthand $(g \ast \dot{n})_i = \sum_j \mathbf{g}_{ij} \ast \dot{n}_j$, and then again suppressing neuron indices, we write the moment in Eq. ~\eqref{eq:generalMomentPathIntegral} as:
\begin{align}
	\langle \dot{n}^p \tilde{n}^q  \rangle = \int {\cal D}\tilde{n} {\cal D} \dot{n} \; \dot{n}^p \tilde{n}^q e^{-S} &=  \int {\cal D}\tilde{n} {\cal D} \dot{n} \; \dot{n}^p \tilde{n}^q e^{\tilde{n} K \dot{n} + \sum_{p,q} \frac{1}{p !}V_{pq} \tilde{n}^{p} (g \ast \dot{n})^{q}} \\
	&= \int {\cal D}\tilde{n} {\cal D} \dot{n} \; \dot{n}^p  \tilde{n}^q \prod^\infty_{\substack{{p, q = 0} \\ {\backslash(p=q=1)}}}  \sum_{l = 0}^\infty \frac{1}{l!} \Bigg( \frac{1}{p!} V_{pq} \tilde{n}^{p} (g \ast \dot{n})^{q} \Bigg )^{l} e^{\tilde{n} K \dot{n}} \\
	&=\Bigg \langle  \dot{n}^p  \tilde{n}^q \prod^\infty_{\substack{{p, q = 0} \\ {\backslash(p=q=1)}}} \sum_{l = 0}^\infty \frac{1}{l!} \Big( \frac{1}{p!} V_{pq} \tilde{n}^{p} (g \ast \dot{n})^{q} \Big)^{l} \Bigg \rangle_0  \label{eq:momentExpVertex}
\end{align}
where we denote the expectation with respect to the free action $S_0[\tilde{n}, \dot{n}]$ by $\langle \rangle_0$. Expectations with respect to the free action are determined by its generating functional, Eq. ~\eqref{eq:freeGenFunc}.  Due to Wick's theorem, any moment will decompose into products of expectation values $\Delta_{ij}(t, t') = \left \langle \dot{n}_i(t) \tilde{n}_j(t') \right \rangle_0$, according to all possible ways of partitioning the operators into $\tilde{n}, \dot{n}$ pairs, i.e. 
\begin{align}
	\langle \dot{n}^p \tilde{n}^q \rangle_0 = \sum_{\substack{{\rm pair-wise} \\{\rm partitions}}} \prod_{\rm pairs} \Delta_{ij}(t,t')
\end{align}
where the indices $i,j, t, t'$ are determined by the partitioning.  For the terms in the expansion (\ref{eq:momentExpVertex}), each term will be decomposed into a sum over ways in which factors of $\tilde{n}$ can be paired with factors of $\dot{n}$ \cite{chow_path_2015}.

\paragraph{}
We can represent each term in this sum diagrammatically by associating each of the $p$ factors of $\dot{n}$ and $q$ factors of $\tilde{n}$ from the moment with external vertices with a single outgoing or exiting line, respectively.  Each vertex factor $V_{pq}$ gets a vertex with $p$ lines exiting to the left (towards the future) and $q$ wavy lines entering from the right (from the past).  The partitions of pairing $\tilde{n}$ and $n$ are determined by connecting outgoing lines to incoming lines.  The terms in the expansion with $l$ powers of a vertex factor will also appear $l!$ times in the partitioning.  As such, the sum over partitions will result in the cancellation of the factor of $l!$ for vertex factor $V_{pq}$.  All such terms from a vertex factor $V_{pq}$ with $p$ outgoing lines will generate $p!$ copies of the same the same term which will cancel the factor of $p!$, justifying our definition. Each vertex factor also carries a sum over neuron indices $i$ and an integral over internal time variable which must be performed to compute the moment; these are the sums and integrals we suppressed in Eq. ~\eqref{eq:actionVertex}.  


\paragraph{}
Thus, in order to compute the terms in the expansion for a moment 1) each factor of $n$ or $\tilde{n}$ gets an external vertex, 2) every graph is formed using the vertices associated with the vertex factors $V_{nm}$ by constructing the available partitions with all possible vertices,  3) For each vertex, contribute a factor of $V_{nm}$, 4) for each line contribute a factor of $\Delta_{ij}$, 5) contribute an operation $g \ast$ for each wavy line (operating on the term associated with the attached incoming line) and finally 6)  all integrals and sums are performed.  Note that some of these terms will produce disconnected graphs.  These correspond to factorizable terms in the moment.  

\paragraph{}
The rules derived using the action above will produce the initial expansions that we demonstrated about the $n=0$ configuration.  The ``resummed'' rules that we present in the Results arise from first performing a slight change of variables in the integrand.  Instead of considering the fluctuations about $n(t)=0$, we shift the configuration and response variables by their mean field solutions.  Defining $\bar{r}_i(t) = \langle \dot{n}_i(t) \rangle_0$ and $\tilde{r}_i(t) = \langle \tilde{n}_i(t) \rangle_0$, these are determined by
\begin{align}
	 \tilde{r}_i(t)  &= 0 \nonumber \\
     \bar{r}_i(t)  &= \phi \left ( \sum_j (\mathbf{g}_{ij}\ast  \bar{r}_j)(t)  + \lambda_i(t) \right ) 
\end{align}
We shift by these solutions by defining
\begin{align}
	\delta \dot{n}_i(t) = \dot{n}_i(t) - \bar{r}_i(t)
\end{align}
This leaves us with the action
\begin{align}
S[\tilde{n}_i, n_i] =\sum_i \int dt \big(\tilde{n}_i\delta \dot{n}_i - \left(e^{\tilde{n}_i (t)}-1\right)\phi \left( \sum_j (\mathbf{g}_{ij} \ast \left (\delta \dot{n}_j + \bar{r}_j \right ))(t)  + \lambda_i(t) \right)  \big) + \tilde{n}_i(t) \bar{r}_i(t)
\end{align}
\paragraph{}
Now we can develop the rules for the expansion we provide in the text using the same procedure outlined above.   The only difference is that $\Delta_{ij}(t,t')$ will be replaced by the linear response around mean-field theory and the vertex factors will be determined by an expansion around the mean field solution.  The rules otherwise remain the same.  The rules so derived are shown in Figure~{\ref{fig:networkPoissonFeynman}}.  An expansion around the true mean $\left \langle \dot{n}(t) \right \rangle$ would lead to the ``effective action", the expansion of which gives rise to the proper vertex factors definiing the different orders of stability correction.

\paragraph{}
Counting powers of the vertex factors allows one to compute a ``weak coupling" expansion.  Alternatively, the fluctuation expansion is determined by the topology of graphs and is equivalent to a steepest descent evaluation of the path integral.  This allows us to truncate according to the number of loops in the associated graphs and is the approach we use in this paper.  The approach here is a standard device in field theory and can be found many texts, for one example see \cite{zinn-justin_quantum_2002}.

\begin{table}[!ht] \label{table:params}
\caption{
\bf{Model parameters (unless otherwise specified in text)}}
\begin{tabular}{|l|c|r|}
\hline
Parameter & Description & Value \\ \hline
$\phi(x)$ & Single-neuron transfer function & $\alpha \lfloor x \rfloor^p$ \\ \hline
$\alpha$ & Gain of single neuron transfer & .1 ms$^{-1}$mV$^{-p}$ \\ \hline
$\lambda(t)$ & Baseline drive & .1 mV \\ \hline
$N_E$ & Number of excitatory neurons & 200 \\ \hline
$N_I$ & Number of inhibitory neurons & 40 \\ \hline
$p_{EE}$ & Excitatory-excitatory connection probability & 0.2 \\ \hline
$p_{EI}$ & Inhibitory-excitatory connection probability & 0.5 \\ \hline
$p_{IE}$ & Excitatory-excitatory connection probability & 0.5 \\ \hline
$p_{II}$ & Excitatory-excitatory connection probability & 0.5 \\ \hline
$\tau$ & Time constant for postsynaptic potentials & 10 ms \\ \hline
$g(t)$ & Shape of postsynaptic potentatials & $(t / \tau^2)\exp{(-t/\tau)}$ \\ \hline
$L_1$ & Left cutoff for power-law degree distribution & 0 \\ \hline
$L_2$ & Right cutoff for power-lay degree distribution & $N_E$ \\ \hline
$\gamma_1$ & Rising exponent for power-law degree distribution & .8 \\ \hline
$\gamma_2$ & Falling exponent for power-law degree distribution & -1.5\\ \hline
$\rho$ & Correlation of Gaussian copula for degree distributions & .8 \\ \hline
\end{tabular}
\begin{flushleft}
\end{flushleft}
\label{tab:params}
 \end{table}



\section*{Correction (2020)} \label{correction}
\paragraph{}
We were recently made aware of a mistake in this article, ``Linking structure and activity in nonlinear spiking network" \cite{ocker_linking_2017}. In Figure 13 we constructed the one-loop correction to the two-point correlation for multivariate generalized linear point process models (also called nonlinear Hawkes processes). Drs. Kordovan and Rotter recently pointed out that we neglected several contributions to that loop correction \cite{kordovan_spike_2020}. 

\paragraph{}
The Feynman diagrams corresponding to these contributions for networks with threshold-quadratic transfer functions are shown in Fig. \ref{fig:2point_1loop_missing_diagrams}. In the networks we studied, these corrections provide small contributions (Fig. \ref{fig:2point_corrected_quadratic}), which can be expected since they are of at least third order in the coupling strength.

\paragraph{}
If the transfer function has higher than second order derivatives, there is an additional contribution to the one-loop correction for the two-point correlation (\cite{kordovan_spike_2020}; Fig. \ref{fig:2point_1loop_missing_diagrams_3}). For the networks with exponential transfer functions, the additional contributions to the one-loop correction for the two-point function were also small (Fig. \ref{fig:2point_corrected_exponential}). Dr. Todorov also pointed out that our simulation code for the networks with exponential transfer functions calculated firing rates using the full simulation length (even when rates diverged before the simulation finished); that bug correction improves the match between theory and simulation for the networks with exponential transfer functions (Fig. \ref{fig:2point_corrected_exponential} vs Figs. \ref{fig:exponential_balanced}, \ref{fig:exponential_unbalanced}).

\paragraph{}
Kordovan \& Rotter also suggested an alternative scheme for constructing diagrams at a given loop order by starting from the loops rather than from external vertices, and discuss the possibility of automated algebraic programs for constructing and solving for loop corrections, comparable to systems that exist for particle physics. We agree that this would be quite useful for the future application of these tools to neuroscience models.

\begin{figure}[ht!]
\begin{adjustwidth}{-1 in}{0in}
\begin{center}
\begin{tikzpicture}
	
	\node[white](1) at (-8, 0) {$t_1$};
	\node[white](2) at (-8, 3) {$t_2$};
	\node[white](3) at (-7, .5) {};
	\node[white](4) at (-7, 2.5) {};
	\node[white](5) at (-7, 1.5) {};
	
	\node[white](6) at (-6, 1) {};
	\node[white](7) at (-6, 2.) {};
	\node[black](8) at (-5, 1.5) {};
		
	\draw[directed] (3)--(1);
	\draw[directed] (4)--(2);
	\draw[synapse] (5)--(4);
	\draw[directed] (3)--(5);
	\draw[synapse] (6)--(3);
	\draw[synapse] (7)--(4);
	\draw[directed] (8)--(6);
	\draw[directed] (8)--(7);
	
	\node[math]() at (-6.5, -1) {$+t_1 \leftrightarrow t_2$};
	
	\node[white](1) at (-3.5, .5) {$t_1$};
	\node[white](2) at (-3.5, 2.5) {$t_2$};
	
	\node[white](3) at (-2.5, 1.5) {};
	\node[white](4) at (-1.5, 1.5) {};
	\node[white](5) at (-.5, 1.5) {};
	
	\node[white](6) at (.5, 2.5) {};
	\node[white](7) at (.5, .5) {};
	\node[black](8) at (1.5, 1.5) {};
	
	\draw[directed] (3)--(2);
	\draw[directed] (3)--(1);
	\draw[synapse] (4)--(3);
	\draw[directed] (5)--(4);
	
	\draw[synapse] (6)--(5);
	\draw[synapse] (7)--(5);
	\draw[directed] (8)--(7);
	\draw[directed] (8)--(6);

	\node[white](1) at (4, 3) {$t_1$};
	\node[white](2) at (4, 0) {$t_2$};
	
	\node[white](3) at (5, 2.5) {};
	\node[white](4) at (5, .5) {};

	\node[white](5) at (6, 2.5) {};
	\node[white](6) at (6, .5) {};

	\node[black](7) at (7.5, 1.5) {};

	\draw[directed] (7)--(6);
	\draw[directed] (7)--(5);
	\draw[synapse] (6)--(4);
	\draw[directed] (4)--(2);
	\draw[synapse] (5)--(3);
	\draw[directed] (3)--(1);

	\node[white](8) at (5.5, 1) {};
	\node[white](9) at (5.5, 2) {};
	\node[black](10) at (6.5, 1.5) {};

	\draw[directed] (10)--(9);
	\draw[directed] (10)--(8);
	\draw[synapse] (9)--(3);
	\draw[synapse] (8)--(4);

	\node[white](1) at (-7, -3) {$t_1$};
	\node[white](2) at (-7, -6.5) {$t_1$};
	\node[white](3) at (-6, -6) {};
	\node[white](4) at (-5, -5.5) {};
	\node[white](5) at (-4, -5) {};
	\node[white](6) at (-3, -4.5) {};
	
	\node[white](7) at (-4, -6) {};
	\node[white](8) at (-5, -6.5) {};
	
	\node[black](9) at (-2, -4) {};
	\node[black](10) at (-4, -7) {};
	
	\draw[directed] (9)--(1);
	\draw[directed] (9)--(6);
	\draw[synapse] (6)--(5);
	\draw[directed] (5)--(4); 
	\draw[synapse] (4)--(3);
	\draw[directed] (3)--(2);
	
	\draw[directed] (10)--(8);
	\draw[directed] (10)--(7);
	\draw[synapse] (8)--(3);
	\draw[synapse] (7)--(5);
	
	\node[math] at (-5, -8) {$+t_1 \leftrightarrow t_2$};

	\node[white](1) at (0, -3) {$t_1$};	
	\node[white](2) at (0, -6.5) {$t_2$};

	\node[white](3) at (1, -4) {};
	\node[white](4) at (2, -5) {};
	
	\node[black](5) at (3, -6) {};
	
	\node[white](6) at (4, -3.5) {};
	\node[white](7) at (4, -4.5) {};
	\node[white](8) at (3, -4) {};
	\node[white](9) at (2, -4) {};
	
	\node[black](10) at (5, -4) {};

	\draw[directed] (10)--(6);
	\draw[directed] (10)--(7);
	\draw[synapse] (7)--(8);
	\draw[synapse] (6)--(8);
	\draw[directed] (8)--(9);
	\draw[synapse] (9)--(3);
	
	\draw[directed] (3)--(1);
	\draw[synapse] (4)--(3);
	\draw[directed] (5)--(4);
	\draw[directed] (5)--(2);

	\node[math] at (1, -8) {$+t_1 \leftrightarrow t_2$};

\end{tikzpicture}
\end{center}
\caption{\label{fig:2point_1loop_missing_diagrams}
Additional contributions to the one-loop correction to the two-point correlation with a quadratic transfer function.}
\end{adjustwidth}
\end{figure}  

\begin{figure}[ht!]
\begin{adjustwidth}{-2 in}{0 in}
\includegraphics{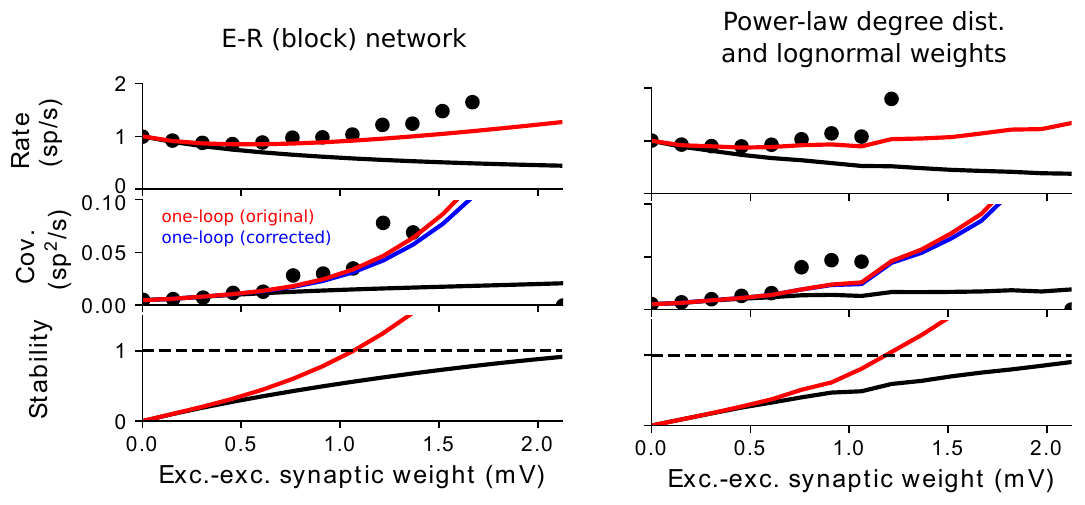}
\caption{ \label{fig:2point_corrected_quadratic}
Corrected one-loop approximation of the two-point correlation for the excitatory-inhibitory networks with threshold-quadratic transfer functions of Fig. 15 (left) and Fig. 17 (right) in \cite{ocker_linking_2017}.}
\end{adjustwidth}
\end{figure}

\begin{figure}[ht!]
\begin{adjustwidth}{-1 in}{0in}
\begin{center}
\begin{tikzpicture}

	\node[white](1) at (0, -3) {$t_1$};	
	\node[white](2) at (0, -6.5) {$t_2$};

	\node[white](3) at (1, -4) {};
	\node[white](4) at (2, -5) {};
	
	\node[black](5) at (3, -6) {};
	
	\node[white](6) at (2, -3) {};
	\node[white](7) at (2, -4) {};	
	\node[black](10) at (3, -3.5) {};

	\draw[directed] (10)--(6);
	\draw[directed] (10)--(7);

	\draw[synapse] (6)--(3);
	\draw[synapse] (7)--(3);

	\draw[directed] (3)--(1);
	\draw[synapse] (4)--(3);
	\draw[directed] (5)--(4);
	\draw[directed] (5)--(2);

	\node[math] at (5, -5) {$+ t_1 \leftrightarrow t_2$};

\end{tikzpicture}
\end{center}
\caption{\label{fig:2point_1loop_missing_diagrams_3}
Additional diagrams for the one-loop correction to the two-point correlation with higher-order transfer functions.}
\end{adjustwidth}
\end{figure}  

\begin{figure}[ht!]
\begin{adjustwidth}{-2 in}{0 in}
\includegraphics{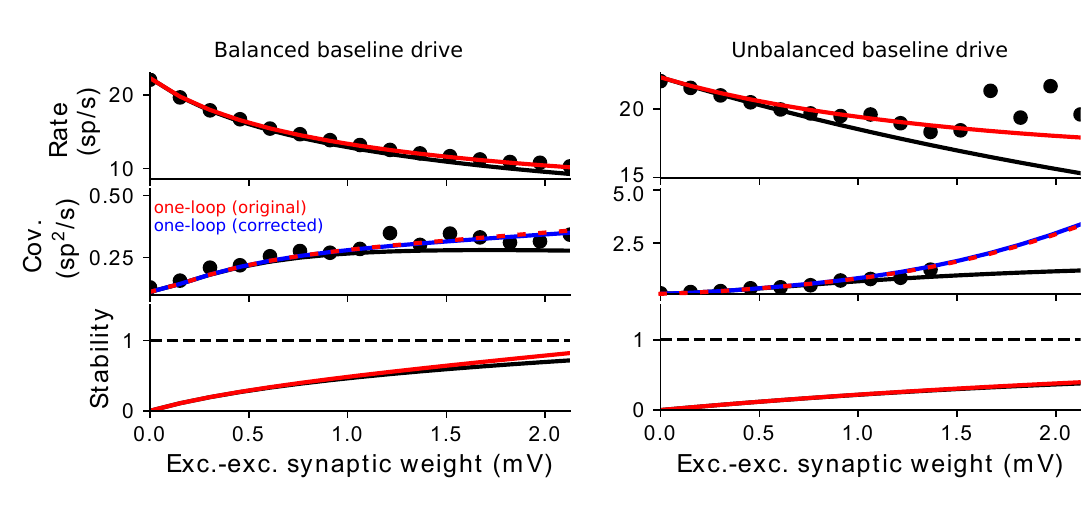}
\caption{ \label{fig:2point_corrected_exponential}
Corrected one-loop approximation of the two-point correlation for the excitatory-inhibitory networks with exponential transfer functions of Fig. 18 (left) and Fig. 19 (right) in \cite{ocker_linking_2017}.}
\end{adjustwidth}
\end{figure}

\section*{Acknowledgments}
We thank Brent Doiron and Robert Rosenbaum for helpful comments on the manuscript. We acknowledge the support of the NSF through grants DMS-1514743 and DMS-1056125 (ESB), DMS-1517629 (KJ),  and NSF/NIGMS-R01GM104974 (KJ), as well as the Simons Fellowships in Mathematics (ESB and KJ).  The authors wish to thank the Allen Institute founders, Paul G. Allen and Jody Allen, for their vision, encouragement and support.

\nolinenumbers

%
%
%
\bibstyle{plos2015.bst}
\bibliography{MyLibrary}

\begin{thebibliography}{10}

\bibitem{sejnowski_storing_1977}
Sejnowski TJ.
\newblock Storing covariance with nonlinearly interacting neurons.
\newblock Journal of Mathematical Biology. 1977;4(4):303--321.
\newblock doi:{10.1007/BF00275079}.

\bibitem{bienenstock_theory_1982}
Bienenstock EL, Cooper LN, Munro PW.
\newblock Theory for the development of neuron selectivity: orientation
  specificity and binocular interaction in visual cortex.
\newblock The Journal of Neuroscience. 1982;2(1):32--48.

\bibitem{gerstner_neuronal_1996}
Gerstner W, Kempter R, van Hemmen JL, Wagner H.
\newblock A neuronal learning rule for sub-millisecond temporal coding.
\newblock Nature. 1996;383(6595):76--78.
\newblock doi:{10.1038/383076a0}.

\bibitem{pfister_triplets_2006}
Pfister JP, Gerstner W.
\newblock Triplets of {Spikes} in a {Model} of {Spike} {Timing}-{Dependent}
  {Plasticity}.
\newblock The Journal of Neuroscience. 2006;26(38):9673--9682.
\newblock doi:{10.1523/JNEUROSCI.1425-06.2006}.

\bibitem{ocker_self-organization_2015}
Ocker GK, Litwin-Kumar A, Doiron B.
\newblock Self-{Organization} of {Microcircuits} in {Networks} of {Spiking}
  {Neurons} with {Plastic} {Synapses}.
\newblock PLoS Comput Biol. 2015;11(8):e1004458.
\newblock doi:{10.1371/journal.pcbi.1004458}.

\bibitem{tannenbaum_shaping_2016}
Tannenbaum NR, Burak Y.
\newblock Shaping {Neural} {Circuits} by {High} {Order} {Synaptic}
  {Interactions}.
\newblock PLOS Comput Biol. 2016;12(8):e1005056.
\newblock doi:{10.1371/journal.pcbi.1005056}.

\bibitem{abeles_role_1982}
Abeles M.
\newblock Role of the cortical neuron: integrator or coincidence detector?
\newblock Israel Journal of Medical Sciences. 1982;18(1):83--92.

\bibitem{usrey_paired-spike_1998}
Usrey WM, Reppas JB, Reid RC.
\newblock Paired-spike interactions and synaptic efficacy of retinal inputs to
  the thalamus.
\newblock Nature. 1998;395(6700):384--387.
\newblock doi:{10.1038/26487}.

\bibitem{bruno_cortex_2006}
Bruno RM, Sakmann B.
\newblock Cortex is driven by weak but synchronously active thalamocortical
  synapses.
\newblock Science (New York, NY). 2006;312(5780):1622--1627.
\newblock doi:{10.1126/science.1124593}.

\bibitem{histed_cortical_2014}
Histed MH, Maunsell JHR.
\newblock Cortical neural populations can guide behavior by integrating inputs
  linearly, independent of synchrony.
\newblock Proceedings of the National Academy of Sciences.
  2014;111(1):E178--E187.
\newblock doi:{10.1073/pnas.1318750111}.

\bibitem{salinas_impact_2000}
Salinas E, Sejnowski TJ.
\newblock Impact of correlated synaptic input on output firing rate and
  variability in simple neuronal models.
\newblock The Journal of Neuroscience: The Official Journal of the Society for
  Neuroscience. 2000;20(16):6193--6209.

\bibitem{averbeck_neural_2006}
Averbeck BB, Latham PE, Pouget A.
\newblock Neural correlations, population coding and computation.
\newblock Nature Reviews Neuroscience. 2006;7(5):358--366.
\newblock doi:{10.1038/nrn1888}.

\bibitem{panzeri_neural_2015}
Panzeri S, Macke JH, Gross J, Kayser C.
\newblock Neural population coding: combining insights from microscopic and
  mass signals.
\newblock Trends in Cognitive Sciences. 2015;19(3):162--172.
\newblock doi:{10.1016/j.tics.2015.01.002}.

\bibitem{series_tuning_2004}
Seriès P, Latham PE, Pouget A.
\newblock Tuning curve sharpening for orientation selectivity: coding
  efficiency and the impact of correlations.
\newblock Nature Neuroscience. 2004;7(10):1129--1135.
\newblock doi:{10.1038/nn1321}.

\bibitem{josic_stimulus-dependent_2009}
Josić K, Shea-Brown E, Doiron B, de~la Rocha J.
\newblock Stimulus-dependent correlations and population codes.
\newblock Neural Comp. 2009;21:2774--2804.

\bibitem{moreno-bote_information-limiting_2014}
Moreno-Bote R, Beck J, Kanitscheider I, Pitkow X, Latham P, Pouget A.
\newblock Information-limiting correlations.
\newblock Nature Neuroscience. 2014;17(10):1410--1417.
\newblock doi:{10.1038/nn.3807}.

\bibitem{zylberberg_direction-selective_2016}
Zylberberg J, Cafaro J, Turner MH, Shea-Brown E, Rieke F.
\newblock Direction-{Selective} {Circuits} {Shape} {Noise} to {Ensure} a
  {Precise} {Population} {Code}.
\newblock Neuron. 2016;89(2):369--383.
\newblock doi:{10.1016/j.neuron.2015.11.019}.

\bibitem{franke_structures_2016}
Franke F, Fiscella M, Sevelev M, Roska B, Hierlemann A, Azeredo da Silveira
  R.
\newblock Structures of {Neural} {Correlation} and {How} {They} {Favor}
  {Coding}.
\newblock Neuron. 2016;89(2):409--422.
\newblock doi:{10.1016/j.neuron.2015.12.037}.

\bibitem{hawkes_spectra_1971}
Hawkes AG.
\newblock Spectra of some self-exciting and mutually exciting point processes.
\newblock Biometrika. 1971;58(1):83--90.
\newblock doi:{10.1093/biomet/58.1.83}.

\bibitem{brillinger_estimation_1976}
Brillinger D.
\newblock Estimation of the second-order intensities of a bivariate stationary
  point process.
\newblock Journal of the Royal Statistical Society Series B (Methodological).
  1976;38(1):60--66.

\bibitem{doiron_oscillatory_2004}
Doiron B, Lindner B, Longtin A, Maler L, Bastian J.
\newblock Oscillatory activity in electrosensory neurons increases with the
  spatial correlation of the stochastic input stimulus.
\newblock Phys Rev Let. 2004;93(4).

\bibitem{trousdale_impact_2012}
Trousdale J, Hu Y, Shea-Brown E, Josić K.
\newblock Impact of {Network} {Structure} and {Cellular} {Response} on {Spike}
  {Time} {Correlations}.
\newblock PLoS Computational Biology. 2012;8(3):e1002408.
\newblock doi:{10.1371/journal.pcbi.1002408}.

\bibitem{ginzburg_theory_1994}
Ginzburg I, Sompolinsky H.
\newblock Theory of correlations in stochastic neural networks.
\newblock Physical Review E. 1994;50(4):3171--3191.
\newblock doi:{10.1103/PhysRevE.50.3171}.

\bibitem{brunel_dynamics_2000}
Brunel N.
\newblock Dynamics of {Sparsely} {Connected} {Networks} of {Excitatory} and
  {Inhibitory} {Spiking} {Neurons}.
\newblock Journal of Computational Neuroscience. 2000;8(3):183--208.
\newblock doi:{10.1023/A:1008925309027}.

\bibitem{mattia_population_2002}
Mattia M, Del~Giudice P.
\newblock Population dynamics of interacting spiking neurons.
\newblock Physical Review E. 2002;66(5):051917.
\newblock doi:{10.1103/PhysRevE.66.051917}.

\bibitem{kass_analysis_2014}
Kass RE, Eden U, Brown E.
\newblock Analysis of neural data.
\newblock 1st ed. Springer series in statistics. Springer-Verlag; 2014.

\bibitem{doiron_mechanics_2016}
Doiron B, Litwin-Kumar A, Rosenbaum R, Ocker GK, Josić K.
\newblock The mechanics of state-dependent neural correlations.
\newblock Nature Neuroscience. 2016;19(3):383--393.
\newblock doi:{10.1038/nn.4242}.

\bibitem{abeles_corticonics:_1991}
Abeles M.
\newblock Corticonics: {Neural} circuits of the cerebral cortex.
\newblock Cambridge University Press; 1991.

\bibitem{diesmann_stable_1999}
Diesmann M, Gewaltig MO, Aertsen A.
\newblock Stable propagation of synchronous spiking in cortical neural
  networks.
\newblock Nature. 1999;402(6761):529--533.
\newblock doi:{10.1038/990101}.

\bibitem{reyes_synchrony-dependent_2003}
Reyes AD.
\newblock Synchrony-dependent propagation of firing rate in iteratively
  constructed networks in vitro.
\newblock Nature Neuroscience. 2003;6(6):593--599.
\newblock doi:{10.1038/nn1056}.

\bibitem{kumar_spiking_2010}
Kumar A, Rotter S, Aertsen A.
\newblock Spiking activity propagation in neuronal networks: reconciling
  different perspectives on neural coding.
\newblock Nat Rev Neurosci. 2010;11:615--627.

\bibitem{bremaud_stability_1996}
Brémaud P, Massoulié L.
\newblock Stability of nonlinear {Hawkes} processes.
\newblock The Annals of Probability. 1996;24(3):1563--1588.
\newblock doi:{10.1214/aop/1065725193}.

\bibitem{gerstner_neuronal_2014}
Gerstner W, Kistler WM, Naud R, Paninski L.
\newblock Neuronal {Dynamics}: {From} {Single} {Neurons} to {Networks} and
  {Models} of {Cognition}.
\newblock Cambridge University Press; 2014.

\bibitem{tsodyks_paradoxical_1997}
Tsodyks MV, Skaggs WE, Sejnowski TJ, McNaughton BL.
\newblock Paradoxical {Effects} of {External} {Modulation} of {Inhibitory}
  {Interneurons}.
\newblock The Journal of Neuroscience. 1997;17(11):4382--4388.

\bibitem{risken_fokker-planck_1996}
Risken H.
\newblock The {Fokker}-{Planck} equation: methods of solution and applications.
\newblock 3rd ed. Haken H, editor. Springer-Verlag; 1996.

\bibitem{hawkes_cluster_1974}
Hawkes AG, Oakes D.
\newblock A cluster process representation of a self-exciting process.
\newblock Journal of Applied Probability. 1974;11:493--503.

\bibitem{saichev_generating_2011}
Saichev AI, Sornette D.
\newblock Generating functions and stability study of multivariate self-excited
  epidemic processes.
\newblock The European Physical Journal B. 2011;83(2):271--282.
\newblock doi:{10.1140/epjb/e2011-20298-3}.

\bibitem{jovanovic_cumulants_2015}
Jovanović S, Hertz J, Rotter S.
\newblock Cumulants of {Hawkes} point processes.
\newblock Physical Review E. 2015;91(4):042802.
\newblock doi:{10.1103/PhysRevE.91.042802}.

\bibitem{zinn-justin_quantum_2002}
Zinn-Justin J.
\newblock Quantum {Field} {Theory} and {Critical} {Phenomena}.
\newblock Clarendon Press; 2002.

\bibitem{buice_field-theoretic_2007}
Buice MA, Cowan JD.
\newblock Field-theoretic approach to fluctuation effects in neural networks.
\newblock Physical Review E. 2007;75(5):051919.
\newblock doi:{10.1103/PhysRevE.75.051919}.

\bibitem{pernice_how_2011}
Pernice V, Staude B, Cardanobile S, Rotter S.
\newblock How structure determines correlations in neuronal networks.
\newblock PLoS Comput Biol. 2011;7(5):e1002059.
\newblock doi:{10.1371/journal.pcbi.1002059}.

\bibitem{ostojic_how_2009}
Ostojic S, Brunel N, Hakim V.
\newblock How connectivity, background activity, and synaptic properties shape
  the cross-correlation between spike trains.
\newblock J Neurosci. 2009;29(333):10234--10253.

\bibitem{heeger_normalization_1992}
Heeger DJ.
\newblock Normalization of cell responses in cat striate cortex.
\newblock Visual Neuroscience. 1992;9(2):181--197.

\bibitem{miller_kenneth_d._neural_2002}
Miller KD, Troyer TW.
\newblock Neural noise can explaine expansive, power-law nonlinearities in
  neural response functions.
\newblock J Neurophysiol. 2002;87(2):653--659.

\bibitem{priebe_contribution_2004}
Priebe NJ, Mechler F, Carandini M, Ferster D.
\newblock The contribution of spike threshold to the dichotomy of cortical
  simple and complex cells.
\newblock Nature Neuroscience. 2004;7(10):1113--1122.
\newblock doi:{10.1038/nn1310}.

\bibitem{priebe_direction_2005}
Priebe NJ, Ferster D.
\newblock Direction {Selectivity} of {Excitation} and {Inhibition} in {Simple}
  {Cells} of the {Cat} {Primary} {Visual} {Cortex}.
\newblock Neuron. 2005;45(1):133--145.
\newblock doi:{10.1016/j.neuron.2004.12.024}.

\bibitem{priebe_mechanisms_2006}
Priebe NJ, Ferster D.
\newblock Mechanisms underlying cross-orientation suppression in cat visual
  cortex.
\newblock Nature Neuroscience. 2006;9(4):552--561.
\newblock doi:{10.1038/nn1660}.

\bibitem{reid_r._clay_divergence_2001}
Reid RC.
\newblock Divergence and reconvergence: multielectrode analysis of feedforward
  connections in the visual system.
\newblock In: Progress in {Brain} {Research}. vol. 130. Elsevier; 2001. p.
  141--154.

\bibitem{song_highly_2005}
Song S, Sjöström PJ, Reigl M, Nelson S, Chklovskii DB.
\newblock Highly nonrandom features of synaptic connectivity in local cortical
  circuits.
\newblock PLoS Biol. 2005;3(3):e68.
\newblock doi:{10.1371/journal.pbio.0030068}.

\bibitem{perin_synaptic_2011}
Perin R, Berger TK, Markram H.
\newblock A synaptic organizing principle for cortical neuronal groups.
\newblock Proceedings of the National Academy of Sciences.
  2011;108(13):5419--5424.
\newblock doi:{10.1073/pnas.1016051108}.

\bibitem{lee_anatomy_2016}
Lee WCA, Bonin V, Reed M, Graham BJ, Hood G, Glattfelder K, et~al.
\newblock Anatomy and function of an excitatory network in the visual cortex.
\newblock Nature. 2016;532(7599):370--374.
\newblock doi:{10.1038/nature17192}.

\bibitem{yoshimura_excitatory_2005}
Yoshimura Y, Dantzker JLM, Callaway EM.
\newblock Excitatory cortical neurons form fine-scale functional networks.
\newblock Nature. 2005;433(7028):868--873.
\newblock doi:{10.1038/nature03252}.

\bibitem{yoshimura_fine-scale_2005}
Yoshimura Y, Callaway EM.
\newblock Fine-scale specificity of cortical networks depends on inhibitory
  cell type and connectivity.
\newblock Nature Neuroscience. 2005;8(11):1552--1559.
\newblock doi:{10.1038/nn1565}.

\bibitem{morgenstern_multilaminar_2016}
Morgenstern NA, Bourg J, Petreanu L.
\newblock Multilaminar networks of cortical neurons integrate common inputs
  from sensory thalamus.
\newblock Nature Neuroscience. 2016;19(8):1034--1040.
\newblock doi:{10.1038/nn.4339}.

\bibitem{iyer_influence_2013}
Iyer R, Menon V, Buice M, Koch C, Mihalas S.
\newblock The {Influence} of {Synaptic} {Weight} {Distribution} on {Neuronal}
  {Population} {Dynamics}.
\newblock PLOS Comput Biol. 2013;9(10):e1003248.
\newblock doi:{10.1371/journal.pcbi.1003248}.

\bibitem{hu_motif_2013}
Hu Y, Trousdale J, Josić K, Shea-Brown E.
\newblock Motif statistics and spike correlations in neuronal networks.
\newblock Journal of Statistical Mechanics: Theory and Experiment.
  2013;2013(03):P03012.
\newblock doi:{10.1088/1742-5468/2013/03/P03012}.

\bibitem{hu_local_2014}
Hu Y, Trousdale J, Josić K, Shea-Brown E.
\newblock Local paths to global coherence: {Cutting} networks down to size.
\newblock Physical Review E. 2014;89(3):032802.
\newblock doi:{10.1103/PhysRevE.89.032802}.

\bibitem{roxin_role_2011}
Roxin A.
\newblock The role of degree distribution in shaping the dynamics in networks
  of sparsely connected spiking neurons.
\newblock Frontiers in Computational Neuroscience. 2011;5:8.
\newblock doi:{10.3389/fncom.2011.00008}.

\bibitem{zhao_synchronization_2011}
Zhao L, Beverlin BI, Netoff T, Nykamp DQ.
\newblock Synchronization from second order network connectivity statistics.
\newblock Frontiers in Computational Neuroscience. 2011;5:28.
\newblock doi:{10.3389/fncom.2011.00028}.

\bibitem{cossell_functional_2015}
Cossell L, Iacaruso MF, Muir DR, Houlton R, Sader EN, Ko H, et~al.
\newblock Functional organization of excitatory synaptic strength in primary
  visual cortex.
\newblock Nature. 2015;518(7539):399--403.
\newblock doi:{10.1038/nature14182}.

\bibitem{jovanovic_interplay_2016}
Jovanović S, Rotter S.
\newblock Interplay between {Graph} {Topology} and {Correlations} of {Third}
  {Order} in {Spiking} {Neuronal} {Networks}.
\newblock PLOS Comput Biol. 2016;12(6):e1004963.
\newblock doi:{10.1371/journal.pcbi.1004963}.

\bibitem{van_vreeswijk_chaotic_1998}
van Vreeswijk C, Sompolinsky H.
\newblock Chaotic balanced state in a model of cortical circuits.
\newblock Neural Computation. 1998;10(6):1321--1371.

\bibitem{sompolinsky_chaos_1988}
Sompolinsky H, Crisanti A, Sommers HJ.
\newblock Chaos in {Random} {Neural} {Networks}.
\newblock Physical Review Letters. 1988;61(3):259--262.
\newblock doi:{10.1103/PhysRevLett.61.259}.

\bibitem{kadmon_transition_2015}
Kadmon J, Sompolinsky H.
\newblock Transition to {Chaos} in {Random} {Neuronal} {Networks}.
\newblock Physical Review X. 2015;5(4):041030.
\newblock doi:{10.1103/PhysRevX.5.041030}.

\bibitem{harish_asynchronous_2015}
Harish O, Hansel D.
\newblock Asynchronous {Rate} {Chaos} in {Spiking} {Neuronal} {Circuits}.
\newblock PLOS Comput Biol. 2015;11(7):e1004266.
\newblock doi:{10.1371/journal.pcbi.1004266}.

\bibitem{mastrogiuseppe_intrinsically-generated_2016}
Mastrogiuseppe F, Ostojic S.
\newblock Intrinsically-generated fluctuating activity in excitatory-inhibitory
  networks.
\newblock arXiv:160504221 [q-bio]. 2016;.

\bibitem{aljadeff_transition_2015}
Aljadeff J, Stern M, Sharpee T.
\newblock Transition to {Chaos} in {Random} {Networks} with
  {Cell}-{Type}-{Specific} {Connectivity}.
\newblock Physical Review Letters. 2015;114(8):088101.
\newblock doi:{10.1103/PhysRevLett.114.088101}.

\bibitem{aljadeff_low-dimensional_2016}
Aljadeff J, Renfrew D, Vegué M, Sharpee TO.
\newblock Low-dimensional dynamics of structured random networks.
\newblock Physical Review E. 2016;93(2):022302.
\newblock doi:{10.1103/PhysRevE.93.022302}.

\bibitem{goedeke_noise_2016}
Goedeke S, Schuecker J, Helias M.
\newblock Noise dynamically suppresses chaos in random neural networks.
\newblock arXiv:160301880 [nlin, q-bio]. 2016;.

\bibitem{buice_beyond_2013}
Buice MA, Chow CC.
\newblock Beyond mean field theory: statistical field theory for neural
  networks.
\newblock Journal of Statistical Mechanics (Online). 2013;2013:P03003.
\newblock doi:{10.1088/1742-5468/2013/03/P03003}.

\bibitem{bressloff_path-integral_2015}
Bressloff PC.
\newblock Path-{Integral} {Methods} for {Analyzing} the {Effects} of
  {Fluctuations} in {Stochastic} {Hybrid} {Neural} {Networks}.
\newblock The Journal of Mathematical Neuroscience. 2015;5(1).
\newblock doi:{10.1186/s13408-014-0016-z}.

\bibitem{ohira_master-equation_1993}
Ohira T, Cowan JD.
\newblock Master-equation approach to stochastic neurodynamics.
\newblock Physical Review E. 1993;48(3):2259--2266.
\newblock doi:{10.1103/PhysRevE.48.2259}.

\bibitem{bressloff_stochastic_2009}
Bressloff P.
\newblock Stochastic {Neural} {Field} {Theory} and the {System}-{Size}
  {Expansion}.
\newblock SIAM Journal on Applied Mathematics. 2009;70(5):1488--1521.
\newblock doi:{10.1137/090756971}.

\bibitem{buice_systematic_2010}
Buice MA, Chow CC, Cowan JD.
\newblock Systematic fluctuation expansion for neural network activity
  equations.
\newblock Neural Comp. 2010;22(377-426).

\bibitem{de_la_rocha_correlation_2007}
de~la Rocha J, Doiron B, Shea-Brown E, Josić K, Reyes A.
\newblock Correlation between neural spike trains increases with firing rate.
\newblock Nature. 2007;448(7155):802--6.
\newblock doi:{10.1038/nature06028}.

\bibitem{fourcaud_dynamics_2002}
Fourcaud N, Brunel N.
\newblock Dynamics of the {Firing} {Probability} of {Noisy}
  {Integrate}-and-{Fire} {Neurons}.
\newblock Neural Computation. 2002;14(9):2057--2110.
\newblock doi:{10.1162/089976602320264015}.

\bibitem{ledoux_dynamics_2011}
Ledoux E, Brunel N.
\newblock Dynamics of networks of excitatory and inhibitory neurons in response
  to time-dependent inputs.
\newblock Frontiers in Computational Neuroscience. 2011;5:25.
\newblock doi:{10.3389/fncom.2011.00025}.

\bibitem{ostojic_two_2014}
Ostojic S.
\newblock Two types of asynchronous activity in networks of excitatory and
  inhibitory spiking neurons.
\newblock Nature Neuroscience. 2014;17(4):594--600.
\newblock doi:{10.1038/nn.3658}.

\bibitem{van_vreeswijk_chaos_1996}
van Vreeswijk C, Sompolinsky H.
\newblock Chaos in neuronal networks with balanced excitatory and inhibitory
  activity.
\newblock Science (New York, NY). 1996;274(5293):1724--1726.

\bibitem{renart_asynchronous_2010}
Renart A, Rocha Jdl, Bartho P, Hollender L, Parga N, Reyes A, et~al.
\newblock The {Asynchronous} {State} in {Cortical} {Circuits}.
\newblock Science. 2010;327(5965):587--590.
\newblock doi:{10.1126/science.1179850}.

\bibitem{ahmadian_analysis_2013}
Ahmadian Y, Rubin DB, Miller KD.
\newblock Analysis of the {Stabilized} {Supralinear} {Network}.
\newblock Neural Computation. 2013;25(8):1994--2037.
\newblock doi:{10.1162/NECOa00472}.

\bibitem{ozeki_inhibitory_2009}
Ozeki H, Finn IM, Schaffer ES, Miller KD, Ferster D.
\newblock Inhibitory {Stabilization} of the {Cortical} {Network} {Underlies}
  {Visual} {Surround} {Suppression}.
\newblock Neuron. 2009;62(4):578--592.
\newblock doi:{10.1016/j.neuron.2009.03.028}.

\bibitem{rubin_stabilized_2015}
Rubin D, Van Hooser S, Miller K.
\newblock The {Stabilized} {Supralinear} {Network}: {A} {Unifying} {Circuit}
  {Motif} {Underlying} {Multi}-{Input} {Integration} in {Sensory} {Cortex}.
\newblock Neuron. 2015;85(2):402--417.
\newblock doi:{10.1016/j.neuron.2014.12.026}.

\bibitem{litwin-kumar_inhibitory_2016}
Litwin-Kumar A, Rosenbaum R, Doiron B.
\newblock Inhibitory stabilization and visual coding in cortical circuits with
  multiple interneuron subtypes.
\newblock Journal of Neurophysiology. 2016;115(3):1399--1409.
\newblock doi:{10.1152/jn.00732.2015}.

\bibitem{hu_sign_2014}
Hu Y, Zylberberg J, Shea-Brown E.
\newblock The {Sign} {Rule} and {Beyond}: {Boundary} {Effects}, {Flexibility},
  and {Noise} {Correlations} in {Neural} {Population} {Codes}.
\newblock PLoS Comput Biol. 2014;10(2):e1003469.
\newblock doi:{10.1371/journal.pcbi.1003469}.

\bibitem{martin_statistical_1973}
Martin PC, Siggia ED, Rose HA.
\newblock Statistical {Dynamics} of {Classical} {Systems}.
\newblock Physical Review A. 1973;8(1):423--437.
\newblock doi:{10.1103/PhysRevA.8.423}.

\bibitem{doi_second_1976}
Doi M.
\newblock Second quantization representation for classical many-particle
  system.
\newblock Journal of Physics A: Mathematical and General. 1976;9(9):1465.
\newblock doi:{10.1088/0305-4470/9/9/008}.

\bibitem{doi_stochastic_1976}
Doi M.
\newblock Stochastic theory of diffusion-controlled reaction.
\newblock Journal of Physics A: Mathematical and General. 1976;9(9):1479.
\newblock doi:{10.1088/0305-4470/9/9/009}.

\bibitem{peliti_path_1985}
Peliti L.
\newblock Path integral approach to birth-death processes on a lattice.
\newblock Journal de Physique. 1985;46(9):1469--1483.
\newblock doi:{10.1051/jphys:019850046090146900}.

\bibitem{tauber_field-theory_2007}
Täuber UC.
\newblock Field-{Theory} {Approaches} to {Nonequilibrium} {Dynamics}.
\newblock In: Henkel M, Pleimling M, Sanctuary R, editors. Ageing and the
  {Glass} {Transition}. No. 716 in Lecture {Notes} in {Physics}. Springer
  Berlin Heidelberg; 2007. p. 295--348.
\newblock Available from:
  \url{http://link.springer.com/chapter/10.1007/3-540-69684-9_7}.

\bibitem{cardy_non-equilibrium_2008}
Cardy J, Falkovich G, Gawadzki K.
\newblock Non-equilibrium {Statistical} {Mechanics} and {Turbulence}.
\newblock Cambridge University Press; 2008.

\bibitem{lefevre_dynamics_2007}
Lefèvre A, Biroli G.
\newblock Dynamics of interacting particle systems: stochastic process and
  field theory.
\newblock Journal of Statistical Mechanics: Theory and Experiment.
  2007;2007(07):P07024.
\newblock doi:{10.1088/1742-5468/2007/07/P07024}.

\bibitem{chow_path_2015}
Chow CC, Buice MA.
\newblock Path {Integral} {Methods} for {Stochastic} {Differential}
  {Equations}.
\newblock The Journal of Mathematical Neuroscience (JMN). 2015;5(1):1--35.
\newblock doi:{10.1186/s13408-015-0018-5}.

\bibitem{ocker_linking_2017}
Ocker GK, Josić K, Shea-Brown E, Buice MA.
\newblock Linking structure and activity in nonlinear spiking networks.
\newblock PLOS Computational Biology. 2017;13(6):e1005583.
\newblock doi:{10.1371/journal.pcbi.1005583}.

\bibitem{kordovan_spike_2020}
Kordovan M, Rotter S.
\newblock Spike {Train} {Cumulants} for {Linear}-{Nonlinear} {Poisson}
  {Cascade} {Models}.
\newblock arXiv:200105057 [physics, q-bio, stat]. 2020;.

\end{thebibliography}

\end{document}